\documentclass[aps,prd,preprintnumbers,nofootinbib,superscriptaddress,10pt,onecolumn]{revtex4-2}

\usepackage[dvipsnames,table]{xcolor}
\usepackage{colortbl}

\usepackage{graphicx, amsmath, bm}  
\usepackage{placeins}
\usepackage{textcomp}
\usepackage{blindtext}
\usepackage{url}
\usepackage{epsfig}
\usepackage{makecell}
\usepackage{amsfonts}
\usepackage{mathtools}
\usepackage{hhline}
\usepackage{array,multirow}
\usepackage[utf8]{inputenc}

\usepackage{docs}%
\usepackage[colorlinks=true,linkcolor=blue]{hyperref}%
\usepackage{float}
\usepackage[caption=false]{subfig}
\usepackage{dcolumn}
\usepackage{amssymb}
\usepackage[toc,page]{appendix}
\usepackage{slashed}
\usepackage{upgreek} 
\usepackage[normalem]{ulem}
\usepackage{tikz}
\usepackage{tikz-feynman}
\usepackage{pifont}
\usetikzlibrary{snakes}
\expandafter\ifx\csname package@font\endcsname\relax\else
 \expandafter\expandafter
 \expandafter\usepackage
 \expandafter\expandafter
 \expandafter{\csname package@font\endcsname}%
\fi
\usepackage[shortlabels]{enumitem}
\usepackage{empheq}

\usepackage{adjustbox}
\usepackage{lipsum}
\usepackage{chngcntr}
\usepackage{bigints}
\usepackage{cleveref}

\tikzfeynmanset{compat=1.1.0}
\usetikzlibrary{arrows}
\tikzset{vertex/.style={circle,draw, minimum size=1.5em}, edge/.style={->,> = latex'}}

\usepackage{silence}
\tikzfeynmanset{warn luatex=false}

\begin{document}

\title{ Flavour and Electroweak Precision Constraints on a Simplified Dark Matter Model with a Light Spin-0 Mediator}

\author{Lipika Kolay}
\email[]{klipika@iitg.ac.in}
\affiliation{Department of Physics, Indian Institute of Technology Guwahati, Assam-781039, India}

\author{Soumitra Nandi}
\email[]{soumitra.nandi@iitg.ac.in}
\affiliation{Department of Physics, Indian Institute of Technology Guwahati, Assam-781039, India}


\begin{abstract} 
This work investigates the allowed parameter spaces of a simplified dark matter (DM) model characterized by a spin-0 mediator with masses in the low to intermediate range ($ < $ 10 GeV). We systematically divide the parameter space into various mass regions of the mediator and constrain the model parameters using a diverse set of observables, including flavour-changing charged and neutral current processes such as rare and semi-leptonic decays of pseudoscalar mesons (B and K), electroweak precision observables, alongside data from fixed-target experiments. Additionally, we explore the model's capability to explain recent Belle-II data on invisible B-meson decays. Our study includes a detailed examination of DM properties and the constraints from Big Bang nucleosynthesis. We present bounds on model parameters through individual and simultaneous analyses of the available inputs and highlight their implications for understanding DM phenomenology. Furthermore, we obtain bounds on the couplings of the possible gauge-invariant dimension-5 operators, leading to the possible interactions between the spin-0 mediator and the SM gauge bosons and fermions. This study comprehensively investigates the constraints and theoretical implications associated with low-mass spin-0 mediator DM models.
\end{abstract}
\maketitle
	
\section{Introduction}
The Standard Model (SM) of particle physics is successful in describing many phenomena of nature at the fundamental level. However, the model falls short in explaining certain key aspects of nature. It lacks a viable candidate for dark matter (DM) and struggles to accommodate the observed baryon asymmetry.
Furthermore, the model does not provide a mechanism for neutrino mass generation. To address these deficiencies, various extensions of the SM have been proposed. All these extensions introduce new degrees of freedom and interactions at higher energy scales and provide suitable DM candidates to explain the observed pattern in the data on DM searches and baryon asymmetry. 
The low-energy observables could play a crucial role in probing the new interactions beyond the SM (BSM), in such cases, we need to look for deviations between the measured and the predicted values (SM) of the relevant observables. 
In experimental searches for DM, it is convenient to study the signatures of this DM candidate in a model-independent way using an effective field theory  (EFT) approach.
This approach involves identifying interactions between DM particles and SM fields through a set of non-renormalizable operators, characterized by an effective scale $ \Lambda $ and the DM mass. 
However, given the limitations of the applicability of the DM EFT at the center-of-mass energies at the LHC, the recent searches at the ATLAS and CMS rely on the simplified DM models (SDMs); for the details, see the analyses of the refs. \cite{Bai:2013iqa,Schmeier:2013kda,Buckley:2014fba,Abdallah:2014hon,Abdallah:2015ter,Berlin:2015wwa,Baek:2015lna,Englert:2016joy,Albert:2016osu,DeSimone:2016fbz,Arcadi:2017kky,Bauer:2017fsw,Arcadi:2017wqi,Abercrombie:2015wmb,LHCDarkMatterWorkingGroup:2018ufk,Arina:2018zcq,Arcadi:2019lka,Arcadi:2020gge,Arcadi:2020jqf}.
Among the various possibilities of the NP particles, the LHCb and ATLAS are continuously searching for scalar and pseudoscalar mediators \cite{Cea:2018tmm, Richard:2020cav, CMS:2018cyk, CMS:2024zqs, ATLAS:2023ofo, Tao:2024syy, Rygaard:2023dlx}, which is our main focus in this study. 

Dark matter models featuring scalar and pseudoscalar mediators are particularly compelling, as they can be naturally embedded within ultraviolet-complete frameworks, offering both theoretical consistency and phenomenological richness.
Prominent examples of such frameworks include Two-Higgs-Doublet-Models (2HDMs)~\cite{Bhattacharyya:2015nca, Bauer:2017ota, Arcadi:2017wqi,Arcadi:2020gge}, supersymmetric extensions~\cite{Hiller:2004ii,Draper:2010ew, March-Russell:2009vla,Chen:2024njd}, and singlet extensions of the SM~\cite{Lopez-Honorez:2012tov, Baek:2014jga, Baek:2017vzd, Ferrer:2006hy, Goncalves:2016iyg}. Beyond these well-motivated frameworks, scalar and pseudoscalar mediators have also been extensively explored in the context of low-mass dark matter scenarios. The sub-10~GeV pseudoscalar regime was analyzed in detail in~\cite{Dolan:2014ska}, incorporating constraints from both flavour physics and the dark sector. Analogous studies for scalar mediators appear in~\cite{Krnjaic:2015mbs, Matsumoto:2018acr}, with the latter allowing mixing with the SM Higgs. Pseudoscalar-mediated dark matter has also been proposed as an explanation for the DAMA modulation signal~\cite{Arina:2014yna, DelNobile:2015lxa, Yang:2016wrl}, while remaining consistent with null results from other direct detection experiments. Finally, scalar and pseudoscalar mediators in loop-induced direct detection scenarios have been studied extensively~\cite{Ertas:2019dew, Kahlhoefer:2017umn}, with the latter focusing on sub-GeV dark matter and self-interactions in the dark sector.

While our focus is on fermionic dark matter coupled to a spin-0 mediator in the low-mass regime, a variety of alternative scenarios for low-mass fermionic dark matter have been explored in the literature. These include models where dark matter carries lepton or quark flavour~\cite{Agrawal:2011ze}, frameworks invoking Minimal Flavour Violation (MFV) to ensure stability~\cite{Batell:2011tc}, and scenarios with top-flavored dark matter~\cite{Kumar:2013hfa}. Higgs-portal fermionic dark matter has also been considered~\cite{Kanemura:2010sh}, while Majorana dark matter with a neutrinophilic scalar mediator was studied in the context of relic density in~\cite{Rajaraman:2018bam}. Collectively, these studies highlight the theoretical diversity and phenomenological richness of low-mass fermionic dark matter models. 
Flavour physics might be important in indirectly detecting new particles or interactions beyond the SM. In this regard, semi-leptonic rare B and K decays may serve as a particularly valuable tool. These decays offer relatively clean experimental signatures, manageable theoretical uncertainties, and suppressed rates in the SM, making them sensitive probes of BSM physics. Recently, the Belle-II collaboration reported an analysis of the rare flavour changing neutral current (FCNC) process $  B^{+} \to K^{+}\nu \bar{\nu} $ \cite{Belle-II:2023esi}, the measured value of the branching fraction shows an excess of $ 2.7 \sigma $ above the SM prediction \cite{Parrott:2022zte}. Similar to this decay, there are other FCNC invisible decays, like $ B^{0} \to K^{(*)} \nu \bar{\nu}, K_{L,S} \to \pi \nu \bar{\nu}, B \to \rho \nu \bar{\nu}, \text{ and } B \to \phi \nu \bar{\nu} $, for which the upper bounds on the respective branching fractions are available \cite{E949:2007xyy, BaBar:2010oqg, Belle:2017oht, KOTO:2020prk, Belle-II:2023esi, BaBar:2013npw}.  In these searches, neutrinos go undetected, and their possible signature is traced via missing energy signals. As we know, the missing energy signals are the revelatory clue of the production of stable neutral particles. Like neutrinos, any other invisible particles will also go undetected and could produce missing energy signals. One of the most common examples of such a neutral particle beyond the SM is the DM.

Apart from the flavour data, electroweak precision observables (EWPOs), dipole moments of leptons and quarks play a crucial role in testing the allowed parameter spaces. These observables include measurements related to the properties of the W and Z bosons, such as their masses and decay rates, as well as other precision electroweak parameters.

In light of recent advancements in experimental data on various flavour-changing neutral current, charged current, and other flavour and electroweak observables from multiple experiments, as well as the increasingly stringent limits on dark matter direct detection cross-sections from several collaborations, a comprehensive global analysis that incorporates all relevant constraints in a correlated manner remains absent. In this work, we aim to fill this gap by systematically exploring the parameter space of the model, taking into account the interplay among these diverse constraints. In dark matter phenomenology, benchmark scenarios are often selected based on limited theoretical guidance, and in collider studies of simplified models, such choices are frequently made in an ad hoc manner. In contrast, our analysis provides a systematically correlated parameter space for the model parameters, offering a more robust foundation for both phenomenological and experimental investigations. This structured approach not only enhances the interpretability of results across different studies but also facilitates meaningful comparisons. Furthermore, our findings indicate that the constraints derived from current data are expected to become significantly stronger with improved precision in low-energy measurements, thereby offering valuable direction for future experimental searches.


This paper is organised as follows: in sec.~\ref{sec:model_description} and sec.~\ref{sec:framework}, we discuss the particle content and possible interactions of the simplified model we have considered and the methodology used to calculate the loop contributions. Sec.~\ref{sec:impacts_on_obs} describes the impact of the NP on various observables, including FCCC, FCNC, and EWPOs, along with cosmological constraints. Finally, in sec.~\ref{sec:analysis_result}, we discuss and analyse our results for all observables, both separately and simultaneously with DM and other constraints. The bounds on a toy model are discussed in sec.~\ref{sec:bounds_toy}. We conclude by summarising our findings in sec.~\ref{sec:summary}.

\section{Working Model: Simplified Dark Matter}\label{sec:model_description}

Among the broad class of SM extensions, scenarios involving scalar and pseudoscalar mediators are especially well motivated. Pseudoscalar mediators, in particular, have gained considerable attention in the context of indirect detection (ID)~\cite{Berlin:2014tja, Izaguirre:2014vva, Ipek:2014gua}, as they couple in a spin-dependent and momentum-suppressed manner in direct detection experiments, thereby evading stringent constraints. These models are especially compelling because they can address long-standing anomalies such as the DAMA annual modulation signal~\cite{Bernabei:2021wkq} and the gamma ray excess in the energy range of $1 - 3$ GeV observed at the Galactic Center~\cite{Daylan:2014rsa}, while also achieving the correct dark matter relic abundance~\cite{Arina:2014yna,Yang:2016wrl}. The motivation for focusing on the low-mass region will be discussed in the latter part of this section.

We have considered an extension of the SM by a singlet Dirac fermion dark matter $ \chi $, which is odd under the discrete symmetry $ \mathbb{Z}_2 $ ($ \chi \to -\chi $). The spin-0 singlet $ S $ is communicating with SM as well as the DM.
The relevant Lagrangian can be written as \cite{Matsumoto:2018acr,Haisch:2015ioa,Englert:2016joy,Kolay:2024wns} : 	
\begin{equation}\label{eq:model}
\begin{split}
\mathcal{L} &=\mathcal{L}_{SM} + \frac{1}{2} \bar{\chi} (i \slashed{\partial} - m_{\chi})\chi  + \frac{1}{2} \partial_{\mu}S \partial^ {\mu} S -\mathcal{L}_{\bar{\chi}\chi S} -\mathcal{L}_{\bar{\psi}\psi S} - \mathcal{L}_{VVS} - V(S,h) \, .
\end{split}
\end{equation}
The interactions of $S$ with the DM ($\chi$) and the SM fermions ($ \psi $) are written as 
\begin{align}\label{eq:fermionS}
\mathcal{L}_{\bar{\chi}\chi S} &= \bar{\chi}(c_{s\chi} + i c_{p\chi} \gamma_5)\chi S, \nonumber \\ 
\mathcal{L}_{\bar{\psi}\psi S} &=m_{\psi} \bar{\psi} (c_s + i c_p \gamma_{5})\psi S.
\end{align} 
with, $ c_s = \frac{\sqrt{2}g_s}{v}, c_p = \frac{\sqrt{2}g_p}{v}  $, $ v $ being the vacuum expectation value (VEV) of SM Higgs and $g_{s(p)}$ are dimensionless couplings. 

 To avoid tree-level FCNCs, we adopt interactions consistent with the principle of Minimal Flavour Violation (MFV) without which the generic couplings of new scalar or pseudoscalar mediators to SM fermions can lead to large FCNCs, which are tightly constrained by precision flavour experiments. The central motivation for employing MFV is to ensure that the mediator couplings in our simplified model are naturally aligned with the flavour and gauge structure of the SM. This alignment is crucial for maintaining consistency with stringent experimental bounds on FCNC processes. MFV provides a systematic framework wherein all flavour-violating interactions are governed by the structure of the SM Yukawa couplings. By ensuring that any new physics respects the same flavour symmetries as the SM. This approach not only enhances the theoretical viability of the model but also simplifies the parameter space by aligning new interactions with known flavour hierarchies. 
Our analysis uses an MFV setup, where flavour dependence comes only from fermion masses. The interactions are strictly flavour‑diagonal, with universal coefficients and no new sources of flavour violation. Thus, the same operator structure also fits non‑MFV scenarios with flavour‑diagonal couplings. As a result, our bounds apply not only to MFV models but also to broader non‑MFV frameworks that share this diagonal flavour structure \footnote{In particular, our bounds are directly applicable to scenarios where the mediator couples to \( t\bar{t} \) and charged lepton pairs (\( e^+e^- \) and \( \mu^+\mu^- \)), as discussed later, since these couplings enter directly at the observable level, allowing the bounds to be used without modification. However, for couplings to other fermions, no direct correspondence can be established in the absence of a flavour-structuring principle such as MFV, and the corresponding constraints must be derived separately from the processes that dominantly probe them.}. 

The interactions of the spin-$0$ boson $S$ with the SM gauge bosons are defined by  
\begin{equation}\label{eq:gaugeint}
\mathcal{L}_{VVS} =  c_W' ~W^+_{\mu}W^{\mu-}S   + c_Z' ~Z_{\mu}Z^{\mu} S.
\end{equation}
Here, the gauge boson mediator couplings are written as  
\begin{equation}\label{eq:gauge_coupling}
c_V' = 2 M_V^2 c_G,
\end{equation}
with $ c_G = \frac{g_{V}}{v} $, so that the NP couplings will be in the same scale as the previous $ S\bar{f}f $ couplings. Before the electroweak symmetry breaking, the scalar potential can be written in terms of the SM higgs doublet $H = \{0, \frac{(v+ h)} {\sqrt{2}}\} $, which is as given below
\begin{equation}\label{eq:modelpotential}
	V(S,H) = \frac{\mu_{S}^2}{2}  S^2 + \frac{\lambda_3}{3!} S^3  + \frac{\lambda_4}{4!} S^4 +  \lambda_1 S H^{\dagger} H + \lambda_2 S^2  H^{\dagger} H + \mu_h^2 H^{\dagger} H + \frac{1}{2} \lambda_H (H^{\dagger}H)^2.
\end{equation}
After the electroweak symmetry breaking, the $V(S,H)$ can be expressed in terms of the physical Higgs field $h$ and the VEV ($v$) and defined as $V(S,h)$ of eq.~\eqref{eq:model}. The relevant expression is given by 
\begin{equation}
V(S,h) = \frac{m_{h}^2}{2} h^2 + \frac{v \lambda_H}{2} h^3 + \frac{\lambda_H}{8} h^4 + \frac{\mu_S^2}{2} S^2 + \frac{\lambda_3}{3!} S^3 + \frac{\lambda_4}{4!} S^4 + v \lambda_{1} \, h S + \frac{\lambda_1}{2} h^2 S + v \lambda_2 \, h S^2  + \frac{\lambda_2}{2} S^2 h^2   \,.
\end{equation}
In principle, the mediator S could mix with the SM Higgs. However, we have kept those mixing parameters ($ \lambda_{1} $) small. A similar simplified model is considered in our previous work \cite{Kolay:2024wns}, where we studied the phenomenology focused on the high-mass region of the mediator, i.e, $ M_S \geq 100$  GeV. In this work, we intend to study the low-mass region of the mediator, i.e., for the region $ M_S \leq 10$ GeV. The corresponding possible higher-dimensional operators and UV completion of this simplified model are discussed in detail in \cite{Kolay:2024wns}.

\paragraph{\underline{Motivation for Low-Mass Spin-0 Mediators}:} 
Inspired by the various motivations for scalar and pseudoscalar extensions, in this work, we focus on a general spin-0 extension of the SM. Such an approach offers advantages over considering purely scalar or pseudoscalar mediators separately. A key motivation is that CP-violating couplings of a spin-0 mediator can contribute to (chromo-)electric dipole moments of fundamental fermions and atomic nuclei, and can introduce new CP-violating phases in meson mixing---effects not possible with purely scalar or pseudoscalar couplings, which affect amplitudes but do not generate new CP-violating phases~\cite{PDG:2024cpv,Batell:2017kty}. Experimentally, a spin-0 mediator with CP-violating couplings can lead to distinctive collider signals, such as angular asymmetries in $\tau^+ \tau^-$ or $t\bar{t}$ decays, and CP-sensitive angular correlations between jets in the transverse plane or among multiple photons~\cite{Haisch:2016gry, Chen:2023ins, Azevedo:2022jnd, ATLAS:2020evk, Chen:2017bff}. These features are absent in the purely scalar or pseudoscalar case and thus serve as clear signatures of CP violation.

 The low-mass regime is particularly compelling among the various extensions of the scalar sector beyond the SM.	The traditional searches have focused on WIMPs in the 10 GeV - 1 TeV mass range. The absence of conclusive evidence from these searches has prompted a paradigm shift toward exploring lower mass candidates. On the contrary, the low-mass region remains relatively less explored than higher masses, making it a fertile ground for potential breakthroughs. At the same time, recent developments in detector sensitivity, background suppression and quantum sensor technologies will help probe low-mass dark matter with a reasonably good precision. In the below items, we will outline a few other motivation for studying dark matter in the sub-10 GeV mass range, including models where both the dark matter particle and the mediator are light. Since for a light dark matter to achieve the correct relic abundance via thermal freeze-out, a light mediator is often required to enhance the annihilation cross-section.	 
\begin{itemize}
	\item  Many well-motivated extensions of the SM, such as
	asymmetric dark matter, hidden sector models, and light dark photons, naturally predict
	dark matter particles in the MeV to GeV range.
	
    \item In the low-mass regions of the DM, the DM-nucleon scattering cross-section is less constraining, allowing a larger viable parameter space \cite{XENON:2024hup, LZ:2023poo, PandaX:2023xgl}.

    \item This region remains accessible to future experiments, especially those designed to probe DM scattering off electrons~\cite{Crisler:2018gci, DAMIC-M:2023gxo, Miriam:2024}. 

    \item The DM can be thermally produced by annihilating with another dark sector particle (mediator), which can provide the required annihilation cross-section to achieve the correct relic density of the Universe~\cite{Okawa:2020jea}.

    \item Indirect signatures from flavour observables also offer sensitivity to light dark sector particles. Processes involving missing energy, such as decays producing neutrinos in the final state, can act as signals for new invisible particles~\cite{Belle-II:2023esi}.

    \item Dedicated searches for such particles are ongoing at experiments like Belle~II~\cite{Belle-II:2022heu, Belle-II:2023ueh, Belle-II:2025bhd, Belle-II:2023ydz} and BES~III~\cite{BESIII:2025cyj, BESIII:2023jji}, as well as in fixed-target and beam-dump setups such as NA64~\cite{NA64:2020qwq, NA64:2021aiq, NA64:2024nwj} and others~\cite{Marsicano:2018vin, Gninenko:2300189, Chen:2017awl}.

    \item Moreover, particles with very small couplings to the SM may behave as long-lived particles, which are actively being searched for across multiple experimental platforms~\cite{Belle-II:2023ueh, Blondel:2022qqo, Alimena:2019zri}.

\end{itemize}
Altogether, the low-mass dark sector presents an exciting frontier for discovering light NP and exploring its phenomenological consequences.

Additionally, we note that the SDM framework has previously been examined in the context of DM phenomenology. However, our work extends significantly beyond the scope of existing studies. While earlier analyses often rely on arbitrarily chosen benchmark points, we present a systematic and correlated exploration of the parameter space. In particular, we incorporate constraints from a wide range of low-energy observables and electroweak precision observables —aspects that have been largely neglected in prior literature. This inclusion is not merely supplementary; it is essential for achieving a robust and realistic assessment of the model.

Furthermore, we establish a direct correlation between the DM couplings and the couplings of $S$ to the SM fermions, and we map the allowed regions in the $(M_S, M_\chi)$ mass plane. This correlated approach represents a critical advancement, as it avoids the shortcomings of disconnected parameter choices and enables a more predictive and testable framework. Our global analysis therefore addresses a key gap in the literature and sets a new benchmark for studies of scalar-mediated DM scenarios.

\paragraph{\underline{Possible higher-dimensional effective operators}:} The dimension-5 SM gauge invariant operators could lead to the interaction given in eq.~\eqref{eq:fermionS} between the SM fermions and the $S$ are as follows \cite{Kolay:2024wns}   
	\begin{equation}\label{eq:lagdim5}
	\mathcal{L}_{\rm dim-5} = \mathcal{L}_{\rm ferm} + \mathcal{L}_{\rm gauge}\,,
	\end{equation} 
	with 
	\begin{equation}\label{eq:toylag_fer}
	\mathcal{L}_{\rm ferm} = -\frac{C}{\Lambda} [\bar{\psi}_L i \gamma_{5} H \psi_{R} P ] - y_f [\bar{\psi}_L H \psi_{R}] + h.c.
	\end{equation}
	and 
	\begin{equation}\label{gauge_higherDim}
	\mathcal{L}_{\rm gauge}=\frac{C'}{\Lambda} P |D_{\mu}H|^2.
	\end{equation}
	Here, the field $P$ is defined as
	\begin{equation}
	P = u + S_1,
	\end{equation}
	with $u$ as the VEV associated with the $P$ field. With the given $H$ and $P$, we expand the above Lagrangian. We note that the mass term of the fermions receives anadditional contribution $\propto$ $\gamma_5$. To remove this $\gamma_5$ dependent term, we use the following chiral rotation to the fermionic fields 
	\begin{equation}
	\psi \to e^{\frac{i \gamma_{5} \alpha}{2}} \psi,  \quad  \psi_L \to e^{-\frac{i \alpha}{2}} \psi_L,    \quad   
	\psi_R \to e^{\frac{i  \alpha}{2}} \psi_R, 
	\end{equation}	 	 
	where  
	\begin{equation}\label{eq:alpha}
	\tan \alpha = \frac{C u v }{y_f \Lambda}.  	
	\end{equation}
	Using this procedure,  we get the interactions between the fermion fields and the spin-0 scalar as
	\begin{equation}\begin{split}\label{higher_dim_ferm_int}
	\mathcal{L}_{\rm ferm} \supset -\mathbb{C}^S_s \, [\bar{\psi}\psi S]\, - \mathbb{C}^S_p \, [\bar{\psi} i \gamma_{5} \psi S]  - \mathbb{C}^{h_1}_s \, [\bar{\psi}\psi h_1]\, - \mathbb{C}^{h_1}_p \, [\bar{\psi} i \gamma_{5} \psi h_1],
	\end{split}
	\end{equation}
	with
	\begin{align}\label{eq:higherdimeq}
	\mathbb{C}^{h_1}_s =  \left( \frac{y_f}{\sqrt{2}} - \frac{C u \alpha }{\Lambda \sqrt{2}} \right) \cos \theta + \frac{C v \alpha }{\Lambda \sqrt{2}}  \sin \theta, \quad \mathbb{C}^{h_1}_{p} =  \left( \frac{C u }{\Lambda \sqrt{2}} + \frac{y_f \alpha}{\sqrt{2}} \right) \cos \theta - \frac{C v }{\Lambda \sqrt{2}} \sin \theta, \nonumber  \\
	\mathbb{C}_s^{S} = 	\left(\frac{y_f}{\sqrt{2}} - \frac{C u \alpha }{\Lambda \sqrt{2} } \right) \sin \theta - \frac{C v \alpha }{\Lambda \sqrt{2}} \cos \theta \quad \text{ and }\quad \mathbb{C}^{S}_p =  \left( \frac{C u }{\Lambda \sqrt{2}} + \frac{y_f \alpha}{\sqrt{2}} \right) \sin \theta  - \frac{C v }{\Lambda \sqrt{2}} \cos \theta.  
	\end{align}
	In these equations, the angle $\theta$ is defined via the following mixing between $S_1$ and $h$ 
	\begin{equation}\begin{split}\label{scalar_mixing}
	h_{1} = \cos \theta ~ h - \sin \theta ~S_1\,, \\
	S = \sin \theta ~h + \cos \theta ~S_1 \,. 
	\end{split}
	\end{equation}
Here, $ h_1 $ and $ S $ are the SM scalar and new scalar defined in their mass basis, respectively. In the above equations, we have obtained the fields $ S  $ and $h_1$ from a mixing of $S_1$ and $h$ with the mixing angle $\theta$. For the details, please see the discussion in ref. \cite{Kolay:2024wns}. We can compare the filed $S$ with the new spin-0 scalar and $h_1$ as the SM Higgs.

Similarly, after expanding the gauge interaction term in eq.~\eqref{gauge_higherDim} we obtain  
	\begin{equation}
	\mathcal{L}_{\rm gauge}=  \mathbb{C}_W^{S}  ~  W^{+\mu} W^{-}_{\mu} S  + \mathbb{C}_Z^{S} ~ Z_{\mu}Z^{\mu} S,
	\end{equation}
	with 
	\begin{align}\label{eq:dim5gaugedcouplings}
	\mathbb{C}_W^{S} = \frac{2 M_W^2}{v} \sin \theta + \frac{C' M_W^2}{\Lambda} \cos \theta \quad \text{and}  \quad \mathbb{C}_Z^{S}  = \frac{M_{Z}^2}{v} \sin \theta + \frac{C'M_{Z^2}}{2\Lambda} \cos \theta . 
	\end{align}
This is the simplest example of a dimension-5 operator from which we can generate the type of interactions we are interested in. There could be more complex scenarios in which one could generate similar structures, but that is beyond the scope of this paper.

\section{Theory Framework}\label{sec:framework}

\subsection*{FCNC and FCCC effective Vertices} 	
In this section, we will focus on the contributions of our model in various FCNC and FCCC vertices. It is well known that FCNC processes are loop-suppressed within the SM, making them particularly sensitive to potential new interactions beyond the SM. Observables related to the neutral-current transition $ b \to  s\mu^{+}\mu^{-}$, such as the branching ratios and the angular observables in $ B \to K^{(*)} \mu^{+}\mu^{-} $ decays have been measured by LHCb \cite{LHCb:2014vgu,LHCb:2014cxe,LHCb:2015svh,LHCb:2020gog} and Belle collaboration \cite{Belle:2016fev}. The measured value of angular observable like $ P_{5}' $ are not in complete agreement with its SM prediction \cite{Bharucha:2015bzk, Biswas:2020uaq,Altmannshofer:2008dz}. Furthermore, the measurement of the branching fraction of the rare $ B_{s} \to \mu^{+}\mu^{-} $ decay is also available \cite{LHCb:2021awg}, which plays an important role in constraining the NP parameter spaces.

In contrast, FCCC processes occur at the tree level in the SM, meaning any contributions from new physics scenarios would be tightly constrained. Given the current level of precision, data from these processes can be instrumental in constraining the parameters of new physics models that contribute at the loop level \cite{Kolay:2024wns}. The interactions described in eqs.~\eqref{eq:fermionS} and \eqref{eq:gaugeint} do not contribute to FCNC and FCCC processes at the tree level, with effects only arising at the loop level. These new interactions will alter the FCNC and FCCC vertices. We will explore these modifications in the following discussion.
\subsection{\bf FCNC Processes}\label{sec:FCNC_loop}	
In the SDM considered above, the contribution to the FCNC processes will be through the one-loop diagrams shown in fig.~\ref{fig:b_to_s} for the $d_i\to d_j S $ vertex.  
\begin{figure}[t]
		\centering 
		\subfloat[]{\begin{tikzpicture}
			\begin{feynman}
			\vertex (a1){\( d_j\)};
			\vertex [below right=1.2cm of a1](a2);
			\vertex [below right=1cm of a2](a3);
			\vertex [below left=1cm of a3](a4);
			\vertex [below left=1cm of a4](a5){\( d_i\)};
			\vertex [right=1.5cm of a3](a6);
			
			\diagram* {
				(a3) -- [fermion, arrow size=1.2pt,edge label'={\( t\)}] (a2) -- [fermion, arrow size=1.2pt] (a1),
				(a5) --[fermion, arrow size=1.2pt] (a4) -- [fermion, arrow size=1.2pt,edge label'={\( t\)}] (a3),
				(a2) --[boson, edge label'={\(W\)}] (a4),
				(a3) --[scalar,edge label={\(S\)}](a6),
				
			};
			\end{feynman}
			\end{tikzpicture}\label{fig:FCNC_loop1}}\hspace{.1cm} 
		\subfloat[]{\begin{tikzpicture}
			\begin{feynman}
			\vertex (a1){\( d_j\)};
			\vertex [below right=1.2cm of a1](a2);
			\vertex [below right=1cm of a2](a3);
			\vertex [below left=1cm of a3](a4);
			\vertex [below left=1cm of a4](a5){\( d_i\)};
			\vertex [right=1.5cm of a3](a6);
			
			\diagram* {
				(a2) -- [boson,edge label={\( W \)}] (a3),
				(a3) --[boson,edge label={\( W \)}] (a4),
				(a5) --[fermion, arrow size=1.2pt](a4) --[fermion, arrow size=1.2pt, edge label={\(t \)}] (a2) -- [fermion, arrow size=1.2pt](a1) ,
				(a3) --[scalar,edge label={\(S\)}](a6),
				
			};
			\end{feynman}
			\end{tikzpicture}\label{fig:FCNC_loop2}}\hspace{.1cm}
		\subfloat[]{\begin{tikzpicture}
			\begin{feynman}
			\vertex (a1){\( d_i\)};
			\vertex [above right=1cm of a1](a2);
			\vertex [above right=0.8cm of a2](a3);
			\vertex [above right=0.8cm of a3](a4);
			\vertex [above left=2cm of a4](a5){\( d_j\)};  
			\vertex [right=1.5cm of a4](a6);
			
			\diagram* {
				(a1) --[fermion, arrow size=1.2pt](a2) --[fermion, arrow size=1.2pt, edge label'={\(t \)}](a3) --[fermion, arrow size=1.2pt,edge label'={\(d_j\)}](a4) --[fermion, arrow size=1.2pt] (a5),	
				(a2) --[boson,half left,  looseness=2, edge label=\( W\)](a3),
				(a4) --[scalar, edge label=\( S\)](a6),		
			};
			\end{feynman}
			\end{tikzpicture}\label{fig:FCNC_loop3}}\hspace{.1cm}
		\subfloat[]{\begin{tikzpicture}
			\begin{feynman}
			\vertex (a1){\( d_i\)};
			\vertex [above right=2.2cm of a1](a2);
			\vertex [above left=0.7cm of a2](a3);
			\vertex [above left=0.7cm of a3](a4);
			\vertex [above left=0.7cm of a4](a5){\( d_j\)};
			\vertex [right=1.5cm of a2](a6);

			\diagram* {
				(a1) --[fermion, arrow size=1.2pt] (a2) --[fermion, arrow size=1.2pt,edge label= {\(d_i\)}] (a3) --[fermion, arrow size=1.2pt, edge label={\(t\)}] (a4) --[fermion, arrow size=1.2pt] (a5),
				(a4) --[boson, half left, looseness=2, edge label = \(W\)] (a3),
				(a2) --[scalar, edge label=\( S\)] (a6),
			};
			\end{feynman}
			\end{tikzpicture}\label{fig:FCNC_loop4}}
\caption{ Feynman diagrams depicting the FCNC vertex correction for the $ d_i \to d_j S $ vertex. In these diagrams, all the internal quarks are shown only with the top quark. Similar diagrams with charm and up quarks will also contribute.}
\label{fig:b_to_s}			
\end{figure}
The contribution from the above diagrams can be expressed as:
	\begin{equation}\label{eq:lbsS}
	\mathcal{L}_{eff}^{bsS}=\frac{ 2\sqrt{2} G_F M_W^2 }{16\pi^2} V_{td_i}  V^*_{td_j}\bigg[ C_1[\bar{d_j} (m_{d_j} P_L + m_{d_i} P_R)d_i]  + C_2 [\bar{d_j} (m_{d_j} P_L - m_{d_i} P_R)d_i] \bigg] S,
	\end{equation}
	where $ C_1 $and $ C_2 $ are the effective coefficients coming from the loop diagrams. Now, the loop contributions contain divergences, which is not surprising. The details on this is discussed in ref. \cite{Kolay:2024wns}. 
	With the leading-log approximation the dominant contributions in $C_1$ and $C_2$ from the diagrams in fig.~\ref{fig:b_to_s} can be expressed as
	\begin{eqnarray}\label{eq:RGE}
	C_{1}(\Lambda) & =& \frac{ m_t^2}{2M_W^2} \left(  (3 c_s) \log \frac{\Lambda^2}{m_t^2}+ \left(3 c_G \right) \log \frac{\Lambda^2}{M_W^2} \right) , \nonumber  \\
	C_{2}(\Lambda) & =&  \frac{m_t^2}{2M_W^2} (ic_p) \log \frac{\Lambda^2}{m_t^2}.
	\end{eqnarray} 
The detailed calculations of the loops are discussed in \cite{Kolay:2024wns}.
\subsection{FCCC Observables} \label{sec:FCCC_processes}
	\begin{figure}[t]
		\centering
		\subfloat[]{\begin{tikzpicture}
			\begin{feynman}
			\vertex (a1){\(u_i\)};
			\vertex [above right=1.1cm of a1](a2);
			\vertex [above right=1.cm of a2](a3);
			\vertex [above left=1.cm of a3](a4);
			\vertex [above left=0.8cm of a4](a5){\(d_j\)};
			\vertex [right=1.3cm of a3](a6){\(W \)};			
			
			\diagram* { 
				(a1) --[fermion, arrow size=1.2pt](a2) --[fermion, arrow size=1.2pt, edge label'=\(u_i \)](a3) --[fermion, arrow size=1.2pt, edge label'=\(d_j \)](a4) --[fermion, arrow size=1.2pt](a5),	
				(a2) --[scalar, edge label=\(S \)](a4),
				(a3) --[photon](a6),
			};	
			\end{feynman}
			\end{tikzpicture}\label{fig:FCCC_vertex_1a}}
        \subfloat[]{\begin{tikzpicture}
			\begin{feynman}
			\vertex (a1){\(u_i\)};
			\vertex [above right=1.1cm of a1](a2);
			\vertex [above right=1.cm of a2](a3);
			\vertex [above left=1.cm of a3](a4);
			\vertex [above left=0.8cm of a4](a5){\(d_j\)};
			\vertex [right=1.3cm of a3](a6){\(W \)};			
			
			\diagram* { 
				(a1) --[fermion, arrow size=1.2pt](a2) --[scalar, edge label'=\(S \)](a3) --[boson, edge label'=\(W \)](a4) --[fermion, arrow size=1.2pt](a5),	
				(a2) --[fermion, arrow size=1.2pt, edge label=\(u_i \)](a4),
				(a3) --[boson](a6),
			};	
			\end{feynman}
			\end{tikzpicture}\label{fig:FCCC_vertex_1b}}
        \subfloat[]{\begin{tikzpicture}
			\begin{feynman}
			\vertex (a1){\(u_i\)};
			\vertex [above right=1.1cm of a1](a2);
			\vertex [above right=1.cm of a2](a3);
			\vertex [above left=1.cm of a3](a4);
			\vertex [above left=0.8cm of a4](a5){\(d_j\)};
			\vertex [right=1.3cm of a3](a6){\(W \)};			
			
			\diagram* { 
				(a1) --[fermion, arrow size=1.2pt](a2) --[boson, edge label'=\(W \)](a3) --[scalar, edge label'=\(S \)](a4) --[fermion, arrow size=1.2pt](a5),	
				(a2) --[fermion, arrow size=1.2pt, edge label=\(d_j \)](a4),
				(a3) --[boson](a6),
			};	
			\end{feynman}
			\end{tikzpicture}\label{fig:FCCC_vertex_1c}}\\
        \subfloat[]{\begin{tikzpicture}
			\begin{feynman}
			\vertex (a1){\(u_i\)};
			\vertex [right=1.2cm of a1](a2);
			\vertex [right=1.5cm of a2](a3);
			\vertex [right=1.2cm of a3](a4){\( u_i\)};
			
			\diagram* { 
				(a1) --[fermion, arrow size=1.2pt](a2) --[fermion, arrow size=1.2pt, edge label'=\(u_i \)](a3) --[fermion, arrow size=1.2pt](a4),
				(a2) --[scalar, half left, looseness=2, edge label = \(S\)] (a3),
			};
			\end{feynman}
			\end{tikzpicture}\label{fig:FCCC_self_1a}}
        \subfloat[]{\begin{tikzpicture}
			\begin{feynman}
			\vertex (a1){\(d_j\)};
			\vertex [right=1.2cm of a1](a2);
			\vertex [right=1.5cm of a2](a3);
			\vertex [right=1.2cm of a3](a4){\( d_j\)};
			
			\diagram* { 
				(a1) --[fermion, arrow size=1.2pt](a2) --[fermion, arrow size=1.2pt, edge label'=\(d_j \)](a3) --[fermion, arrow size=1.2pt](a4),
				(a2) --[scalar, half left, looseness=2, edge label = \(S\)] (a3),
			};
			\end{feynman}
			\end{tikzpicture}\label{fig:FCCC_self_1b}}
		\caption{Feynman diagrams contributing as a correction to $u_i\to d_j W$ vertex and self-energy corrections to the up and down type quark fields which are relevant for the wave function renormalizatios. }
		\label{fig:FCCC_feyn}
	\end{figure}
	
We will also have contribution to the FCCC process via one-loop penguin diagram along with the counter-term diagrams. All the relevant diagrams are shown in fig.~\ref{fig:FCCC_feyn}. The general contribution considering all the loops will be given by the following:
	\begin{equation}
	{\cal L}^{eff}_{u_i\to d_j W} = -\frac{g}{2\sqrt{2}}V^*_{ij}\left[ C_{VL} {\bar d_j}\gamma_{\mu}(1-\gamma_5) u_i  + C_{VR} {\bar d_j} \gamma_{\mu}(1 +\gamma_5)u_i \right] W^{\mu}.
	\label{eq:effvertex}
	\end{equation} 
	Here, the impact of new physics arising from loop corrections is reflected in the coefficients $ C_{VL} $ and $ C_{VR} $. Thus, in the tree-level scenario (purely within the SM), $ C_{VL}=1 $ and $ C_{VR}=0 $. The expression for $ C_{VL} $ and $ C_{VR} $, diagram-wise, can be found in ref. \cite{Kolay:2024wns}. 
	Hence, the total contributions to $C_{VL}$ and $C_{VR}$ in eq.~\eqref{eq:effvertex} could be written as
	\begin{equation}
	C_{VL} = C^{\ref{fig:FCCC_vertex_1b}}_{VL} +  C^{\ref{fig:FCCC_vertex_1c}}_{VL} + C^{\rm counter}_{VL}, \ \ \ \ \  C_{VR} = C^{\ref{fig:FCCC_vertex_1a}}_{VR} +  C^{\ref{fig:FCCC_vertex_1b}}_{VR} + C^{\ref{fig:FCCC_vertex_1c}}_{VR}.
	\end{equation} 
	In the following section, we will discuss the various observables potentially sensitive to these $d_i \to d_j S$ transitions and the new contributions to the FCCC vertices.    
	
\section{Implications on  Various Observables} \label{sec:impacts_on_obs}
This section will examine how our model affects various observables, with new contributions emerging through the 1-loop corrections in FCNC and FCCC vertices discussed earlier. First, we will address observables associated with the FCNC and FCCC processes. Furthermore, we will discuss the impact of our model on the electroweak observables in separate subsections.
\subsection{Observables Related to FCNC Processes}\label{sec:FCNC_processes}
\begin{figure}[t]
		\centering
		\subfloat[]{\begin{tikzpicture}
			\begin{feynman}
			\vertex [crossed dot] (b){};
			\vertex [above left=2cm of b](a){\( d_j\)};
			\vertex [below left=2cm of b](c){\(d_i\)};
			\vertex [crossed dot,right=1.5cm of b](d){};
			\vertex [above right=2cm of d](e){\(d_i\)};
			\vertex [below right=2cm of d](f){\( d_j\)};
			
			\diagram* {
				(c) -- [fermion, arrow size=1.1pt] (b) -- [fermion, arrow size=1.1pt] (a),
				(b) --[scalar,edge label={\(S\)}] (d),
				(e) --[fermion, arrow size=1.1pt] (d) --[fermion, arrow size=1.1pt] (f),
			};
			\end{feynman}
			\end{tikzpicture}\label{fig:mixing_diagrams}}~~~~~~~~
		\subfloat[]{\begin{tikzpicture}
			\begin{feynman}
			\vertex (a1){\(d_i\)};
			\vertex [crossed dot, above right=2cm of a1](a2){};
			\vertex [above left=2cm of a2](a3){\(d_j\)};
			\vertex [right=2cm of a2](a6);	
			\vertex [above right=1.5cm of a6](a7){\(\ell \)};
			\vertex [below right=1.5cm of a6](a8){\(\ell \)};		
			
			\diagram* { 
				(a1) --[fermion, arrow size=1.1pt](a2) --[fermion, arrow size=1.1pt](a3),
				(a8) --[fermion, arrow size=1.1pt](a6) --[fermion, arrow size=1.1pt](a7),
				(a2) --[scalar, edge label=\( S\)](a6),
			};	
			\end{feynman}
			\end{tikzpicture}\label{fig:btosll}}
		\caption{Feynman diagram contributing to the quark level processes of neutral meson mixing and $ b \to s \ell \ell, ~b \to d \ell \ell  $ and $ s \to d \ell \ell $ process for this model. The crossed dots represent the effective $d_i \to d_jS$ vertex discussed in sec.~\ref{sec:FCNC_loop}. Similar diagrams can be drawn for up-quark FCNC processes.} 
	\end{figure}
\paragraph{\underline{\bf Neutral Meson Mixing :} }
The mixing between neutral pseudoscalar mesons is important for constraining NP models since the data here are very precise. In our model, the contribution to the meson mixing amplitude will come from the Feynman diagram shown in fig.~\ref{fig:mixing_diagrams}. In the last section, we have shown the $d_i \to d_j S$ effective vertex calculation. In this study, we focus on the mediator's low mass region, where the mediator's mass ranges are considered in the GeV and sub-GeV regions. For decays like $ P \to P' \ell \ell $, the di-lepton invariant mass square $ q^2 $ will be in the range $ (2 m_{\ell})^2 \leq q^2 \leq (M_{P} - M_{P'})^2 $. So, we cannot Taylor expand the $S$ propagator in powers of $ \frac{q^2}{M_S^2} $. In such cases, we have to consider the Breit-Wigner (BW) propagator for $ S $ which is given by 
	\begin{equation}\label{eq:prop}
	\frac{i}{q^2-M_S^2 + i \Gamma_s M_S}.
	\end{equation}
	In this form of the propagator, we get a finite contribution from the propagator even at the resonance region. This is because, when $ q^2 \approx M_S^2 $, the leading effects will be from the imaginary part proportional to the $ S  $ decay width. The imaginary part will receive contributions from every particle into which $ S $ can decay. Generally, without full knowledge of all possible decay channels for $ S $, predicting its complete decay rate becomes challenging. Since here, the mass of the $ S $ is in GeV and sub-GeV range, the leading contribution to decay width comes from decay to leptons, light quarks, and DM $ \chi $. It cannot decay to heavy bosons like $ W $, $ Z $ bosons, and also to the top-quark and Higgs bosons. For the decay of $S$ to the SM fermions $f$, the decay widths $ \Gamma(S \to f\bar{f})$ is defined as   
	\begin{equation}\label{eq:Sdecayff}
	\Gamma(S\to \bar{f}f) = N_c \frac{m_f^2 M_S}{8 \pi} \biggl( 1-\frac{4m_{f}^2}{M_S^2}  \biggr)^{1/2} \biggl[  (c_s^2 + c_p^2) - \frac{4 c_s^2 m_{f}^2}{M_S^2}  \biggr] \, ,
	\end{equation}
	where, $ N_c=3 $ for decay to quarks and $ N_c=1 $ for decay to leptons. The decay width of $S$ to the dark fermion pair is given by : 
	\begin{equation}\label{eq:SdecayXX}
	\Gamma(S\to \bar{\chi}\chi) = \frac{ M_S}{8 \pi} \biggl( 1-\frac{4M_{\chi}^2}{M_S^2}  \biggr)^{1/2} \biggl[  (c_{s\chi}^2 + c_{p\chi}^2) - \frac{4 c_{s\chi}^2 M_{\chi}^2}{M_S^2}  \biggr]\, .
	\end{equation}
	
In this work, we have considered the contribution of our model to $ B^0-\bar{B}^0,~ B_s^0 -\bar{B}_s^0 $ mixings. The contributions of our model to $ D_{0}-\bar{D}_{0} $ and $ K_0 -\bar{K}_0 $ mixings are CKM suppressed. Also, the contributions are loop-suppressed due to the small masses of the particle running in the loop. The relevant observable $ \Delta M_{q'}  $ (with $q' = d,s$) is defined as (for the mixing of $ B_{q'}^{0}$ and $\bar{B}_{q'}^{0} $ mesons):
	\begin{equation}
	\Delta M_{q'} = 2 | M_{12}^{q'}|, 
	\end{equation}
	where $  M_{12}^{q'} $ is the amplitude of the mixing process and $ M_{B_{q'}} $ is the mass of the neutral meson. 
	
The frequencies of oscillations $ \Delta M_s $ and $\Delta M_d$ have been measured by various experimental groups, the averages of which are given by \cite{HFLAV:2022, ParticleDataGroup:2024cfk}:
	\begin{subequations}
		\begin{align}
		&\Delta M_{s} = 17.765 \pm 0.006 ~~ps^{-1}, \\&
		\Delta M_d = 0.5065 \pm 0.0019 ~~ps^{-1}.
		\end{align}	
	\end{subequations}  
	The predictions in the SM are given by \cite{DiLuzio:2019jyq}:
	\begin{subequations}\begin{align}
		&\Delta M_{s}^{SM} =  2 |(M_{12}^s)_{SM}| =  18.23 \pm  0.63 ~~ps^{-1}, \nonumber \\&
		\Delta M_{d}^{SM} = 2 |(M_{12}^d)_{SM}| =  0.535 \pm 0.021 ~~ps^{-1}.
		\end{align}
	\end{subequations}
	The SM predictions have relatively large errors, and they are fully consistent with their respective measured values. Given the error in the estimates, there is still room for NP effects in these observables.  
	
	Taking into account the contributions from BSM physics, we can write the total oscillation frequency as
	\begin{equation}\label{eq:NP_mixing}
	\begin{split}
	\Delta M_{q,tot} =2 \big| M_{12}^{q',SM}+ M_{12}^{q',NP}| =\Delta M_{q',SM} |\Delta_{q'}|,
	\end{split}\end{equation}
	with
	\begin{equation}
	|\Delta_{q'}| = \frac{\Delta M_{q,tot}}{\Delta M_{q',SM}} = \Big|1 + \frac{M_{12}^{q',NP}}{M_{12}^{q',SM}}\Big|.
	\end{equation}
	In our model, the contribution in $(M_{12}^{q'})_{NP}$ is given by  
	\begin{align}\label{eq:BsBsbarmix}
	M_{12}^{q',NP} =& 4 G_F^2 M_W^2 (V_{tb}V^*_{t{q'}})^2 \biggl(  \frac{1}{16 \pi^2 } \biggr)^2  \frac{M_{B_{q'}}^2 }{(m_b + m_{q'})^2} M_{B_{q'}}^2 f_{B_{q'}}^2 \eta_B \frac{1}{q^2-M_S^2 + i \Gamma_s M_S}\times \nonumber \\ 
	& \left( -\frac{5}{12}B_{2} ( C_1^2 m_{q'}^2 + C_2^2 m_b^2)  + \frac{1}{2}B_{4} m_b m_{q'} (C_1 C_2  + C_2 C_1 )   \right)\,.
	\end{align}
	$ B_2, ~ B_4$ can be found from \cite{Kolay:2024wns}. $ C_{1} $, $ C_{2} $ are the loop contribution given in eq.~\eqref{eq:RGE}. In the $\Delta_{q'}$, the sources of the mixing phases in the SM and the BSM scenarios are identical, which is the CKM matrix, and it cancels. 
	However, due to the imaginary component of the BW propagator, $\Delta_{q'}$ will have a phase. As we will see later, this phase will be negligibly small since the total decay width $\Gamma_S$ will be small due to the small allowed values of $c_s$ and $c_p$. 
	
	Using the measured and the SM predicted values of $\Delta M_{q}$, we have obtained the following allowed values of $\Delta_{q^{\prime}}$: 
	\begin{equation}
	|\Delta_d| = 0.9467 \pm 0.0373, \ \ \ \ \ \ \ \ 	|\Delta_s| = 0.9746 \pm 0.0337.
	\end{equation}
	We have used these two inputs to find the bounds on the coefficients $c_s$, $c_p$, and $c_{G}$.  
	
\paragraph{\underline{\bf Rare Decays of Mesons :}
\label{para:raredecays} }
Rare di-leptonic decays of neutral pseudoscalar mesons are important for probing new physics since the rate of these decays is suppressed in the SM. In this study, we primarily focus on FCNC processes, like $b\to s \mu^+\mu^-$, $b\to d \mu^+\mu^-$, $s\to d \mu^+\mu^-$. Examples of such decays include $ B_s^0 \to \mu^+ \mu^-, ~B^0 \to \mu^+ \mu^-, ~ K_{L,S} \to \mu^+ \mu^- $. Decay to other leptons, like electrons and taus, is also possible depending on the kinematics. For decay to tau, only the upper bound of the branching ratio is available, whereas for electron, the branching ratio for our case will be very suppressed due to the effect of considering MFV-type coupling. Our model contributes to these processes via the diagram shown in fig.~\ref{fig:btosll}. Experimental data on the respective branching ratios are available from various collaborations, including Belle-II, LHCb, ATLAS, and CMS. The following are the experimental data of the branching ratios \cite{ PDG:2022, LHCb:2021trn, LHCb:2020ycd}: 
\begin{align}\label{eq:rare_exp}
	&\mathcal{B}(B_s^{0} \to \mu^+ \mu^-) = (3.09^{+0.46 ~ +0.15}_{-0.43~-0.11}) \times 10^{-9},\nonumber \\&
	\mathcal{B}(B^{0} \to \mu^+ \mu^-) = (0.12 ^{+0.08}_{-0.07} \pm 0.01) \times 10^{-9},\nonumber \\&
	\mathcal{B}(K_L \to \mu^+ \mu^-) = (6.84 \pm 0.11)\times 10^{-9}, \\&
	\mathcal{B}(K_S \to \mu^+ \mu^-) < 2.1 \times 10^{-10}. \nonumber 
	\end{align}
	The SM contribution will be dependent only on $C_{10}$, including the QED corrections the SM predictions are given as \cite{Beneke:2019slt}:
	\begin{subequations}
		\begin{align}
		\mathcal{B}(B_s^{0} \rightarrow \mu^+ \mu^-) = (3.66 \pm 0.14)\times 10^{-9}\,,\\  \mathcal{B}(B^{0} \rightarrow \mu^+ \mu^-) = (1.03 \pm 0.05)\times 10^{-10}\,.
		\end{align}
	\end{subequations}
Note that these SM predictions are fully consistent with the respective measurements shown in eq.~\eqref{eq:rare_exp}. The expression of the branching ratios of the $B$ and $K$ mesons can be found in our previous work \cite{Kolay:2024wns}. We have excluded processes such as \(c \to u \mu^+ \mu^-\) from our analysis due to CKM factors' suppression of the new physics amplitude and the masses of loop particles, as mentioned previously.
Furthermore, we have the available data in the processes related to $ b \to s \ell \ell $ transitions. Plenty of data is available in $B\to K \mu^+\mu^-$, $B\to K^*\mu^+\mu^-$ and $B_s\to \phi\mu^+\mu^-$ decays; for example, see the refs. \cite{CDF:2011tds,LHCb:2013lvw,Belle:2016fev,CMS:2017rzx,LHCb:2020gog,LHCb:2022qnv}. We have utilized all these data to constrain the new couplings. The detailed methodology and the inputs of this global analysis can be seen in our earlier publications \cite{Bhattacharya:2019dot,Biswas:2020uaq,Kolay:2024wns}.	
	
The low energy effective Hamiltonian describing the $b\to s \ell^+\ell^-$ transitions is \cite{Buras:1995iy,Becirevic:2012fy}
\begin{equation} \label{eq:Heff}
	{\cal H}_{eff} = - \frac{4\,G_F}{\sqrt{2}} V_{tb}V_{ts}^\ast
	\left[  \sum_{i=1}^{6} C_i (\mu)
	\mathcal O_i(\mu) + \sum_{i=7,8,9,10,P,S} \biggl(C_i (\mu) \mathcal O_i + C'_i (\mu) \mathcal
	O'_i\biggr)\right] \,,
	\end{equation}
	The parameters, operators and Wilson coefficients are described in the above references. 
	
As previously discussed, the contribution to \( b \to s \ell^+ \ell^- \) processes in our model arises from the diagrams shown in fig.~\ref{fig:btosll}, where \( d_i \) is replaced by \( b \) and \( d_j \) by \( s \). The amplitude of these diagrams depends on the effective \( b \to s S \) vertex defined earlier and the lepton vertex factor, which is determined through the interaction.
\begin{equation}\label{eq:leptonS}
	\mathcal{L}^{int}_{\ell\ell S} = -m_{\ell}\bar{\ell} (  c_s  + i c_p\gamma_{5}) \ell \,.
	\end{equation}
To estimate the NP contributions in $b\to s \ell^+\ell^-$ decays, we need to use eqs.~\eqref{eq:lbsS} and \eqref{eq:leptonS}, respectively. The contribution we obtain from the diagram of fig.~\ref{fig:btosll}: 
\begin{eqnarray}\label{eq:effHbtosll}
	\mathcal{H}_{eff}^{NP} &= \frac{g^2 V^*_{ts}V_{tb}  }{32 \pi^2} \biggl (  C_1 [\bar{s} (m_s P_L + m_b P_R)b ] + C_2 [\bar{s} (m_s P_L - m_b P_R)b]  \biggr )\times \\&     
	\frac{ m_{\ell}}{q^2-M_S^2 + i \Gamma_s M_S}\biggl (  c_s [\bar{\ell}\ell] + i c_p[\bar{\ell} \gamma_{5}\ell]  \biggr )\,. \nonumber \\
\end{eqnarray}
A direct comparison of this equation with the most general effective Hamiltonian given in eq.~\eqref{eq:Heff}, we will get 
	\begin{align}
	C'_S &= \frac{ M_W^2 m_{\ell} m_s }{e^2 m_b }\frac{ c_s \left(C_1(\Lambda) + C_2(\Lambda)\right) }{q^2-M_S^2 + i \Gamma_s M_S }\,,\quad & C_S =  \frac{M_W^2 m_{\ell} }{e^2} \frac{c_s (C_1(\Lambda) - C_2(\Lambda))}{ q^2-M_S^2 + i \Gamma_s M_S}\,, \nonumber \\
	C'_P &= \frac{M_W^2 m_{\ell} m_s }{e^2 m_b} \frac{(i c_p) (C_1(\Lambda) + C_2(\Lambda))}{q^2-M_S^2 + i \Gamma_s M_S}\,, \quad   &C_P = \frac{M_W^2 m_{\ell} }{e^2 } \frac{(i c_p) (C_1(\Lambda) - C_2(\Lambda))}{q^2-M_S^2 + i \Gamma_s M_S}\,. 
	\end{align}
Therefore, in our working model, the contributions in $b\to s \ell^+\ell^-$ will be via the four operators $\mathcal{O}_{S,P}$ and $\mathcal{O}'_{S, P}$. Note that all these four operators will contribute to the rates of $B\to K^{(*)}\mu^+\mu^-$ and $B_s \to \phi\mu^+\mu^-$ decays. In the next section, we will discuss the constraints from a global analysis of all the data available in $b\to s\mu^+\mu^-$ processes.

\paragraph{\underline{\bf Fix Target Experiment :}}
\label{subsec:fixtarget}
Light spin-0 particles can be produced in fixed-target experiments directly (through gluon fusion) and indirectly (from meson decays). Several proton beam-dump experiments, such as CHARM \cite{CHARM:1985anb}, NuCal \cite{Blumlein:1990ay}, and E613 \cite{Duffy:1988rw}, have provided relevant constraints on such scenarios. In this work, we focus on the bounds from CHARM and consider production via meson decays. This spin-0 particle can be produced from either a K-meson or a B-meson (production from D-meson decays is not considered here due to CKM and loop particle mass suppression). The production cross-section of the spin-0 particle is given by : 
\begin{equation}\label{eq:fix_target_sig}
	\sigma_S \sim \sigma_{pp}M_{pp} \biggl ( \frac{1}{14} \mathcal{B}(K^+ \to \pi^+ + S) +  \frac{1}{28} \mathcal{B}(K_L \to \pi^0 + S) + 3\times 10^{-8} \mathcal{B}(B \to X + S) \biggr ).
\end{equation}
The expressions for branching ratios are given in the previous discussion in eq.~\eqref{eq:Br_P2PS}. Here, the proton-proton cross-section is given by $ \sigma_{pp} \simeq 40 $ mbarn and the average hadron multiplicity $ M_{pp} =13,$ for the CNGS beam. We can normalise the cross-section to the neutral pion yield by using $ \sigma_{\pi_0}\approx \sigma_{pp}M_{pp}/3 $ \cite{Dolan:2014ska}. Note that in the above equation $X$ represents a hadronic resonace, and in our case the dominating contribution will for $X=K$. In the mass range $ M_S \leq (M_K - M_{\pi}) \sim 0.36 \rm{~GeV}, $ production from kaon decays dominate. For $ 0.36 \rm{~GeV} \leq M_S \leq M_{B}-M_K $ mass range,  decays dominate from B-meson. When the spin-0 particle is produced from $ K_L, K^+, B $ mesons and decays into $ e^+e^-, \mu^+\mu^- $ and $ \gamma \gamma  $, then the constraint from CHARM is important. Originally, CHARM had not detected any event. So, we have taken a bound at $ 90\% $ confidence level of $ N_{det} <2.3 $ events \cite{Clarke:2013aya}.

The number of spin-0 particle produced in the detector solid angle is expressed as $ N_S \approx 2.9 \times 10^{17} \frac{\sigma_S}{\sigma_{\pi_0}} $ \cite{Bezrukov:2009yw}. The number of events in the detector region is 
	\begin{equation}\label{eq:fix_target_N}
	N_{det} \sim N_S\left[\text{exp}\left(\frac{-L_1 }{\gamma\, \beta\, c\, \tau_S}\right) - \text{exp}\left( \frac{-L_2}{\gamma\, \beta\, c\, \tau_S}\right) \right] \sum_{X=e,\mu,\gamma}Br(S \to XX)\,.
	\end{equation}
	Here,  $ L_1=480~m  $, the distance of the detector from the beam dump, and the difference $ L_2-L_1 = 35~m$ is the detector's length. The typical relativistic factor is $ \gamma \beta c  = \frac{10}{M_S}$ and the lifetime of $S$ is given by $ \tau_S = \frac{1}{\Gamma_S}$. The expressions for branching ratios of $S$ are given in the previous discussion in eqs. \eqref{eq:SdecayXX} and \eqref{eq:Sdecayff}. The allowed parameter space for this case will be discussed in the next section.

\vspace{0.5cm}
\paragraph{\underline{\bf Invisible Decays of Mesons :}} \label{section_invisible}
Recently, the Belle-II collaboration has provided data on the branching ratio of the decay $B^{+} \to K^{+} \nu \bar{\nu}$, which has $2.7 \sigma$ discrepancies with the SM. This significance was obtained under the assumption of heavy NP \cite{Belle-II:2023esi}. Similarly, data or upper bounds are available on a few other invisible decay channels like $ B \to K^{(*)} \nu \bar{\nu}, B \to \rho \nu \bar{\nu}, K \to \pi \nu \bar{\nu} $. We have presented the available data or the upper limits on the respective channels in table~\ref{tab:rareinputs}. 

 Decays like $ B \to K \nu \bar{\nu}, ~ K \to \pi \nu \bar{\nu} $, which contain neutrinos as final state particles, are treated as invisible decays in the SM since they are not detectable, and experimental bounds come from treating them as missing energy. So, instead of $ \nu $, if we have any other long-lived particle, we will have the same experimental signature. It is to be noted that non-standard interactions of neutrinos may also contribute to the above rare decays or to any other missing energy signal. However, this work will consider the missing energy signals via the DM in the final state of the rare decays mentioned above. We have not considered the coupling of the SM neutrino with the mediator, as they are generally massless in the SM. We consider SDM with a spin-0 mediator $S$ and the fermionic DM $\chi$. Therefore, depending on the kinematics (masses of the DM and the mediator), the additional contributions from the SDM to the rare FCNC processes will be either via the three body $d_i \to d_j \chi\bar{\chi}$ ($d_{i/j}$ down type quarks) transitions or via the two body $d_i \to d_j S$ decays, with $ S $ an undetected particle. Inputs are available for the branching ratios of the decays $K_L^{(+)} \to \pi^{0(+)} S$ \cite{KOTO:2020prk,NA62:2021zjw} and for $B\to K S$ \cite{Altmannshofer:2023hkn}. So the signature of the decay $ P \to P' \chi \bar{\chi} $ will have a similar experimental signature as $ P \to P' \nu \bar{\nu}$. Here $P$ and $P'$ stand for pseudoscalar or vector mesons. The DM $ \chi $ will have a mass in the kinematically allowed region: $ M_{\chi} \leq \dfrac{M_{P} - M_{P'}}{2} $. Several NP scenarios in the context of scalar and pseudoscalar mediators have been studied recently and can be found from \cite{Altmannshofer:2023hkn, Calibbi:2025rpx, Berezhnoy:2025tiw, Abdughani:2023dlr, He:2023bnk, Berezhnoy:2023rxx, Datta:2023iln, Davoudiasl:2024cee, Bolton:2024egx, McKeen:2023uzo, He:2025jfc}.
	  
The measured branching fraction of $P \to P'\nu\bar{\nu}$ decay can be expressed as:
\begin{equation}
Br[P \to P'\nu\bar{\nu} ]_{Exp} = Br[P \to P' \nu \bar{\nu}]_{SM} + Br[P \to P' + \chi \bar{\chi}]_{NP} \,, 
\end{equation}   
Here, the NP and SM contributions will be added in the decay width order, as the available phase space for these two decays is different.

	Apart from the $ P \to P' \chi \bar{\chi} $ three-body decay, the analysis of $ P \to P' \nu \bar{\nu}$ decays are potentially sensitive to the process $ P \to P' + S $. Here, $ S $ is our spin-0 mediator, have mass in the region $ M_{S} \leq (M_{P}-M_{P'})$. Therefore, $S$ can be produced on-shell and can decay almost entirely to DM particle. In that case, the experimental measurement of invisible decay can be seen as:
	\begin{equation}\label{eq:invisible_resonance}
	Br[P \to P'\nu\bar{\nu} ]_{Exp} = Br[P \to P' \nu \bar{\nu}]_{SM} + Br[P \to P' + S ] \,, 
	\end{equation}
	So, which kind of decay we get, will actually depend on whether the mediator can be produced on-shell or not. If it can be produced off-shell, then we get the signature of $ P \to P' \bar{\chi} \chi $ three-body decay. However, if it is on-shell, then this decay can be considered as $ P \to P' S (\to \chi \bar{\chi}) $. The Belle-II experimental data on $B\to K\nu\bar{\nu}$ is provided under the assumption of a heavy new particle, which will be accurate for the first scenario. Below, we will discuss both type of decay and their branching ratio in this simplified model.

	\paragraph{$ \underline{\rm Decay~ via~ Virtual ~Mediator ~S} $ :} First, we discuss the three-body decay $  P \to P' \bar{\chi} \chi $ where the spin-0 mediator $ S $ will be an off-shell propagator. A representative Feynman diagram of such a process is shown in fig.~\ref{fig:inv_feyn_low1}. For the decay of B meson (both charged and neutral), the underlying quark interaction process is  $ b\to s(d) \bar{\chi} \chi $, and the same for K-meson is $ s\to d \chi \bar{\chi} $. For the decay $ P\to P'\chi \bar{\chi} $ to be kinematically allowed, we will have the bound on the fermionic dark matter mass: $ 2 m_{\chi}\leq (M_P - M_{P'}) $. 
	
	\begin{figure}[t]
		\centering
		\subfloat[]{\begin{tikzpicture}
			\begin{feynman}
			\vertex (a1){\(d_i\)};
			\vertex [crossed dot, above right=2cm of a1](a2){};
			\vertex [above left=2cm of a2](a3){\(d_j\)};
			\vertex [right=2cm of a2](a6);	
			\vertex [above right=1.5cm of a6](a7){\(\chi \)};
			\vertex [below right=1.5cm of a6](a8){\(\chi \)};		
			
			\diagram* { 
				(a1) --[fermion, arrow size=1.1pt](a2) --[fermion, arrow size=1.1pt](a3),
				(a8) --[fermion, arrow size=1.1pt](a6) --[fermion, arrow size=1.1pt](a7),
				(a2) --[scalar, edge label=\( S\)](a6),
			};	
			\end{feynman}
			\end{tikzpicture}\label{fig:inv_feyn_low1}}~~~~
		\subfloat[]{\begin{tikzpicture}
			\begin{feynman}
			\vertex (a1){\(d_i\)};
			\vertex [crossed dot, above right=2cm of a1](a2){};
			\vertex [above left=2cm of a2](a3){\(d_j\)};
			\vertex [right=1.4cm of a2](a6){ };
			\vertex [right=0.8cm of a6](a9);
			\vertex [right=1.1cm of a9](a10);
			\vertex [above right=1.5cm of a10](a7){\(\chi \)};
			\vertex [below right=1.5cm of a10](a8){\(\chi \)};		
			
			\diagram* { 
				(a3) --[fermion, arrow size=1.1pt](a2) --[fermion, arrow size=1.1pt](a1),
				(a9) --[scalar, edge label=\(S\)](a10),
				(a8) --[fermion, arrow size=1.1pt](a10) --[fermion, arrow size=1.1pt](a7),
				(a2) --[scalar, edge label=\( S\)](a6),
			};	
			\end{feynman}
			\end{tikzpicture}\label{fig:inv_feyn_low2}}
		\caption{(a) Feynman diagram contributing to the three body $ P \to P' \bar{\chi}\chi$ decays. The crossed dot denotes the effective vertex coming from the 1-loop correction. (b) Feynman diagram contributing to the invisible decay process $ P \to P' S (S \to \bar{\chi}\chi)$ via on-shell $ S$. }		
	\end{figure}

	For the decay $ B \to K \chi \bar{\chi} $, the amplitude will be calculated from the diagram of fig.~\ref{fig:inv_feyn_low1} which is given by 
	\begin{align}	
	\mathcal{M} &=	\frac{g^2 V_{tb} V^*_{ts} }{32 \pi^2 } \frac{\left(  C_1(\Lambda) (m_b+m_s) + C_2(\Lambda) (m_b-m_s)\right) }{\sqrt{(M_B^2/2 -M_S^2)^2+\Gamma_S^2 M_S^2}}\langle K|\bar{s}b|B \rangle \left(\bar{\chi}(c_{s\chi} + ic_{p\chi}\gamma_{5})\chi\right), \nonumber \\
	& =(C_L + C_R)  \langle K|\bar{s}b|B \rangle \left(\bar{\chi}(c_{s\chi} + ic_{p\chi}\gamma_{5})\chi\right).
	\end{align}
	with 
\begin{equation}
C_L = 	\frac{g^2  V^*_{ts} V_{tb} (m_b+m_s) }{32 \pi^2 \sqrt{(M_B^2/2 -M_S^2)^2+\Gamma_S^2 M_S^2}} C_1(\Lambda) ,\ \  C_R = \frac{g^2  V^*_{ts} V_{tb} (m_b-m_s) }{32 \pi^2 \sqrt{(M_B^2/2 -M_S^2)^2+\Gamma_S^2 M_S^2}}   C_2(\Lambda) . 
\end{equation}
Here, $C_1$ and $C_2$ are the effective vertex factors for $b\to s S$ vertex, given in sec.~\ref{sec:FCNC_loop}.  The $B\to K$ transition matrix element is defined via the QCD form factor $f_0(q^2)$
\begin{equation}
\langle K|\bar{s}b|B \rangle  = \frac{M_{B}^2 - M_{K}^2}{m_b-m_s} f_0(q^2)\,. 
\end{equation}	
Similarly, we can express the amplitude of $B\to K^* \chi \bar{\chi} $ decay using the QCD form factors and $C_{L}$ and $C_R$. 
	\begin{table}[t]
		\begin{center}
			\rowcolors{1}{Emerald!15}{SpringGreen!10}
			\renewcommand{\arraystretch}{1.9}
			\begin{tabular}{|ccc|}
				\hline
				Branching Ratio & SM Value & Experimental Value \\
				\hline
				\hline
				$ B^+ \to K^+ \nu\bar{\nu} $ & $ \left((5.06 \pm 0.14 \pm 0.28)\right) \times 10^{-6} $ \cite{Becirevic:2023aov} & $ \left(2.3 \pm 0.5^{+0.5}_{-0.4}\right) \times 10^{-5} $ \cite{Belle-II:2023esi} \\
				
				$ B_0 \to K_{S}^0 \nu\bar{\nu} $ & $ \left( 2.05 \pm 0.07 \pm 0.12 \right) \times 10^{-6} $ \cite{Becirevic:2023aov} & $ < 1.3 \times 10^{-5} $ \cite{Belle:2017oht}\\
				
				$ B^+ \to K^{*+} \nu \bar{\nu} $ & $ \left( (10.86 \pm 1.30 \pm 0.59) \right) \times 10^{-6} $ \cite{Becirevic:2023aov} & $< 6.1 \times 10^{-5} $ \cite{Belle:2017oht} \\

				$ B_0 \to K^{*0} \nu \bar{\nu} $ & $ \left( 9.05 \pm  1.25 \pm  0.55 \right) \times 10^{-6} $ \cite{Becirevic:2023aov} & \shortstack{$ \left( 3.8^{+2.9}_{-2.6} \right) \times 10^{-5} $  \cite{BaBar:2013npw} \\ $ < 1.8 \times 10^{-5} $ \cite{Belle:2017oht}} \\
				
				$ B^+ \to \pi^+ \nu\bar{\nu} $ & $ \left(1.57 \pm 0.44 \right) \times 10^{-7} $ \cite{Straub:2018kue} & $ < 1.4  \times 10^{-5} $ \cite{Belle:2017oht} \\
				
				$ B_0 \to \pi^{0} \nu\bar{\nu} $ & $ (0.73\pm 0.21) \times 10^{-7} $ \cite{Straub:2018kue} & $ < 0.9 \times 10^{-5}  $ \cite{Belle:2017oht} \\
				
				$ B^+ \to \rho^{+} \nu\bar{\nu} $ & $ (3.92\pm 0.79) \times 10^{-7} $ \cite{Straub:2018kue} & $ < 3 \times 10^{-5} $ \cite{Belle:2017oht} \\
				
				$ B_0 \to \rho^0 \nu\bar{\nu} $ & $ \left(1.82 \pm 0.31 \right) \times 10^{-7} $ \cite{Straub:2018kue} & $ < 4 \times 10^{-5} $ \cite{Belle:2017oht}\\
				
				$ K^+ \rightarrow \pi^+ \nu\bar{\nu}  $ & $ (8.60 \pm 0.42)\times 10^{-11} $ \cite{Buras:2024ewl} & $ \left(1.06^{+0.40}_{-0.34} \pm 0.09 \right) \times 10^{-10} $ \cite{NA62:2021zjw} \\
				
				$ K_L \to \pi^0 \nu\bar{\nu} $ & $ \left(2.94 \pm 0.15 \right) \times 10^{-11} $ \cite{Buras:2024ewl} & $ <2.2 \times 10^{-9} $ \cite{KOTO:2024zbl} \\			
				\hline
				\hline
			\end{tabular}
		\end{center}
		\caption{Updated SM predictions and the measured values or upper limit on the branching ratios of the invisible decays of $B$ and $K$ mesons.}\label{tab:rareinputs}
	\end{table}

	The differential decay rate distribution for $ B \to K^{(*)}\chi  \bar{\chi} $ decay can be expressed as 
	\begin{eqnarray}\label{eq:Br_inv_virtual}
	\frac{d\Gamma_{B \to K^{(*)}  \chi \bar{\chi}  }  }{d q^2} = \frac{1}{32 M_{B}^3} \frac{1}{(2 \pi)^3} \frac{\lambda^{1/2}(M_{\chi}^2,M_{\chi}^2, q^2) }{q^2} \lambda^{1/2} (M_{B}^2,M_{K^{*}}^2,q^2) \left|\mathcal{M}_{B \to K^{(*)}   \chi \bar{\chi} }\right|^2\,,
	\end{eqnarray}
	with
	\begin{subequations}
		\begin{eqnarray}
		\left| \mathcal{M}_{B \to K^{*} \chi \bar{\chi} } (q^2) \right|^2& =  & \frac{(C_{R} - C_{L})^2}{(m_{b} + m_{s} )^2} \lambda(M_{B}^2, M^2_{K^{(*)}}, q^2) A^2_{0}(q^2) \left[ 2 q^2 (c_{s\chi}^2 + c_{p\chi}^2) - 8 c_{s\chi}^2 M_{\chi}^2\right] \,,\nonumber  \\
		\\
		\left| \mathcal{M}_{B \to K   \chi \bar{\chi} } (q^2) \right|^2 & = & (C_{L} + C_{R})^2 \frac{(M_{B}^2 - M_{K}^2 )^2 }{(m_{b}-m_{s})^2} f^2_{0}(q^2)\left[ 2 q^2 (c_{s\chi}^2 + c_{p\chi}^2) - 8 c_{s\chi}^2 M_{\chi}^2\right]\,.
		\end{eqnarray}
	\end{subequations}
	
	Here, $ A_0 $ is the vector form factor defined as
	\begin{equation}\label{eq:B2V_FF}
	\langle K^{*}(p,\eta) | \bar{s} \gamma_{5} b | \bar{B}(p_{B})\rangle = \left( \frac{2 M_{K^*} (\eta^*.q)}{i (m_{s}+m_{b})}\right)A_{0}(q^2)\,.
	\end{equation}

	The form factors $ f_{0}(q^2), A_{0}(q^2) $ are discussed in detail in next subsection. 	
	The Kallen function $ \lambda $ can be given by: 
	\begin{equation}\label{eq:kallen}
	\lambda(a,b,c) = a^2 + b^2 +c^2-2ab-2ac-2bc.
	\end{equation}
	The branching ratios depend on the couplings $ c_s$,  $c_p $, and $c_{G}$ via $C_1$ and $C_2$. Furthermore, these rates are sensitive to the coupling of DM-mediator interaction $ c_{s\chi}$, $c_{p\chi}$ along with the DM mass, which has to be in the region mentioned above. 
	
	In order to obtain the decay rate distributions or the branching fractions, we need knowledge of the $q^2$ shapes of the associated form factors (FF). For the $ B \to K  $ process, the knowledge of $f_0(q^2)$ is required. Also, the symmetry relation among the matrix elements can be written as
	\begin{equation}\label{eq:FF_symm}
	\langle K_{S} | \bar{b} \gamma_{\mu} s | B^{0} \rangle  = -\langle K_{S} | \bar{s} \gamma_{\mu} b | \bar{B}^{0} \rangle = \frac{1}{\sqrt{2}} \langle K^{+} | \bar{s} \gamma_{\mu} b | \bar{B}^{+} \rangle\,.
	\end{equation} 
	In BCL (Bourreley-Caprini-Lellouch) parametrization, the FF can be written as \cite{Bourrely:2008za, Parrott:2022rgu}: 
	\begin{equation}\label{eq:FF_expand}
	f_{0} (q^2) = \frac{\mathcal{L}}{1-\frac{q^2}{M^2_{H_{s0}^{*}}}} \sum_{n=0}^{N-1} a_{n}^{0}z^{n}.
	\end{equation}
	This uses a mapping of $ q^2 $ to $ z $ so that the physical $ q^2 $ range $ 0 \leq q^2 \leq (M_{B} -M_{K})^2 $ is mapped to a region of unit circle in $ z $
	\begin{equation}\label{eq:Z_expand}
	z(q^2, t) = \frac{\sqrt{t_{+} - q^{2}} - \sqrt{t_{+} - t_0}}{\sqrt{t_{+} - q^2} + \sqrt{t_{+}-t_{0} }},
	\end{equation}
	with $ t_{+} = (M_{B} + M_{K})^2 $. Here $ t_{0}=0 $ is set \cite{Parrott:2022rgu}. Here, ${M^2_{H_{s0}^{*}}} = M_{B} + \Delta  $, with $ \Delta = 0.45 \rm GeV $. Expanding up to $ N=3, $ fit results of the expansion coefficients and correlation among them are given in \cite{Parrott:2022rgu}. The coefficients are given by:
	\begin{table}[H]
		\centering
		\renewcommand{\arraystretch}{1.2}
		\begin{tabular}{|c|c|}
			\hline
			\multicolumn{2}{|c|}{$ B \to K  $ }\\
			\hline
			$ a_{0}^{0} $ & 0.2545(90) \\
			$ a_{1}^{0} $ & 0.210(76) \\
			$ a_{2}^{0} $ & 0.02(17) \\
			\hline
		\end{tabular}
	\end{table}
	
	The FFs of $ B \to K^{*} $ is obtained as
	\begin{equation}\label{eq:B2Kst_FF}
	\langle K^{*} (p,\eta) | \bar{s} \gamma_{5} b | \bar{B}(p_{B}) \rangle = \left( \frac{2 M_{K^{*}} (\eta^{*}.q )}{i (m_s + m_b)}\right) A_{0}(q^2).
	\end{equation}
	Using a simplified series expansion, we have written the respective FFs as \cite{Horgan:2013hoa} : 
	\begin{equation}\label{eq:B2Kst_FF_expand}
	F(t) = \frac{1}{P(t)} \left[ a_0 (1+c_{01} \Delta x + c_{01s} \Delta x_s )+a_1 z\right] 
	\end{equation}
	where $ z $ has the same expression as in eq.~\eqref{eq:Z_expand} and written as $ z(t,t_0) $, with $ t_{\pm} = (M_{B} \pm M_{K^{*}})^2 $, and $ q^2 =t. $ The pole factor is given as 
	\begin{equation}\label{eq:B2Kst_pole}
	P(t;\Delta m) = 1 - \frac{t}{(M_B + \Delta m )^2}. 
	\end{equation}
	We also have
	\begin{subequations}
		\begin{eqnarray}\label{eq:B2Kst_deltax}
		\Delta x &=& \dfrac{M_{\pi}^2 - m_{\pi, \rm phys}^2}{(4 \pi f_{\pi})^2}\,, \\
		\Delta x_s& =& \frac{(M_{\eta_s}^2 -M_{\eta_s, \rm phys}^2 )}{(4 \pi f_{\pi})^2}\, .
		\end{eqnarray}
	\end{subequations}
	Here, $ f_{\pi} = 132 \rm MeV $ is used. Masses of the pseudoscalar and the physical masses are available in ref. \cite{Horgan:2013hoa}. The values of the coefficients $a_{0}\,,a_{1}\,, c_{01}\,, c_{01s} $ are also taken from \cite{Horgan:2013hoa}. 
	
	For $ B \to \rho $ FFs can be written in Simplified Series Expansion (SSE) \cite{Bharucha:2015bzk}:
	\begin{equation}\label{eq:B2rho_FF}
	F(q^2) = P(q^2) \sum_{k} a_{k} \left[ z(q^2) - z(0)\right]^{k}\,,
	\end{equation}
	with $ P(q^2) = (1-q^2/M_{R,i}^2)^{-1}$, is a simple pole corresponding to the first resonance in the spectrum. For this case $ M_{R,i} = 5.279 \rm ~GeV. $ The value of the coefficients $ a_i $ can be found from \cite{Bharucha:2015bzk}. Furthermore, we have taken the inputs on $ B \to \pi $ FFs from \cite{Biswas:2021qyq}. To obtain the  $ K \to \pi $ form factor, we have used the analysis given in ref. \cite{Carrasco:2016kpy}.

	In the numerical analyses, for the three body decays $ P \to P' \chi \bar{\chi}$, we have determined the allowed values of the branching ratios $\mathcal{B}(P \to P' \chi \bar{\chi})_{\rm NP}$ from the differences of the respective measured and the SM predicted values, such as 
	\begin{equation}\label{eq:br_inv_virtual}
	\mathcal{B}(P \to P' \chi \bar{\chi})_{\rm NP} = \mathcal{B}(P \to P' \nu \bar{\nu})_{\rm Exp} - \mathcal{B}(P \to P' \nu \bar{\nu})_{\rm SM}.
	\end{equation}
	The respective values of $\mathcal{B}(P \to P' \nu \bar{\nu})_{\rm SM}$ and $\mathcal{B}(P \to P' \nu \bar{\nu})_{\rm Exp}$ are given in table~\ref{tab:rareinputs}. Using these inputs, we have obtained the respective values of $\mathcal{B}(P \to P' \chi \bar{\chi})_{\rm NP}$ which we have used further to constraints the new parameters of our simplified model. 
	
	\paragraph{$\underline{ \text{Decay via Resonance  S} } $ : }
	We have mentioned that the analysis of $ P \to P' \nu \bar{\nu}$ could be sensitive to the on-shell production of the new light boson $S$ through the decay $ P \to P' + S (\to \bar{\chi}\chi)$. The experimental upper limit of such processes is available for a few channels of $ B $ and $ K $ meson. Technically, the bound shall depend on the lifetime of the particle $ S $. Hence, the branching ratio will be a function of the mass of $ S $. The upper limit on branching ratios $ \mathcal{B}(K^+ \to \pi^+ + S)$ and $ \mathcal{B}(K_L \to \pi^0 + S)$ as a function of the mass of the invisible particle is given by KOTO \cite{KOTO:2018dsc} and Na62 \cite{NA62:2021zjw}, respectively. For $ B \to K + S $ decay the branching ratio is provided by the experiment CLEO \cite{CLEO:2001acz}, and their result is valid only for the case when $ M_{S}=0$. 
	
	In ref. \cite{Altmannshofer:2023hkn}, the authors have reinterpreted the $B \to K^{(*)} \nu\bar{\nu}$ measurement as a search for the two-body $B \to K^{(*)} S$ decay with the assumption that the undetected particle $S$ is stable (approximately) or decays invisibly. They used the differential distributions of the
	$B^{0,+} \to K^{0,+}\nu\bar{\nu}$ measurements of Belle II \cite{Belle-II:2023esi} and
	BaBar \cite{BaBar:2013npw} and fitted the NP signal to reconstruct the data by
	modeling the resonance $S$ with a Gaussian distribution. They have repeated this exercise for $B\to K^*\nu\bar{\nu}$ measurement by BaBar \cite{BaBar:2013npw}. Finally, they have obtained a $M_S$ dependent limits on the branching fractions $\mathcal{B}(B\to K S)$ and $\mathcal{B}(B\to K^* S)$, respectively.  They have given $ 1 \sigma  $ and $ 2 \sigma  $ error with the best-fit value of the branching ratios, depending upon the mass of $ S $. We have used these results to constrain the SDM parameters. Depending upon whether we take $ 1\sigma $ or $ 2 \sigma  $ error of the branching ratio, we get very different bounds on the parameter space while considering the decay with other constraints, which will be discussed in the next section. So, we will analyse by considering both $ 1 \sigma $ or $ 2 \sigma $ error of this decay. They have provided the branching ratio for $ M_{S} \leq 3  $ GeV. We will show later that these invisible channels greatly impact the parameter space. We will divide the parameter space into two regions viz $ M_{S} \leq 3  $ GeV and $ M_{S} > 3 $ GeV.

	Allowing the mediator to decay to the dark fermion pair, we get the invisible signature from the decay $ P \to P' \chi \bar{\chi} $. The branching ratio can be written as 
	\begin{equation}\label{eq:br_invisible_low3}
	\mathcal{B}\left(P \to P'  \chi \bar{\chi} \right) = \mathcal{B}(P \to P' S) \times \mathcal{B}\left( S \to \bar{\chi} \chi\right).
	\end{equation}
	Here, $ P' $ could be either a pseudoscalar or a vector meson with the following branching fractions 
	\begin{subequations}
		\begin{eqnarray} \label{eq:Br_P2PS}
		\mathcal{B}(P \to M S) &  = &  \frac{\tau_{P}}{16 \pi M_{P}^3} \lambda^{1/2}(M_P^2,M_{M}^2, M_S^2) \left| \mathcal{M}_{P \to M  S } (M_S^2)\right|^2 \\
		\mathcal{B}(P \to V S) & = &  \frac{\tau_{P}}{16 \pi M_{P}^3} \lambda^{1/2}(M_P^2,M_{V}^2, M_S^2) \left| \mathcal{M}_{P \to V S } (M_S^2)\right|^2 
		\end{eqnarray}
	\end{subequations}
	where $M$ is a pseudoscalar and $V$ is a vector meson. In our case, the amplitudes take the following forms
	\begin{subequations}
		\begin{eqnarray}
		\left|\mathcal{M}_{P \to M  S }(q^2) \right|  & = &  ( C'_{R} + C'_{L} ) \left(\frac{M_{P}^2 - M_{M}^2}{m_{d_{i}}-m_{d_{j}}} \right) f_{0}(q^2) \,, \\
		\left| \mathcal{M}_{P \to V  S }(q^2) \right|  & = & (C'_{R} - C'_{L})  \, \,  \frac{\lambda^{1/2}(M_{P}^2, M_{V}^2, q^2)}{(m_{d_{i}} + m_{d_{j}})} A_{0}(q^2)\,.
		\end{eqnarray}
	\end{subequations}	
	where, 
	\begin{equation}
	C'_{L} = \frac{g^2  V^*_{ts} V_{tb} (m_b+m_s) }{32 \pi^2 } C_1(\Lambda) \,, \quad C'_{R} = \frac{g^2  V^*_{ts} V_{tb} (m_b-m_s) }{32 \pi^2 }   C_2(\Lambda).
	\end{equation} 
	
	Here, 
	\begin{equation}\label{eq:inv_StoXX}
	\mathcal{B}\left( S \to \chi  \bar{\chi} \right) = \frac{\Gamma(S \to \chi \bar{\chi} )}{\Gamma(S \to \chi \bar{\chi} )+ \Gamma(S \to f \bar{f} )_{\rm SM}}
	\end{equation}
	$ \Gamma(S \to f \bar{f} )_{\rm SM} $ denotes the sum of the decay width of $ S $ to all the possible SM fermions. The expression of the decay widths of mediator $ S $ to $f \bar{f} $ and to $\chi \bar{\chi}$ are given in eq.~\eqref{eq:Sdecayff} and eq.~\eqref{eq:SdecayXX}, respectively.

	\subsection{FCCC Observables}\label{subsec:fccc}
	In sec.~\ref{subsec:fccc}, we discussed how our model influences the effective vertices of FCCC. These corrections are significant for processes where the SM contribution arises from tree-level diagrams, such as semileptonic and leptonic decays involving \(B\), \(B_s\), and \(K\) mesons. As we will see later, for low mediator mass, the constraints from  FCCC processes are relaxed as compared to the rare FCNC and invisible decays. So here, we will just name the possible observables to which this simplified model can contribute. The detailed discussion can be found in \cite{Kolay:2024wns}.
\paragraph{\underline{\bf Anomalous Couplings of  $ tbW $  Vertex :}}
	The general Lagrangian for $ t \to b W_{\mu}^- $ decay is given by \cite{CMS:2020ezf,CMS:2020vac}:
	\begin{equation}\label{t_to_bW_4}
	\mathcal{L}_{tWb} = - \frac{ g}{\sqrt{2}} \bar{b} \gamma_{\mu} \left(V_L P_L + V_R P_R\right) t W_{\mu}^{-} - \frac{ g}{\sqrt{2}} \bar{b} ~\frac{i \sigma_{\mu \nu}q_{\nu} }{M_W} \left(g_LP_L + g_R P_R\right) t W_{\mu}^{-} + h.c.
	\end{equation}
	$ V_{L(R)},~ g_{L(R)} $ are known as the anomalous couplings. In SM, we have $ V_L=V^*_{tb} $ and $ V_R,~g_L,~g_R=0 $. The experimental constraints on these couplings are provided by ATLAS \cite{ATLAS:2016fbc} and CMS \cite{CMS:2016asd,CMS:2020ezf,CMS:2020vac}, as given in table~3 of ref. \cite{Kolay:2024wns}. 	In this model we get, 
	\begin{equation}\begin{split}
	& V_R = V^*_{tb}~ C_{VR}\, , \\&
	V_L = V^*_{tb}~(1+ C_{VL})\,. 
	\end{split} \end{equation}
	$ C_{VL(R)} $ are shown in eq.~\eqref{eq:effvertex}. The contributions to $g_L$ and $g_R$ will be negligibly small compared to $V_R$ or $V_L$, hence we ignored that.
    \vspace{0.5cm}
	\paragraph{\underline{\bf Semi-leptonic and Leptonic Decays of Mesons :}}
    The corrections to the SM charged-current vertices will affect the semileptonic and leptonic decays of $B$, $B_s$, $K$, $D$, and $D_s$ mesons.  The most general Hamiltonian that contains all possible four-fermion operators of lowest dimension for $ d_j \to u_i \ell \bar{\nu} $ can be written as   
	\begin{equation}\label{eq:utodceff}
	\mathcal{H}_{eff}=\frac{4 G_F}{\sqrt{2}} V_{u_i d_j}\left[ (1+C_{VL})\mathcal{O}_{VL}  + C_{VR} \mathcal{O}_{VR} \right].
	\end{equation}
	The four-fermi operators are given by  
	\begin{align}
	\mathcal{O}_{VL} &=  (\bar{u}_{iL} \gamma^{\mu} d_{jL})(\bar{\ell}_{L} \gamma_{\mu} \nu_{\ell L})\,, \nonumber \\
	\mathcal{O}_{VR} &=  (\bar{u}_{iR} \gamma^{\mu} d_{jR})(\bar{\ell}_{L} \gamma_{\mu} \nu_{\ell L}) \,. 
	\end{align}
	For a detailed discussion, please refer to \cite{Kolay:2024wns}. The contribution of NP to the branching ratio of the decays $ P \to M \ell \nu $ and $ P \to \ell \nu $ can be considered as the modification of the CKM matrix elements: $|V_{u_id_j}^\prime|= |V_{u_id_j} (1+ C_{VL} \pm C_{VR})|$. From the experimental data, CKM elements can be extracted, i.e., NP can be constrained \cite{Biswas:2021pic}. Since, for the low mass of the mediator, FCCC constraints will not play a vital role, we will skip that part here. 
	
\subsection{Electroweak Precision Observables}
The precise measurements of EWPOs at the \(W\) and \(Z\) poles allow us to impose constraints on new physics scenarios. This is achieved by analyzing how new physics contributions at the loop level affect these electroweak observables. Here, we will just summarise the observables we have taken. A detailed discussion of the loop contribution and observables is given in ref. \cite{Kolay:2024wns}. 
\vspace{0.5cm}
\paragraph{\underline{\bf Oblique Parameters :}} 
The most general expression for the gauge boson $(V = W
\text{ or } Z)$ self-energy can be written as 
\begin{equation}\label{eq:selfV}
\Sigma_V(q^2) = \bigg(g^{\mu\nu} - \frac{q^{\mu} q^{\nu}}{q^2}\bigg)\Sigma_{V,T}(q^2) + \frac{q^{\mu} q^{\nu}}{q^2}\Sigma_{V,L}(q^2)\,.
\end{equation}
The transverse component of the self-energy correction ($\Sigma_{V,T}$) contributes to the oblique parameters $S, T, U$. Another important observable, $\Delta\rho$, is the deviation in the ratio of charge and neutral current (for a detailed discussion, please check ref.~\cite{Kolay:2024wns} ). The dependency of the oblique parameters on the self-energy correction of the vector bosons can be found from \cite{Kolay:2024wns, ParticleDataGroup:2024cfk}.  Now, one of the main observables in EWPO is the mass of the $W$ boson $M_{W}$.  Different experimental collaborations give the measurement of $ M_{W} $ such as LHCb~\cite{LHCb:2021bjt}, ATLAS \cite{ATLAS:2023fsi}, D0 \cite{D0:2012kms} , CDF  \cite{CDF:2022hxs} , and  CMS \cite{CMS-PAS-SMP-23-002}. Although most of the data agree with the SM prediction within $1\sigma$ error, the CDF data \cite{CDF:2022hxs} has a deviation of $\sim 7 \sigma$ from the SM prediction. The mass of the $W$ boson can be written as: 
\begin{equation}\label{eq:delrMW}
M_W^2 \left(1- \frac{M_W^2}{M_Z^2} \right) = \frac{\pi \alpha_{em}}{\sqrt{2} G_F} \frac{1}{1- \Delta r}.
\end{equation}
In the oblique approximation, the NP contributions in $\Delta r$  can be expressed in terms of the $[S,T,U]$ parameters via the equation
\begin{equation}\label{eq:delr}
\Delta r = \frac{\alpha_{em}}{s_W^2} \left( - \frac{1}{2} \Delta S + c_W^2 \Delta T + \frac{c_W^2-s_W^2}{4 s_W^2} \Delta U  \right).
\end{equation}
In this work, we have taken the observable as $\delta (\Delta r)$, which can be calculated from the experimental and SM data of $M_{W}$ and is defined as:
\begin{equation}\label{eq:deltar}
\delta(\Delta r) = (\Delta r)_{exp} - (\Delta r)_{SM},
\end{equation} 
In our analysis, we have presented results based on all the above-mentioned data, excluding the CDF measurement. 
While compiling constraints on the W boson mass, we have chosen not to include the recent measurement reported by the CDF collaboration, which exhibits a significant deviation from the SM prediction and from other experimental results. This decision is motivated by the fact that the CDF result stands as a clear statistical outlier, with a central value that is in strong tension with global electroweak fits and with measurements from ATLAS, CMS, and LHCb.

Given the absence of corroborating evidence and the potential for unresolved systematic uncertainties, we do not find a compelling reason to treat this measurement on equal footing with the broader, more consistent body of data. In the absence of independent confirmation, incorporating the CDF value would disproportionately skew the global fit and potentially bias the interpretation of new physics scenarios which may be misleading. Until further experimental validation is available, we adopt a conservative approach by relying on the more robust and mutually consistent set of W boson mass measurements. 

We emphasize that the exclusion of the CDF W boson mass measurement from our analysis should not be interpreted as a reason to question the validity or significance of our results. Our decision is based on a conservative and methodologically consistent approach, aiming to avoid the undue influence of a single, currently unconfirmed outlier on the global parameter space. As discussed, the CDF result remains in significant tension with other high-precision measurements and lacks independent experimental verification.

Furthermore, for readers or reviewers interested in understanding the potential implications of including this measurement on the parameter space—particularly in the higher mass regions of $ M_S $ and $ M_{\chi}$, can be found in our earlier analysis presented in Ref.~\cite{Kolay:2024wns}. There, we explored the implications of incorporating this measurement and discussed its impact on the viable regions of the model parameter space. 
\vspace{0.5cm}
\paragraph{\underline{\bf $Z $ Pole Observables :}}
This model will contribute to the decay of $ Z $ boson via a penguin loop diagram; the corresponding diagrams can be found in fig.~9 of \cite{Kolay:2024wns}. $ Z $ bosons interact with the SM fermions via vector and axial couplings. After inserting the loop diagrams, the couplings will be modified: 	
	\begin{gather}
	\nonumber g_{af} \to a_f + \Delta a_{f}^{NP} \,, \\  g_{vf} \to v_f + \Delta v_f^{NP} \,. \end{gather}
	LEP and SLAC measured various branching ratios to have better control over systematic uncertainties. These measurements include different observables related to this process, commonly referred to as the \(Z\)-pole observables, which have been determined with fairly good accuracy \cite{PDG:2022}. Among these observables, we have focused on the following decay rate ratios \cite{Bernabeu:1996zh,Papavassiliou:2000pq}: 
	\begin{equation}\label{eq:RZobs}
	R_{\ell} = \frac{\Gamma_{had}}{\Gamma_{\ell}} \quad, ~~R_{c} = \frac{\Gamma_{c}}{\Gamma_{had}} \quad, ~~R_{b} = \frac{\Gamma_{b}}{\Gamma_{had}}. 
	\end{equation}  
	We have  
	$ \Gamma_{had} = \Gamma_{u}+\Gamma_{d} + \Gamma_{s} + \Gamma_{c}+\Gamma_{b}$.
	The expression in terms of our model parameters is given in sec.~3 of \cite{Kolay:2024wns}. The SM values are given below \cite{Freitas:2014hra,PDG:2022} :
	\begin{eqnarray}
	& R^{SM}_{e} = 20.736 \pm 0.010, \quad   R^{SM}_{\mu} = 20.736 \pm 0.010, \quad R^{SM}_{\tau }=20.781 \pm 0.010, \nonumber \\&
	R^{SM}_{b} = 0.21582 \pm 0.00002, \quad  R^{SM}_{c}= 0.17221 \pm 0.00003.
	\end{eqnarray}
	
	The experimental values corresponding to these observables can be found in  \cite{PDG:2022}, also in table~\ref{tab:CKM-updated-obs}. We also have considered asymmetry observables like $A_{\tau}$, $A_{\mu}$, $A_{e}$, $A_{b}$ and $A_{c}$ with value: 
	\begin{equation}
	A_{f} = 2 \frac{g_{af} ~g_{vf} }{ g_{af}^2 + g_{vf}^2 }\,,
	\end{equation}
	More details on these are discussed in \cite{Kolay:2024wns}.
\subsection{Electric and Magnetic Dipole Moments}\label{subsec:lepton_sector}	
Dipole moments, both electric and magnetic, as well as their chromo counterparts for quarks, serve as powerful probes of physics BSM. The anomalous magnetic moment of the muon shows a $4.2\sigma$ discrepancy from the SM prediction \cite{Muong-2:2021ojo}. Although the very recent experiment and SM prediction are consistent within $1\sigma$ \cite{Muong-2:2025xyk, Aliberti:2025beg}, which helps to constrain any NP contributing to that sector. Also, recent measurements of the electron magnetic moment \cite{Morel:2020dww,Hanneke:2008tm,Parker:2018vye} also indicate possible deviations. These tensions suggest the presence of contributions from new particles or interactions.
Electric dipole moments (EDMs), in contrast, are predicted to be extremely small within the SM, many orders of magnitude below current experimental sensitivities. This makes them especially promising observables for detecting CP-violating effects from BSM physics.
In our scenario involving a spin-0 mediator, there are additional contributions to the dipole moments of fermions. The corresponding Feynman diagrams are shown in fig.~\ref{fig:Feynamn_AMM}.
\begin{figure}[htb!]
    \centering
	\subfloat[]{\begin{tikzpicture}
	\begin{feynman}
	\vertex (a1){\(f\)};
	\vertex [above right=1.3cm of a1](a2);
	\vertex [above right=1.1cm of a2](a3);
	\vertex [above left=1.1cm of a3](a4);
	\vertex [above left=1.1cm of a4](a5){\(f\)};
	\vertex [right=1.7cm of a3](a6){\(\gamma \)};			
		
	\diagram* { 
	(a1) --[fermion, arrow size=1.1pt](a2) --[fermion, arrow size=1.1pt, edge label'=\(f \)](a3) --[fermion, arrow size=1.1pt, edge label'=\(f \)](a4) --[fermion, arrow size=1.1pt](a5),	
	(a2) --[scalar, edge label=\(S \)](a4),
	(a3) --[photon](a6),
	};	
	\end{feynman}
	\end{tikzpicture}\label{fig:Feyn_EDM}}~~~
    \subfloat[]{\begin{tikzpicture}
	\begin{feynman}
	\vertex (a1){\(q\)};
	\vertex [above right=1.3cm of a1](a2);
	\vertex [above right=1.1cm of a2](a3);
	\vertex [above left=1.1cm of a3](a4);
	\vertex [above left=1.1cm of a4](a5){\(q\)};
	\vertex [right=1.7cm of a3](a6){\(g \)};			
		
	\diagram* { 
	(a1) --[fermion](a2) --[fermion, edge label'=\(q \)](a3) --[fermion, edge            label'=\(q \)](a4) --[fermion](a5),	
	(a2) --[scalar, edge label=\(S \)](a4),
	(a3) --[gluon](a6),
	};	
	\end{feynman}
	\end{tikzpicture}\label{fig:Feyn_cEDM}}
    \caption{Feynman diagram contributing to dipole moments of quarks and leptons. The diagram in fig.~\ref{fig:Feyn_EDM} is contributing to the electric and magnetic dipole moments of both leptons and quarks. The diagram in fig.~\ref{fig:Feyn_cEDM} contributes to the chromo-electric and chromo-magnetic dipole moments and is only applicable for quarks.  }
    \label{fig:Feynamn_AMM}
\end{figure}
To calculate the electric and magnetic dipole moments, the corresponding general effective Lagrangian can be written as~\cite{Lindner:2016bgg}:
\begin{equation}\label{eq:eff_Hamiltonian_anomalous}
\mathcal{L}_{\rm EDM, MDM} = \frac{\mu_{f}^{M}}{2} \, \overline{f} \sigma^{\mu\nu} f F_{\mu\nu}
+ \frac{\mu_{f}^{E}}{2} \, \overline{f} i  \sigma^{\mu\nu} \gamma^5 f F_{\mu\nu}.
\end{equation}
From the above effective Lagrangian, the anomalous magnetic dipole moment (MDM) $a_{i}$ is given by \cite{Lindner:2016bgg}
    \begin{gather}
     a_{f}  = \frac{ 2 m_{f}}{e} \mu_{f}^{M} \,.
    \end{gather}
The electric dipole moment is given by: $\mu_{f}^{E}$. The experimental value of the anomalous magnetic moment of the muon is positive~\cite{Muong-2:2021ojo}, while for the electron, different experiments report different signs for $a_{e}$~\cite{Morel:2020dww, Hanneke:2008tm, Parker:2018vye}. In our model, the scalar coupling ($c_{s}$) leads to a positive contribution to the anomalous magnetic moment, whereas the pseudoscalar coupling ($c_{p}$) yields a negative contribution.

Among the EDMs of the leptons, the electron EDM currently provides the most stringent experimental constraint. For the top quark, no direct measurements of the EDM and MDM exist. The indirect bounds are summarized in table~\ref{tab:EDM_MDM_values}, along with other relevant observables, their SM predictions, and experimental limits. In our analysis, we will later demonstrate that only the electron EDM plays a significant role in constraining the parameter space.

\begin{table}[t]
	\centering
	\rowcolors{1}{cyan!10}{lime!5}
	\renewcommand{\arraystretch}{1.7}
	 \resizebox{0.7\textwidth}{!}{%
		\begin{tabular}{|c|c|c|c|c|}
			\hline
			\textbf{Observables} & \textbf{SM Values} & \textbf{Experimental Values}  \\
			\hline
			$a_{e}$ & $1159652180.252(95) \times 10^{-12} $ \cite{Morel:2020dww}  & $1159652180.73 (28) \times 10^{-12}$ \cite{Hanneke:2008tm}  \\
			$a_{\mu}$ & $116592033 (62) \times 10^{-11}$ \cite{Aliberti:2025beg} & $1166592071.5 (14.5) \times 10^{-11}$ \cite{Muong-2:2025xyk} \\
			$a_{\tau}$ & $1.17721 (5)  \times 10^{-3} $ \cite{Eidelman:2007sb} & $-0.057 < a_{\tau} < 0.024 $ \cite{ATLAS:2022ryk} \\
			$a_{t}$ & $ 0.02 $ \cite{Bernreuther:2005gq} & $< 0.16 $ \cite{Etesami:2016rwu} \\
			\hline
			$|\mu^{E}_e| $ & $\lesssim 1.0\times 10^{-35}$ e-cm \cite{Ema:2022yra} & $< 4.1 \times 10^{-30}$ e-cm \cite{Roussy:2022cmp}  \\
			$\mu^{E}_{\mu}$ & $\sim 1.4 \times  10^{-38}$ e-cm \cite{Yamaguchi:2020eub} & $< 1.9 \times 10^{-19}$ e-cm \cite{Muong-2:2008ebm} \\
			$\mu^{E}_{\tau}$ & $\sim -7.3 \times 10^{-38}$ e-cm \cite{Yamaguchi:2020eub} & $ < 1.0\times 10^{-17}$ e-cm \cite{Belle:2021ybo}  \\
			$\mu^{E}_{t}$ & $< 10^{-30} $ e-cm \cite{Soni:1992tn}& $< 5 \times 10^{-20}$ e-cm \cite{Cirigliano:2016njn} \\
			\hline 
			$|d_{n}|$& $\sim  10^{-32} $ e-cm \cite{Dar:2000tn}& $ < 1.8 \times 10^{-26}$ e-cm \cite{Abel:2020pzs}  \\
			$ \mu_{t} $ & $ 3.39  \times 10^{-18} \, g_{s} \text{-cm}$ \cite{Hernandez-Juarez:2020drn,Aranda:2020tox} & $ 2.73 (1.93) \times 10^{-18} \,  g_s\text{-cm}
			$ \cite{CMS:2019kzp}  \\
			$d_{t}$ & $ \sim 10^{-30} \, g_{s}$-cm \cite{Soni:1992tn} & $ < 1.71  \times 10^{-18} \, g_s$-cm \cite{CMS:2022voq,CMS:2022quh} \\
			\hline
	\end{tabular}
    }
	\caption{SM Predictions and experimental values (upper limits) of (c-)EDMs and (c-)MDMs of a few SM fermions and neutrons.}
	\label{tab:EDM_MDM_values}
\end{table}

Additionally, we obtain contributions to the chromo-magnetic dipole moment (cMDM) and chromo-electric dipole moment (cEDM) for quarks through the diagram shown on the right panel of fig.~\ref{fig:Feynamn_AMM}. These effects can be described by a general effective Lagrangian of the form \cite{CMS:2022quh}:
\begin{equation}\label{eq:lag_cMDM_cEDM}
    \mathcal{L}_{\text{cMDM, cEDM}} =  g_{s} \left[ i \frac{\hat{d}_{q}}{2 m_{q}} \, \bar{q} \sigma^{\mu\nu} \gamma_{5} T^a q\, G^a_{\mu\nu} + \frac{\hat{\mu}_{q}}{2 m_{q}} \, \bar{q} \sigma^{\mu\nu} T^a q\, G^a_{\mu\nu} \right]\,.
\end{equation}
In this parameterization, $\hat{d}_{q}$ and $\hat{\mu}_{q}$ are dimensionless coefficients that encode the strength of the cEDM and cMDM, respectively. The operators are suppressed by the quark mass $m_{q}$, as expected for dipole-type interactions.
The physical chromo-electric and chromo-magnetic dipole moments are then defined as
\begin{gather}
    d_{q} = \frac{g_{s} \hat{d}_{q}}{m_{q}}\,, \\
    \mu_{q} = \frac{g_{s} \hat{\mu}_{q}}{m_{q}}\,.
\end{gather}
This Lagrangian is written in such a form because experimental and theoretical studies often express bounds in terms of the dimensionless parameters $\hat{d}_{q}$ and $\hat{\mu}_{q}$. 
The corresponding values from experiments, along with their SM predictions, are listed in table~\ref{tab:EDM_MDM_values}. 
For the top-quark EDM, we have given the indirect upper bound. The bound or constraint on top EDM is derived in the framework of effective operators. The value listed in table~\ref{tab:EDM_MDM_values} is at the scale $\Lambda = 1$ TeV. The prediction at $\Lambda=10$ TeV is also available in literature and gives a bound almost one order smaller \cite{Fuyuto:2017xup}.  
The (c-)EDMs are particularly important observables for our case, since to get a non-zero contribution to the (chromo-)electric, we need both the couplings $c_s$ and $c_p$ to be non-zero. 
At the hadronic level, quark-level interactions lead to observable effects such as the neutron EDM. 
The main contributions to the neutron EDM come from short-range QCD interactions involving quark EDMs and cEDMs \cite{Altmannshofer:2020ywf}:
\begin{equation}
    d_{n} = - \frac{v}{\sqrt{2}} \left[ \beta_{n}^{uG} d_{u} + \beta_{n}^{dG} d_{d} + \beta_{n}^{sG} d_{s}  + \beta_{n}^{u\gamma} \mu^{E}_{u} + \beta_{n}^{d\gamma} \mu^{E}_{d} + \beta_{n}^{s\gamma} \mu^{E}_{s} \right]\,.
\end{equation}
Here, $\beta_{i}^{(k)}$ are the hadronic matrix elements. The values are given by \cite{Engel:2013lsa, Bhattacharya:2015esa, Bhattacharya:2016zcn}:
$\beta_n^{dG}  \approx 8^{+10}_{-6} \times 10^{-4} \, e\, \text{fm}, \quad
-\dfrac{v}{\sqrt{2}} \beta_n^{u\gamma} \approx -0.233(28), 
-\dfrac{v}{\sqrt{2}} \beta_n^{d\gamma} \approx 0.776(66), \quad
-\dfrac{v}{\sqrt{2}} \beta_n^{s\gamma} \approx 0.008(9) $.


\begin{figure}[t]
	\centering
	\subfloat[]{\begin{tikzpicture}
			\begin{feynman}
				\vertex (a1){\(f\)};
				\vertex [right=1cm of a1](a2);
				\vertex [right=2cm of a2](a3);
				\vertex [right=1cm of a3](a4){\(f\)};
				\vertex [draw, circle, inner sep=0pt, minimum size=30pt, above right=1cm of a2](a5){\(f^{\prime}\)};
				\vertex [above=1.5cm of a5](a6);
				
				\diagram*{
					(a1) --[fermion, arrow size=1pt](a2) --[fermion, arrow size=1pt, edge label'={\(f\)}](a3) --[fermion, arrow size=1pt](a4),
					(a2) --[thick, scalar, edge label={\(S\)}](a5),
					(a3) --[boson, edge label'={\(\gamma/Z\)}](a5),
					(a5) --[boson, edge label={\(\gamma\)}](a6),
				};
			\end{feynman}
	\end{tikzpicture}}~
	\subfloat[]{\begin{tikzpicture}
			\begin{feynman}
				\vertex (a1){\(f\)};
				\vertex [right=1cm of a1](a2);
				\vertex [right=2cm of a2](a3);
				\vertex [right=1cm of a3](a4){\(f\)};
				\vertex [above right=1.cm of a2](a5){\(W\)};
				\draw[thick, decorate, decoration={snake, amplitude=2pt, segment length=6pt}] (a5) circle (15pt);
				\vertex [above=0.5cm of a5](a6);
				\vertex [above=1cm of a6](a7);
				
				\diagram*{
					(a1) --[fermion, arrow size=1pt](a2) --[fermion, arrow size=1pt, edge label'={\(f\)}](a3) --[fermion, arrow size=1pt](a4),
					(a2) --[thick, scalar, edge label={\(S\)}](a5),
					(a3) --[boson, edge label'={\(\gamma/Z\)}](a5),
					(a6) --[boson, edge label={\(\gamma\)}](a7),
				};
			\end{feynman}
	\end{tikzpicture}}\\
	\subfloat[]{\begin{tikzpicture}
			\begin{feynman}
				\vertex (a1){\(f\)};
				\vertex [right=1cm of a1](a2);
				\vertex [right=0.6cm of a2](a3);
				\vertex [right=0.6cm of a3](a4);
				\vertex [right=1.2cm of a4](a5);
				\vertex [right=1cm of a5](a6){\(f\)};
				\vertex [below=1cm of a3](a7);
				\vertex [above=1.3cm of a4](a8);
				
				\diagram*{
					(a1) --[fermion, arrow size=1pt](a2) --[fermion, arrow size=1pt, edge label={\(\ \ \ \ \ f\)}](a3)--[fermion, arrow size=1pt](a4) --[fermion, arrow size=1pt, edge label={\(\ \ \ \ \ f\)}](a5) --[fermion, arrow size=1pt](a6),
					(a2) --[thick, boson, half left, looseness=0.8, edge label={\(Z\)}](a8) --[thick, boson, half left, looseness=0.8, edge label={\(Z\)}](a5),
					
					(a3) --[thick, photon, edge label={\(\gamma\)}](a7),
					(a4) --[scalar, edge label'={\(S\)}](a8),
				};
			\end{feynman}
	\end{tikzpicture}}~
	\subfloat[]{\begin{tikzpicture}
			\begin{feynman}
				\vertex (a1){\(q\)};
				\vertex [right=1cm of a1](a2);
				\vertex [right=0.6cm of a2](a3);
				\vertex [right=0.6cm of a3](a4);
				\vertex [right=1.2cm of a4](a5);
				\vertex [right=1cm of a5](a6){\(q\)};
				\vertex [below=1cm of a3](a7);
				\vertex [above=1.3cm of a4](a8);
				
				\diagram*{
					(a1) --[fermion, arrow size=1pt](a2) --[fermion, arrow size=1pt, edge label={\(\ \ \ \ \ q'\)}](a3)--[fermion, arrow size=1pt](a4) --[fermion, arrow size=1pt, edge label={\( \ \ q\)}](a5) --[fermion, arrow size=1pt](a6),
					(a2) --[thick, boson, half left, looseness=0.7, edge label={\(W\)}](a8) --[thick, scalar, half left, looseness=0.7, edge label={\(S\)}](a5),
					
					(a3) --[thick, photon, edge label={\(\gamma\)}](a7),
					(a4) --[thick, boson, edge label'={\(W\)}](a8),
				};
			\end{feynman}
	\end{tikzpicture}}\\
	\subfloat[]{\begin{tikzpicture}
			\begin{feynman}
				\vertex (a1){\(q\)};
				\vertex [right=1cm of a1](a2);
				\vertex [right=2cm of a2](a3);
				\vertex [right=1cm of a3](a4){\(q\)};
				\vertex [draw, circle, inner sep=0pt, minimum size=30pt, above right=1cm of a2](a5){\(q^{\prime}\)};
				\vertex [above=1.5cm of a5](a6);
				\vertex [below=0.9cm of a1](a100){};
				
				\diagram*{
					(a1) --[fermion, arrow size=1pt](a2) --[fermion, arrow size=1pt, edge label'={\(f\)}](a3) --[fermion, arrow size=1pt](a4),
					(a2) --[thick, scalar, edge label={\(S\)}](a5),
					(a3) --[thick, gluon, edge label'={\(g\)}](a5),
					(a5) --[thick, gluon, edge label={\(g\)}](a6),
					
				};
			\end{feynman}
	\end{tikzpicture}}~
	\subfloat[]{\begin{tikzpicture}
			\begin{feynman}
				\vertex (a1){\(q\)};
				\vertex [right=1cm of a1](a2);
				\vertex [right=0.6cm of a2](a3);
				\vertex [right=0.6cm of a3](a4);
				\vertex [right=1.2cm of a4](a5);
				\vertex [right=1cm of a5](a6){\(q\)};
				\vertex [below=1cm of a3](a7);
				\vertex [above=1.3cm of a4](a8);
				
				\diagram*{
					(a1) --[fermion, arrow size=1pt](a2) --[fermion, arrow size=1pt, edge label={\(\ \ \ \ \ q\)}](a3)--[fermion, arrow size=1pt](a4) --[fermion, arrow size=1pt, edge label={\(\ \ \ \ \ q\)}](a5) --[fermion, arrow size=1pt](a6),
					(a2) --[thick, boson, half left, looseness=0.8, edge label={\(Z\)}](a8) --[thick, boson, half left, looseness=0.8, edge label={\(Z\)}](a5),
					
					(a3) --[thick, gluon, edge label={\(g\)}](a7),
					(a4) --[scalar, edge label'={\(S\)}](a8),

				};
			\end{feynman}
	\end{tikzpicture}}~
	\subfloat[]{\begin{tikzpicture}
			\begin{feynman}
				\vertex (a1){\(q\)};
				\vertex [right=1cm of a1](a2);
				\vertex [right=0.6cm of a2](a3);
				\vertex [right=0.6cm of a3](a4);
				\vertex [right=1.2cm of a4](a5);
				\vertex [right=1cm of a5](a6){\(q\)};
				\vertex [below=1cm of a3](a7);
				\vertex [above=1.3cm of a4](a8);
				
				\diagram*{
					(a1) --[fermion, arrow size=1pt](a2) --[fermion, arrow size=1pt, edge label={\(\ \ \ \ \ q'\)}](a3)--[fermion, arrow size=1pt](a4) --[fermion, arrow size=1pt, edge label={\( \ \ q\)}](a5) --[fermion, arrow size=1pt](a6),
					(a2) --[thick, boson, half left, looseness=0.7, edge label={\(W\)}](a8) --[thick, scalar, half left, looseness=0.7, edge label={\(S\)}](a5),
					
					(a3) --[thick, gluon, edge label={\(g\)}](a7),
					(a4) --[thick, boson, edge label'={\(W\)}](a8),

				};
			\end{feynman}
	\end{tikzpicture}}
	\caption{Two-loop diagrams contributing to the EDMs of the quarks and leptons, relevant to our scenario. The first figures in the first and third panels are of the Barr-Zee type diagrams and dominantly contribute to the (c-)EDM.}
	\label{fig:Barr_zee}
\end{figure}
\vspace{0.5cm}
\paragraph{\underline{\bf One-Loop Contributions in EDMs :}}
As shown in fig.~\ref{fig:Feynamn_AMM}, the dipole moment receives contributions through a triangular loop correction in our scenario. Additionally, the process $f \to f \gamma (g)$ can also receive corrections from the external fermion legs, which enter through wavefunction renormalization. However, since only tensor-type operators contribute to the dipole moments, these external leg corrections do not affect the dipole moment itself.
The loop contributions to the EDM can be written as:

{ \footnotesize \begin{equation}
\mu^{E}_{f}= \frac{c_p \, c_s \, e \, Q_f \, M_S}{16 \, \pi^2  \, x \, (1 - 4x^2)}
\left[
(1 - 4x^2)(2x^2 + \log x^2) + 2 \sqrt{1 - 4x^2}(1 - 2x^2)  \log\left( \frac{1 + \sqrt{1 - 4x^2}}{2x} \right) 
\right]\,.
\end{equation}}

The loop contributions to the MDM can be written as:

{\footnotesize \begin{equation}
\begin{aligned}
\mu^{M}_{f} = & \frac{e \, Q_f \, M_S \, (1 - 4x^2)}{32 \pi^2 x^3} \Bigg[ \;
 c_s^2 \Big( x^2 (2 - 3x^2) 
+ (1 - 3x^2) \log(x^2) 
+ 2 \sqrt{1 - 4x^2}(5x^2 - 1 - 4x^4) 
\log\left( \frac{1 + \sqrt{1 - 4x^2}}{2x} \right) \Big) \\
& \quad +  c_p^2 \Big( x^2 (2 + x^2) 
+ (1 - x^2) \log(x^2) 
+ 2 \sqrt{1 - 4x^2}(3x^2 - 1) 
\log\left( \frac{1 + \sqrt{1 - 4x^2}}{2x} \right) \Big) 
\Bigg]\,, 
\end{aligned}
\end{equation}}
with $x=m_f/M_S$. 

\begin{gather}
    \hat{\mu}_{q} = \frac{m_{q} \, g_{s}}{e Q_{f} } \mu^{M}_{f} \,,  \\
    \hat{d}_{q} = \frac{m_{q}\, g_{s}}{e Q_{f} } \mu^{E}_{f}  \,.
\end{gather}
\vspace{0.5cm}
\paragraph{\underline{\bf Two-loop level contributions in EDMs :}}
In our working scenario, in addition to the one-loop contributions to the EDMs, we will also obtain non-negligible two-loop contributions to the lepton and quark EDMs. We will get these two-loop level contributions via the diagrams shown in fig.~\ref{fig:Barr_zee}. All these diagrams have non-negligible contributions to the EDM. The first figures in the first and third panels are of the Barr-Zee type diagrams and dominantly contribute to the (c-)EDM. Among these diagrams, the first, with the top quark in the loop, will dominate. The diagram-wise contributions to the EDM can be found in Ref.~\cite{Brod:2018lbf}. All three couplings related to SM and the spin-0 mediator, $c_s$, $c_p$, and $c_G$, will contribute to the EDM. Since our mediator interacts with a very heavy quark, such as the top, these two-loop contributions will be dominant. Very stringent constraints on the parameter space can be derived from the experimental upper limit on the electron EDM. The EDM constraints of other particles are not as strong as those of the electron. In the later sections, we will discuss the constraints from the available inputs on EDMs and (c-)EDMs.

\subsection{Cosmological Constraints}\label{subsec:cosmoinputs}
Apart from the constraints coming from flavour physics and the $W$ and $Z$-pole observables, we have also studied cosmological constraints. Below, we discuss constraints coming from Big Bang Nucleosynthesis and the condition of thermal equilibrium, which sets a lower limit on the couplings in parameter space. 
\vspace{0.5cm}
\paragraph{\underline{\bf Big Bang Nucleosynthesis :}}
If the dominant annihilation of the DM particles is into spin-0 particles, these particles need to decay rapidly enough before Big Bang Nucleosynthesis (BBN) to prevent changes in the expansion rate and the entropy density during BBN or the destruction of certain elements. Being long-lived, the spin-0 particle $S$ may affect the universe's early history. If $S$ had been in thermal equilibrium and decayed around or later than about a second after the Big Bang, the produced particles would affect BBN \cite{Steigman:2012ve,Kaplinghat:2013yxa}. Although calculating precise constraints is challenging, such constraints become less significant if the average lifetime of the spin-0 particle is less than 1 second \cite{Kolb:1990vq, Kahlhoefer:2017umn, Dolan:2014ska}. However, the constraints are severe for very small masses of $S$ and become less constraining for higher masses. 

The light mediator can decay to the SM fermions ($f$) as well as dark matter particles ($\chi$). The decay width of $S \to f \bar{f}$ is $\Gamma_{S \to f \bar{f}}$, given in eq.\eqref{eq:Sdecayff} and the decay width of $S \to \chi \bar{\chi}$ is $\Gamma_{S \to \chi \bar{\chi}}$, given in eq.~\eqref{eq:SdecayXX}. The total decay width of $S$ is given as:
\begin{equation}
    \Gamma_{S} = \sum_{f} \Gamma_{S \to f \bar{f} } + \Gamma_{S \to \chi \bar{\chi}} \,. 
\end{equation}
The lifetime of the spin-0 mediator is given by:
\begin{equation}
    \tau_{S} = \frac{1}{\Gamma_{S}}\,.
\end{equation}

We can conservatively state that if the spin-0 mediator has a lifetime 
\(\tau_S \lesssim 1~\text{second}\), it will have no significant impact on the light element abundances in the Universe \cite{Rubakov:2005tx}. This constraint can also be verified by comparing the decay width \(\Gamma_S\) with the Hubble expansion rate at \(T = 1~\text{MeV}\). To avoid disrupting BBN, the decay rate must exceed the Hubble rate at that time. 
The constraint can be expressed as follows \cite{Kolb:1990vq,BBN_PDG_2024}:
\begin{equation}
    \Gamma_{S} > H(T= 1~\text{MeV})
\end{equation}
The Hubble parameter in the radiation-dominated era is given by:
\begin{equation}
H(T) = 1.66 \, \frac{\sqrt{g_*(T)}}{M_{\rm Pl}} \, T^2 \,, 
\end{equation}
 where, $H(T)$ is the Hubble expansion rate at temperature $T$, $g_*(T)$ is the effective number of relativistic degrees of freedom \cite{Huang:2021dba} contributing to the energy density at temperature $T$,  $M_{\rm Pl} = 1.22 \times 10^{19}~\text{GeV}$ is the Planck mass. 
This constraint will give us a lower limit on the couplings of our interest. As can be seen from the expression of decay width, with increasing the mass of the mediator, the decay width increases. That is why constraint is more important in the lower-mass region of the mediator. 

\vspace{0.5cm}
\paragraph{\underline{\bf Thermal Equilibrium :}} 
In our working model, we have considered a fermion $ \chi $, which is a DM candidate and is expected to reproduce the correct relic density of the Universe via the freeze-out method. For the DM to freeze out, it should be primarily in thermal equilibrium with the SM in the early universe. So, if the interaction rate of the DM with the SM particles is very low, it cannot be in thermal equilibrium. The interaction of the DM with SM particles is governed by the couplings $ c_s, c_p $ in our working model. These couplings can not be very small for the DM to be in thermal equilibrium. Here, we will discuss the lower limit on the couplings. Certainly, the dark sector could undergo thermalisation if the couplings $ c_{s\chi} $ and $ c_{p\chi} $ are large. However, it is likely to possess a temperature different from that of the visible sector. The DM could still undergo freeze-out into spin-0 particles, but the resulting abundance becomes dependent on the specifics of reheating. If all the couplings are very small, DM can no longer produce the observed relic density via freeze-out; it can be obtained through the freeze-in method.

For the DM to be in thermal equilibrium, we know, the interaction rate has to be larger than the expansion rate of the universe : 
\begin{equation}\label{eq:therm_equi}
	\langle \sigma v \rangle n_{\rm eq} > H, 
\end{equation}
where $ H $ is the Houble rate: $ H=1.66 \sqrt{g_*} T^2 /M_P $, $M_P  $ being the Planck mass and $ g_* $  is the d.o.f and $ n_{\rm eq} $ is the equilibrium number density. $ \langle \sigma v \rangle $ can be given by either the production of SM particles from SM annihilation or vice versa. DM can annihilate to various particles of SM. 
\begin{equation}\label{eq:therm_equi2}
	\langle \sigma v \rangle  n_{\rm eq} = \sum_{X=f, V} \langle \sigma v \rangle_{\bar{\chi} \chi \to XX }n^{\chi}_{\rm eq}
\end{equation} 
Here, $ X $ stands for SM fermions and vector bosons. We can write:
\begin{equation}\label{eq:therm_equi3}
	\langle \sigma v \rangle_{\bar{ \chi} \chi \to XX}\,n^{\chi}_{\rm eq} = \frac{1}{8 \pi^2 M_{\chi}^2 K_2(M_{\chi}/T) } \int_{s_{\rm min}}^{\infty} ds\, \sigma_{\bar{\chi}\chi \to XX}\, \left(s-4 M_{\chi}^2\right)\, \sqrt{s} \,K_{1}(\sqrt{s}/T), 
\end{equation}
$ K_{i} $'s are the modified Bessel functions of the second kind, and the expression for  $ \sigma_{\bar{\chi} \chi \to \bar{f} f} $ can be found in appendix E of  \cite{Kolay:2024wns}. 

\vspace{0.5cm}
\paragraph{\underline{\bf Bounds from BBN and thermal equilibrium :}}
Since we are mainly interested in the low-mass region of the DM, the cosmological constraints will be important here to set a lower bound on the parameter space. Below, we will discuss the BBN constraint and the condition for the WIMP to be in thermal equilibrium.    
	
As discussed in the previous paragraph, if the lifetime of the spin-0 mediator satisfies $\tau_{S} > 1\,\mathrm{s}$, BBN imposes a significant constraint to prevent excessive entropy injection. To avoid this issue, we require $\tau_{S} < 1\,\mathrm{s}$, or equivalently $\Gamma_{S} > H(T = 1~\mathrm{MeV})$, which provides a lower bound on the couplings $c_{s}$, $c_{p}$, and $c_{G}$. Furthermore, these couplings cannot be arbitrarily small if dark matter is to remain in thermal equilibrium with the SM and reproduce the observed relic abundance via freeze-out.

In the low-mass regime of the mediator, the coupling $c_{G}$ is largely irrelevant for cosmological constraints, since a very light mediator cannot decay into SM boson pairs. Consequently, $c_{G}$ is practically insensitive to these constraints. In what follows, we discuss the bounds on the coupling plane arising from BBN requirements and the condition of thermal equilibrium.

\begin{figure}[t]
	\begin{center}
		\subfloat[]{\includegraphics[scale=0.22]{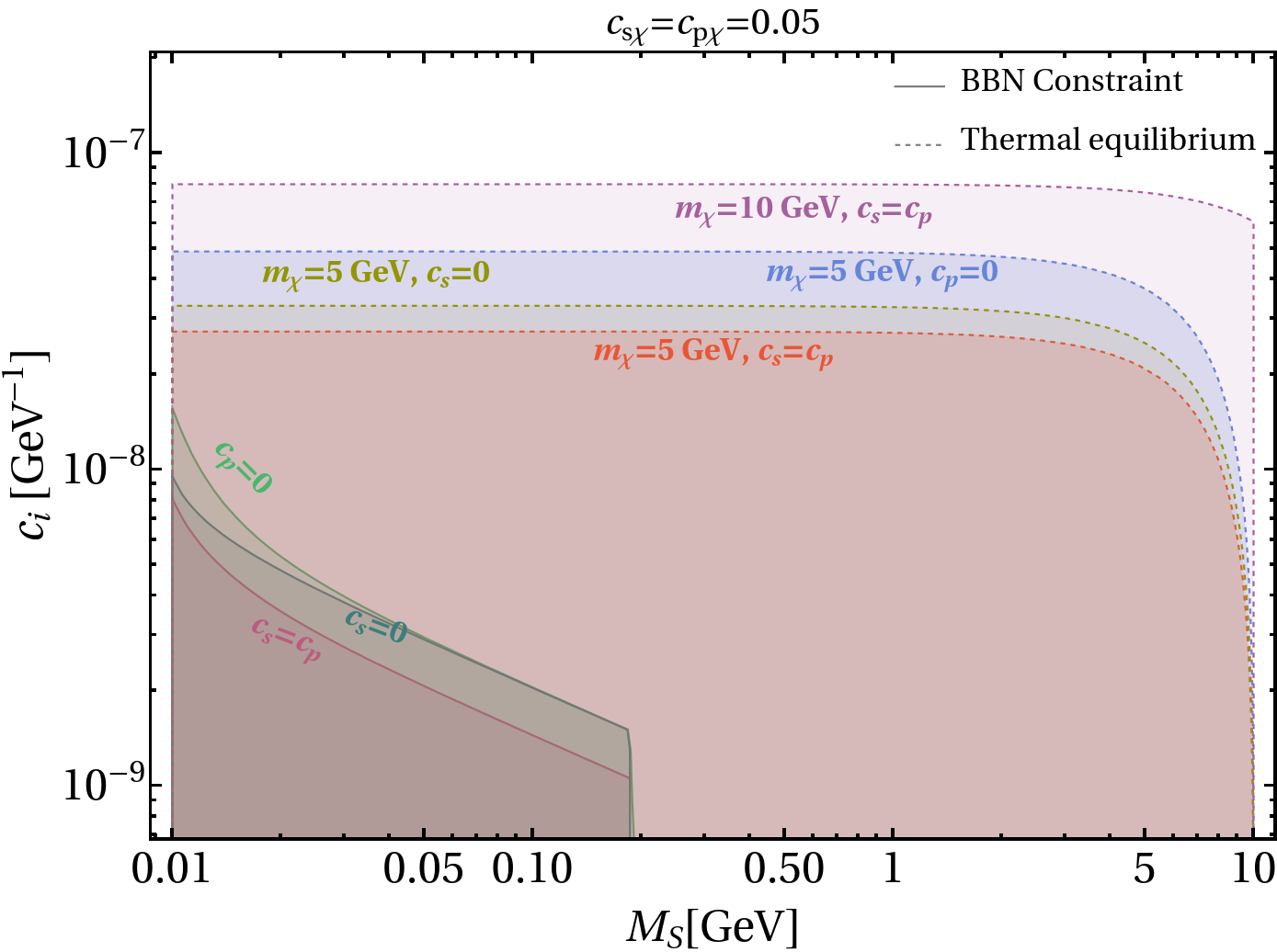}} ~~
		\subfloat[]{\includegraphics[scale=0.22]{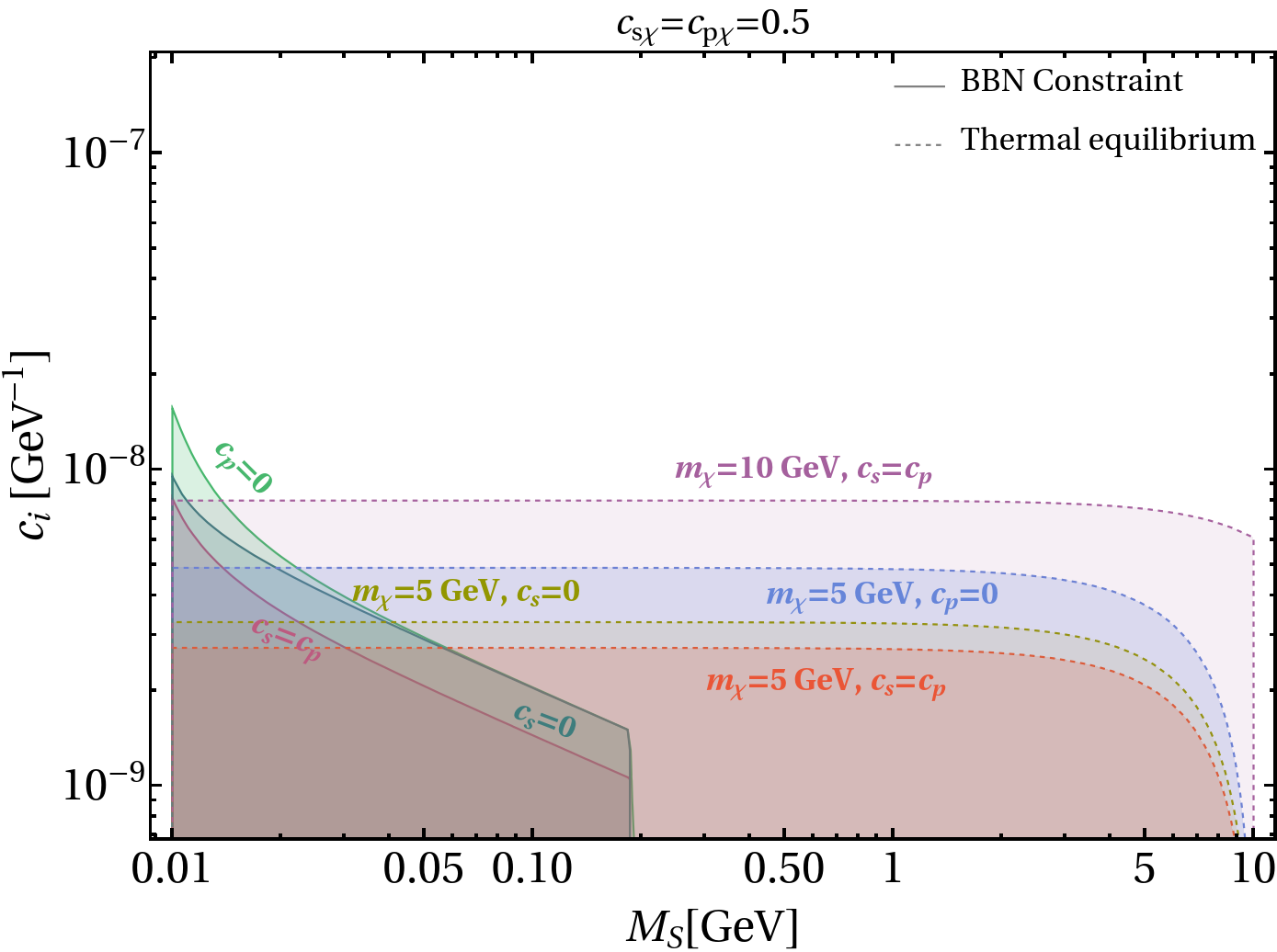}} \\
	\end{center}
	\caption{Excluded parameter space in mass-coupling ($ c_{i}-M_S $) plane, from BBN constraint and condition for the DM to be in thermal equilibrium. The region with the dashed boundary is excluded from the condition of the DM to be in thermal equilibrium, whereas the region with solid lines is excluded from the BBN constraints. Different colors correspond to different benchmark scenarios, depending on the combinations of couplings and mass of the DM ($M_{\chi}$). The left panel is shown for the DM-mediator couplings $c_{s\chi,p\chi} = 0.05$, while the right panel plot is done for the couplings $c_{s\chi,p \chi} = 0.05$.} \label{fig:BBN_therm}   
\end{figure}

Fig.~\ref{fig:BBN_therm} shows the excluded parameter space from both the constraint from BBN and thermal equilibrium discussed in sec.~\ref{subsec:cosmoinputs}. 
The plot in the left panel is done for the DM-mediator coupling $c_{s(p) \chi} = 0.05 $, while the right one is plotted for $c_{s(p)\chi} = 0.5$\,.
The boundary with solid lines shows the excluded regions from the BBN constraint. In our working model, the parameters $ c_s, ~c_p$ are the parameters that will be more sensitive to BBN constraints. Showing the variation with all these couplings is not possible in a single diagram, so we have plotted some coupling combinations. The dashed boundaries show the region where DM cannot be in thermal equilibrium with SM. Here, the couplings of DM with the mediator will also play a role since we are studying the annihilation $\bar{\chi} \chi \to \bar{f} f $. To do the plot, we have fixed the couplings at $ c_{s(p)\chi}  = 0.5  $ (right), and $0.05$ (left). 
Increasing the couplings will increase $ \langle \sigma v \rangle  $, which will exclude the parameter space for a smaller coupling $ (c_i) $ value. The parameter space will also depend on the mass of the DM $ M_{\chi}. $  We have shown for two values of mass $ M_{\chi} = 5  $ GeV (blue, olive, red) and $ M_{\chi} = 10 $ GeV (violet). With increasing mass, the excluded region will be more. Also, for $ M_{\chi} = 5 $ GeV,  we have plotted for purely scalar (blue), purely pseudoscalar (olive), and both equal (red) couplings.  

The BBN constraint is coming from the decay-width of the spin-0 mediator $ S $.  The solid green region is for purely scalar coupling, teal is for purely pseudoscalar coupling, and pink is when both couplings are present and equal. 
Note that both the $ c_s$ and $c_p$ are tightly constrained by BBN, and their numerical values $\lesssim 10^{-8}$ are not allowed for $M_S < 0.2$ GeV. However, the BBN constraints are relatively relaxed for $M_S > 0.2$ GeV, and important constraints will be from the thermal equilibrium.
Note that, for the calculation of the BBN constraints, we have fixed the dark matter mass to $M_\chi = 5\,\text{GeV}$. To allow the mediator to decay into dark matter, the kinematic condition $M_S \geq 2 M_\chi$ must be satisfied. In the plots shown in fig.~\ref{fig:BBN_therm}, where $M_\chi = 5\,\text{GeV}$ is chosen, this condition is not met for most of the parameter space, and thus the decay of the mediator into DM does not play a role. We have also checked cases with lower DM mass (plots not shown here), and as expected, for $M_S > 2 M_\chi$, the mediator can decay into DM, which significantly increases the total decay width. This, in turn, relaxes the bounds on the SM mediator couplings $c_{s,p}$.

As mentioned, the BBN constraint does not depend on the couplings $c_{s\chi}, \, c_{p\chi}$. On the other hand, the couplings $ c_{s\chi}, c_{p\chi} $ are sensitive to the constraints of thermal equilibrium, which are also visible in the variations shown in fig.~\ref{fig:BBN_therm}. For small values of $c_{s(p)\chi}$, the allowed values for both couplings are $> 10^{-7} \, \rm GeV^{-1} $. Hence, for any value of the couplings and mass range, we are interested in the coupling region $ c_{s,p} \gtrsim  10^{-7} \, \rm GeV^{-1}$, which is safe from these two constraints. So, in further analysis, where the couplings are $ > 10^{-7} \, \rm GeV^{-1} $, we will not show these two constraints explicitly.

\subsection{Model Contributions to the Dark Matter Phenomenology}\label{sec:DM_theory}
	
	As mentioned in sec.~\ref{sec:model_description}, we have considered a fermionic DM $ \chi $ which is communicating with the SM particles via a spin-0 mediator $ S $. Here we have taken $ \chi $ to be a WIMP type DM, whose stability can be described by imposing a $ \mathbb{Z}_{2} $ symmetry on it to prevent it from decaying. The DM generates the observed relic density \cite{Planck:2018vyg} via s-channel and t-channel annihilation diagrams. The dominating annihilation channels are $ \chi \bar{\chi} \to {f \bar{f}\,, SS}$, where $f$ stands for SM fermions. Another loop-generated annihilation channel will be $ \chi \bar{\chi}  \to g g,$ which is coming from a top-quark mediated penguin loop. All the relevant diagrams are the same as in \cite{Kolay:2024wns}.
	Since, in this study, we are mainly interested in comparatively lower mass DM with mass $ M_{\chi} \leq 10  $ GeV, the annihilation process $ \chi \bar{\chi}  \to {t \bar{t} \,, V V} $, where $ V $ stands for SM vector bosons, will not contribute to relic density. So, for $ M_{\chi} \leq M_{S}, $ the dominating channel will be $ \chi \bar{\chi} \to b \bar{b} $ and for $ M_{\chi} \geq M_{S} $, the dominating channel will be $ \chi \bar{\chi} \to SS $ via both s-channel and t-channel diagrams.  Also, for very small $ c_{s}\,, c_{p} $, annihilation to the spin-0 mediator will be the only process that contributes significantly to relic density.

	\begin{figure}[t]
		\centering
		\subfloat[]{\begin{tikzpicture}
			\begin{feynman}
			\vertex (a1){\(\chi \)};
			\vertex [above right=1.5cm of a1](a2);
			\vertex [above left=1cm of a2](a3){\(\chi \)};
			\vertex [right=1cm of a2](a6);	
			\vertex [above right=1cm of a6](a7){\(f \)};
			\vertex [below right=1cm of a6](a8){\(f \)};		
			
			\diagram* { 
				(a1) --[fermion, arrow size=1.1pt](a2) --[fermion, arrow size=1.1pt](a3),
				(a8) --[fermion, arrow size=1.1pt](a6) --[fermion, arrow size=1.1pt](a7),
				(a2) --[scalar, edge label=\( S\)](a6),
			};	
			\end{feynman}
			\end{tikzpicture}\label{fig:DM_ann1}}
		\subfloat[]{\begin{tikzpicture}
			\begin{feynman}
			\vertex (a1){\(\chi \)};
			\vertex [above right=1.5cm of a1](a2);
			\vertex [above left=1cm of a2](a3){\(\chi \)};
			\vertex [right=1cm of a2](a6);	
			\vertex [above right=1cm of a6](a7){\(S \)};
			\vertex [below right=1cm of a6](a8){\(S\)};		
			
			\diagram* { 
				(a1) --[fermion, arrow size=1.1pt](a2) --[fermion, arrow size=1.1pt](a3),
				(a8) --[scalar](a6) --[scalar](a7),
				(a2) --[scalar, edge label=\( S\)](a6),
			};	
			\end{feynman}
			\end{tikzpicture}\label{fig:DM_ann2}}
		\subfloat[]{\begin{tikzpicture}
			\begin{feynman}
			\vertex (a1){\( \chi\)};
			\vertex [right=1.6cm of a1](a2);
			\vertex [right=1.3cm of a2](a3){\( S\)};
			\vertex [below=1.9cm of a2](a4);
			\vertex [left=1.3cm of a4](a5){\(\chi \)};
			\vertex [right=1.3cm of a4](a6){\(S\)};
			
			\diagram* { 
				(a1) --[fermion, arrow size=1.1pt](a2),
				(a2) --[scalar](a3),
				(a6) --[scalar](a4) --[fermion, arrow size=1.1pt](a5),
				(a2) --[fermion, arrow size=1.1pt, edge label={\(\chi \)}](a4),
				
			};
			\end{feynman}
			\end{tikzpicture}\label{fig:DM_ann3}}
		\subfloat[]{\begin{tikzpicture}
			\begin{feynman}
			\vertex (a1){\(\chi \)};
			\vertex [above right=1.6cm of a1](a2);
			\vertex [above left=1.3cm of a2](a3){\(\chi \)};
			\vertex [right=1.cm of a2](a6);	
			\vertex [above right=0.7cm of a6](a7);
			\vertex [below right=0.7cm of a6](a8);	
			\vertex [above right=0.7cm of a7](a9){\(g \)};
			\vertex [below right=0.7cm of a8](a10){\(g \)};
			
			\diagram* { 
				(a1) --[fermion, arrow size=1.1pt](a2) --[fermion, arrow size=1.1pt](a3),
				(a8) --[fermion, arrow size=1.1pt, edge label=\( t\)](a6) --[fermion, arrow size=1.1pt, edge label=\( t\)](a7),
				(a7) --[fermion, arrow size=1.1pt, edge label=\( t\)](a8) --[gluon](a10),
				(a7) --[gluon](a9),
				(a2) --[scalar, edge label=\( S\)](a6),
			};	
			\end{feynman}
			\end{tikzpicture}\label{fig:DM_ann4}}
		\caption{Feynman diagrams depicting the annihilation channels of the DM are significant for determining relic density.}
		\label{fig:Feyn_DM}
	\end{figure}
	
	\begin{figure}[t]
		\centering
		\subfloat[]{\begin{tikzpicture}
			\begin{feynman}
			\vertex (a1){\( \chi\)};
			\vertex [right=1.6cm of a1](a2);
			\vertex [right=1.3cm of a2](a3){\( \chi\)};
			\vertex [below=1.9cm of a2](a4);
			\vertex [left=1.3cm of a4](a5){\(N \)};
			\vertex [right=1.3cm of a4](a6){\(N\)};
			
			\diagram* { 
				(a1) --[fermion, arrow size=1.1pt](a2)--[fermion, arrow size=1.1pt](a3),
				(a5) --[fermion, arrow size=1.1pt](a4) --[fermion, arrow size=1.1pt](a6),
				(a2) --[scalar, edge label={\(S\)}](a4),
				
			};
			\end{feynman}
			\end{tikzpicture}\label{fig:DM_DD}}
		\subfloat[]{\begin{tikzpicture}
			\begin{feynman}
			\vertex (a1){\( \chi\)};
			\vertex [right=1.6cm of a1](a2);
			\vertex [right=1.3cm of a2](a3){\( \chi\)};
			\vertex [below=1.9cm of a2](a4);
			\vertex [left=1.3cm of a4](a5){\(e \)};
			\vertex [right=1.3cm of a4](a6){\(e\)};
			
			\diagram* { 
				(a1) --[fermion, arrow size=1.1pt](a2)--[fermion, arrow size=1.1pt](a3),
				(a5) --[fermion, arrow size=1.1pt](a4) --[fermion, arrow size=1.1pt](a6),
				(a2) --[scalar, edge label={\(S\)}](a4),
				
			};
			\end{feynman}
			\end{tikzpicture}\label{fig:DM_ee}}
		\subfloat[]{\begin{tikzpicture}
			\begin{feynman}
			\vertex (a1){\(\chi \)};
			\vertex [above right=1.5cm of a1](a2);
			\vertex [above left=1cm of a2](a3){\(\chi \)};
			\vertex [right=1cm of a2](a6);	
			\vertex [above right=1cm of a6](a7){\(b \)};
			\vertex [below right=1cm of a6](a8){\(b \)};		
			
			\diagram* { 
				(a1) --[fermion, arrow size=1.1pt](a2) --[fermion, arrow size=1.1pt](a3),
				(a8) --[fermion, arrow size=1.1pt](a6) --[fermion, arrow size=1.1pt](a7),
				(a2) --[scalar, edge label=\( S\)](a6),
			};	
			\end{feynman}
			\end{tikzpicture}\label{fig:DM_ID}}	
		\caption{Feynman diagrams contribute to the direct detection process, DM-electron scattering process, and to the indirect detection for annihilation to $ b\bar{b} $ (right).}
	\end{figure}\label{DM_detection}

	Fig.~\ref{fig:Feyn_DM} shows the Feynman diagrams of annihilation of the DM, which can annihilate to SM or another dark sector particle via s-channel or t-channel annihilation diagrams. It will mainly annihilate to the SM fermion pair $ \bar{f} f $ via an s-channel diagram for $ M_{\chi} \geq m_{f} $. Hence, it cannot annihilate into a top-quark pair. The next dominating process will be annihilation to $ b \bar{b} $ pair. Although in our simplified model, we have couplings of the spin-0 mediator with the SM gauge boson, here, DM cannot annihilate to the SM gauge boson due to the kinematical constraint. There will be a large annihilation cross-section of $ \chi \bar{\chi} \to g g $ via a top-penguin diagram, resulting in a considerable contribution to $ \langle \sigma v \rangle $. 
	The effective vertex of this process is given as follows: 
	\begin{equation}\label{gluon_coup}
	\mathcal{L}_{gluon} = \frac{\alpha_s}{8 \pi} \biggl (  c_s \tau [1+ (1-\tau)f(\tau) ]G^{\mu \nu} G_{\mu \nu} +  2 c_p \tau f(\tau) G^{\mu \nu} \tilde{G}_{\mu \nu} \biggr) S \,,
	\end{equation}
	Where, $ \tau = \frac{4 m_t^2}{M_S^2} $ and 
	\[ f(\tau) = \begin{cases} 
	\arcsin^2 \frac{1}{\sqrt{\tau}} & \text{if } ~\tau \geq 1 \\
	-\frac{1}{4} (\log \frac{1+\sqrt{1-\tau}}{1-\sqrt{1-\tau}} - i \pi )^2\,  & \text{if } ~\tau < 1
	\end{cases} \]
	
	The processes mentioned above apply to any mass  $ M_{S} $ of the spin-0 mediator. But for the case $ M_{S} \leq M_{\chi}$, another channel will open up  $ \chi \bar{\chi} \to S S$ via an s-channel as well as a t-channel diagram, as shown in figs.~\ref{fig:DM_ann2} and \ref{fig:DM_ann3}. The t-channel diagram will have a significant contribution to the relic, especially in the case of very small $ c_{s} $ and $ c_{p} $; it will be governed by the couplings $ c_{s(p) \chi} $.

	In our scenario, the parameter space could also be constrained from the t-channel DM-nucleon scattering process $ \chi \, N \to \chi \, N $. The upper bounds on the DM-nucleon scattering cross-section are provided by various experiments: XENONnT \cite{XENON:2025vwd, XENON:2023cxc}, LZ \cite{LZ:2024zvo, LZ:2022lsv}, and PandaX-4T \cite{PandaX:2024qfu, PandaX:2023ejt}, etc. The direct detection bound given by the above experiments is mostly effective for DM mass $ M_{\chi} \gtrsim 5 $ GeV, and the bound is sensitive to the DM and mediator masses. The most stringent bound comes from LUX-ZEPLIN \cite{LZ:2024zvo}, which gives $\langle \sigma v \rangle \sim 5 \times  10^{-47} \, \rm cm^2 $ for $M_{\chi} \sim 10 $ GeV while for $M_{\chi} \sim 5 $ the cross-section is of order $10^{-44}\rm cm^2$   \cite{LZ:2022lsv}. 
	The DD bound on nucleon scattering cross-section for the sub-GeV DM is given by the experiments as: LZ \cite{LZ:2023poo}, XENONnT \cite{ XENON:2024hup}, PandaX-4T \cite{PandaX:2023xgl}, DarkSide-50 \cite{DarkSide-50:2022qzh}, SENSEI \cite{SENSEI:2023zdf},    SuperCDMS \cite{SuperCDMS:2023sql}, LUX \cite{LUX:2018akb}. For $ 2 \lesssim M_{\chi} \lesssim 5 $ GeV the cross-section $\langle \sigma v \rangle \sim 1.4 \times 10^{-42} \, \rm cm^{2}$ which is obtained from DarkSide-50 \cite{DarkSide-50:2022qzh,ParticleDataGroup:2024cfk}. While for $  M_{\chi} < 2 $ GeV, the bounds are relatively relaxed, and the cross-section $\langle \sigma v \rangle$ will be in the range $10^{-37}\rm cm^2$ to $10^{-39}\rm cm^2$. In our analysis, we have used these mass-dependent bounds. The expression for the spin-independent direct detection cross-section is given in Appendix E of \cite{Kolay:2024wns}, where the dependency of the cross-section can be seen as (at $ t \to 0  $ limit): 
	\begin{equation}\label{eq:DD_DM}
	\sigma_{\chi N}^{SI} \propto \frac{c_{s\chi}^2 c_{s}^2}{M_{S}^4}\,.
	\end{equation}
	
An alternative way to search for a light DM candidate is through the scattering of electrons, i.e., $ DM + e \to DM + e $ \cite{Essig:2011nj, Essig:2017kqs}. The experimental and the proposed bounds on the cross-section for this process are given by various proposed and ongoing experiments like: XENON100 \cite{XENON:2016jmt}, 
    CRESST-III \cite{CRESST:2019jnq}, 
    DAMIC-M \cite{DAMIC-M:2023gxo}, 
    DarkSide-50 \cite{DarkSide:2022knj}, 
    ALETHEIA \cite{Liao:2022thr}, 
    DarkSide-20k \cite{DarkSide-20k:2017zyg}, 
    LDMX  \cite{LDMX:2018cma}, 
    PandaX-II \cite{PandaX-II:2021nsg}, 
    SENSEI \cite{SENSEI:2024yyt}, 
    Oscura \cite{Oscura:2023qik} etc. They have given an upper bound on the DM-e scattering cross-section, which can also describe the detection possibility of low-mass DM. The details of the scattering cross-section are given in \ref{appendix:DM_electron}. In our simplified model, since we have chosen MFV-type couplings, the coupling of an electron with the mediator is proportional to the mass of the electron, so we will get a suppression in the cross-section. Hence, this channel will not be relevant in our phenomenological study.   
	
	The investigation of gamma-ray annihilation spectra in indirect detection (ID) places constraints on the dark matter annihilation cross-section rates into SM particle pairs such as $b\bar{b}$, $\tau^+ \tau^-$ etc. Collaborations such as Fermi-LAT \cite{Fermi-LAT:2015att, Fermi-LAT:2016afa}, High Energy Stereoscopic System (H.E.S.S) \cite{HESS:2016mib}, and Cherenkov Telescope Array (CTA) \cite{Silverwood:2014yza} have contributed to establishing these bounds. More details on these bounds can be found at \cite{Kolay:2024wns, Bhattacharya:2024nla}.

    In addition to direct and indirect detection bounds, light dark matter is also subject to cosmological constraints. Residual annihilation processes around the time of recombination can inject energy into the thermal plasma, thereby altering the ionization history of the Universe and leaving observable imprints on the CMB power spectrum. This effect is particularly relevant for light dark matter, where significant annihilation into SM fermions persists at late times, with Planck data constraining the corresponding annihilation cross-section \cite{Planck:2018vyg}. In the low-mass region, where bounds from direct searches via nucleon recoil are less effective, the CMB limits play an important role in constraining the parameter space.

	\section{Analysis and Results}\label{sec:analysis_result}
	
	Our simplified model impacts various observables associated with the FCNC and FCCC heavy flavour decays as well as $W$- and $Z$-pole observables. Furthermore, the model parameters can be constrained from BBN bounds and the condition of DM thermal equilibrium. It also contributes to the present relic density of the universe and can be constrained by the DD and ID bounds of the DM.	This section aims to examine the constraints on the model parameters derived from the available data on these processes.
	
	Given the vast amount of data to analyze, which could complicate the analysis, separately, we will focus on inputs that provide weaker constraints on the model parameters. This approach allows us to discard less relevant data sets in the later stages of the analysis.
	We have separately studied the impact of the data on the semi-leptonic FCNC processes $B \to K^{(*)}\mu^+\mu^-$,  $B_s\to \phi\mu^+\mu^-$. In addition, we have analyzed constraints from meson mixing amplitudes $\Delta M_d$, $\Delta M_s$, and rare decays such as $B^{0}_{(s)} \to \mu^+\mu^-$ and $K_{L,S} \to \mu^+\mu^-$. Constraints from CHARM experiments and invisible decays of pseudoscalar mesons were also included in the analysis. Furthermore, we have evaluated the impact of FCCC processes, $W$- and $Z$-pole observables, and imposed BBN and thermal equilibrium conditions to establish lower bounds on the couplings. DM constraints were analyzed separately and in conjunction with these other observables. We will also discuss the constraint from the CHARM experiment separately, as that constraint is important for our analysis.
	
	In this analysis, we focus on the low-mass limit of the mediator $S$. The constraints on the model parameters are expected to depend strongly on the mediator mass $M_S$. As discussed, we study several processes where $S$ is a virtual particle and some decays where $S$ is produced on-shell. To analyze the phenomenology and constraints on the model parameters, we have divided the parameter space into two regions based on the mass of the spin-0 mediator:
	\begin{itemize}
		\item $ M_S  > 3 $ GeV.
		\item $ M_S  \leq 3 $ GeV.		
	\end{itemize}
	Note that for $ M_S  \leq 3 $ GeV, in our model, we will have a contribution to $B\to K^{(*)} S$ decays which is not possible for $ M_S  > 3 $ GeV. Therefore, we constrain the parameters separately for $ M_S  \leq 3 $ GeV and the data set includes the data on $\mathcal{B}(B\to K^{(*)} S)$. In comparison, for $ M_S  > 3 $ GeV, we will drop the data related to $\mathcal{B}(B\to K^{(*)} S)$.  
	
	
	We have also divided our analysis into three parts. First, we show the constraints from flavour and electroweak observables. Next, we studied the constraints from DM analysis, considering relic density, direct detection limits, electron scattering bounds, and indirect detection constraints. Finally, we will present the correlated parameter space by analyzing both the flavour and electroweak constraints and the dark sector constraints.

\subsection{Constraints from Flavour, EWPOs and Dipole moments}
In the following sections, we study the constraints on the new physics parameter space. This part of the analysis incorporates the available data on FCNC, FCCC, and EWPO observables discussed earlier, along with the constraints from dipole moments. The constraints are derived for various choices of the mediator mass $M_S$, as outlined previously. At the beginning of this subsection, we specifically highlight the constraints arising from the extensive data on semileptonic decays of the form $B(B_S) \rightarrow K^{(*)}(\phi), \ell^+ \ell^-$. The rationale for this focus will become evident in the subsequent discussion. 

\subsubsection{ \bf Bounds from Semileptonic $ B(B_s) \rightarrow K^{(*)}(\phi)\, \ell^+ \ell^- $ Decays} 	
\begin{table}[t]
		\begin{center}
			\rowcolors{1}{blue!5}{blue!3!red!6!green!4}
			\renewcommand{\arraystretch}{1.9}
			\begin{tabular}{|c|c|c|c|c|}
				\hline
				\hline
				$\Lambda \left[\rm TeV \right]$  &  $M_S\left[ \rm GeV \right]$  &  $c_s\left[ \rm GeV^{-1}\right]$  &  $c_p\left[ \rm GeV^{-1}\right] $  &  $c_G\left[ \rm GeV^{-1}\right]$  \\
				\hline
				\hline
				\cellcolor{blue!5}   &  $10$  &  $\text{0.00078(3894)}$  &  $\text{-0.0167(45)}$  &  $\text{-0.00051(2651)}$  \\
				\cellcolor{blue!5}  &  $5$  &  $\text{0.00040(2106)}$  &  $\text{-0.0087(23)}$  &  $\text{-0.00024(1304)}$  \\
				\cellcolor{blue!5}   &  $1$  &  $\text{0.00023(2404)}$  &  $\text{-0.0050(14)}$  &  $\text{-0.00014(1428)}$  \\
				\multirow{-4}{*}{\cellcolor{blue!5}1}    &  $\text{0.1}$  &  $\text{0.00025(1888)}$  &  $\text{-0.0054(15)}$  &  $\text{-0.00015(1135)}$  \\
				\hline
				\hline
				\cellcolor{blue!5}  &  $10$  &  $\text{0.089(33)}$  &  $\text{-0.0053(57)}$  &  $\text{-0.067(24)}$  \\
				\cellcolor{blue!5}  &  $5$  &  $\text{0.0397799(64)}$  &  $\text{-0.0023080(26)}$  &  $\text{-0.0300400(37)}$  \\
				\cellcolor{blue!5}  &  $1$  &  $\text{0.000099(12893)}$  &  $\text{-0.0039(13)}$  &  $\text{-0.000065(8617)}$  \\
				\multirow{-4}{*}{\cellcolor{blue!5}2}  &  $\text{0.1 }$  &  $\text{-0.00021(2602)}$  &  $\text{0.0046(13)}$  &  $\text{0.00014(1713)}$  \\
				\hline
			\end{tabular}
			\caption{Fit results of the couplings $ c_s, c_p \text{ and } c_G $ from a fit to the available data on $B\to K^{(*)}\mu^+\mu^-$ and $B_s\to \phi\mu^+\mu^-$ decays and on $R_{K^{(*)}}$. The $p$-value, which represents the probability of the fit, is approximately 2\%. }
			\label{tab:b2s_global}
		\end{center}
	\end{table}

We have taken into account the constraints obtained from the available data on the differential rates and angular observables in $B \to K^{(*)}\mu^+\mu^-$ and $B_s \to \phi\mu^+\mu^-$ decays. The NP contributions to these decays from our simplified model are discussed in the later part of the sec.~\ref{para:raredecays}. Additionally, we have considered inputs on $R_{K^{(*)}}$, as provided in \cite{LHCb:2022qnv}. In this part of the analysis, we have not included the inputs on the branching fractions of the rare leptonic decays (eq.~\eqref{eq:rare_exp}). 
	
As discussed earlier, the new physics contributions to these decays arise through the operators $\mathcal{O}_s$, $\mathcal{O}_s^{'}$, $\mathcal{O}_p$, and $\mathcal{O}_p^{'}$. We performed a $\chi^2$ minimisation procedure using the  \emph{Mathematica\textsuperscript{\textregistered}} package~\cite{optex}, incorporating the available data on differential rates, isospin asymmetry, and angular observables for the decays $B \to K\mu^+\mu^-$, $B \to K^*\mu^+\mu^-$, and $B_s \to \phi\mu^+\mu^-$. In this analysis, we focused solely on the muon sector, as the new physics effects in $b \to s e^+e^-$ transitions are negligible. The relevant inputs for the differential rates and angular observables were taken from several references~\cite{CDF:2011tds, LHCb:2013lvw, LHCb:2014cxe, LHCb:2014vgu, LHCb:2015svh, Belle:2016fev, CMS:2017rzx, ATLAS:2018gqc, LHCb:2020gog, LHCb:2021zwz}, and we follow the approach detailed in~\cite{Biswas:2021pic}.
	
The fit results for various values of the scalar mass $M_S$ are presented in table~\ref{tab:b2s_global}. To examine the dependence on the cut-off scale $\Lambda$, we have performed the fits for $\Lambda = 2$ TeV and $\Lambda = 1$ TeV, respectively. From these results, it is evident that the constraints arising from the semileptonic decays $B \to K^{(*)}\mu^+\mu^-$ and $B_s \to \phi\mu^+\mu^-$ are relatively weak. Although some of the data included in the fit exhibit tension with the corresponding SM predictions, the resulting bounds on the new physics (NP) parameters remain quite relaxed, and in some cases, practically unconstrained.

This indicates that the current measurements of angular observables and decay rates in these channels are not sufficiently sensitive to the Wilson coefficients $C_{S}^{(\prime)}$ and $C_{P}^{(\prime)}$ to impose stringent limits. Consequently, the allowed values of the couplings $c_s$, $c_p$, and $c_G$ can be as large as $\mathcal{O}(10^{-2})$ GeV$^{-1}$, with a $p$-value of approximately 2\%. As we will demonstrate in the following sections, other datasets discussed earlier provide significantly tighter constraints on the NP parameter space. Therefore, in the remainder of this work, we will refer to flavour observables excluding the data on semileptonic $B \to K^{(*)}\mu^+\mu^-$ and $B_s \to \phi\mu^+\mu^-$ decays.   
\subsubsection{ \bf A combined study of Flavour data, EWPOs and Dipole moments} \label{sec:flavour_all_paramspcs}
\begin{table}[t]
	\footnotesize
	\centering
	\renewcommand{\arraystretch}{1.25}
	\begin{tabular}{|c|c|c|c|}
		\hline
		Observable & Value & Observable  & Value \\
		\hline
		$|V_{ud}|$ (nucl) & $0.97373 \pm 0.00009 \pm 0.00053 $ \cite{Xayavong:2021pkp} &	$|\varepsilon_K|$ & $(2.228 \pm 0.011) \times 10^{-3}$ \cite{PDG:2022}  \\
		$|V_{us}|f_+^{K \to \pi}(0)$ & $0.2165 \pm 0.0004$ \cite{ParticleDataGroup:2022pth} & 	sin~$2\beta$ & $0.699 \pm 0.017$ \cite{HFLAV:2022} \\
		$|V_{cd}|_{\nu N}$ & $0.230 \pm 0.011$ \cite{PDG:2022} & 	$\phi_s$ & $-0.057 \pm 0.021$ \cite{CKMFitter:2021}    \\
		$|V_{cs}|_{W \to c\bar{s}}$ & $0.94^{+0.32}_{-0.26} \pm 0.13$ \cite{ParticleDataGroup:2022pth} & $\alpha$ & $(85.2^{+4.8}_{-4.3})^\circ$ \cite{HFLAV:2022}  \\
		$|V_{ub}|_{excl}$ & $(3.91 \pm 0.13)\times 10^{-3}$ \cite{Biswas:2022yvh} & $\gamma$ & $(66.2^{+3.4}_{3.6} )^\circ $ \cite{HFLAV:2022}  \\
		$|V_{cb}|_{B\to D}$ & $(40.84 \pm 1.15) \times 10^{-3}$ \cite{Jaiswal:2017rve} &  	$ V_L $ & $ 0.995 \pm 0.021 $ \cite{PDG:2022}  \\
		$\frac{\mathcal{B}(\Lambda_p \to p \mu^- \bar{\nu}_\mu)_{q^2 > 15}}{\mathcal{B}(\Lambda_p \to \Lambda_c \mu^- \bar{\nu}_\mu)_{q^2 > 7}}$ & $(0.947 \pm 0.081)\times 10^{-2}$ \cite{LHCb:2015eia} & $ \Delta_s $ & $ -0.0345 \pm 0.0498 $ \cite{DiLuzio:2019jyq, LHCb:2021moh}    \\
		$\mathcal{B}(B^- \to \tau^- \bar{\nu}_\tau)$ & $(1.09 \pm 0.24) \times 10^{-4}$ \cite{HFLAV:2019} & $ \Delta_d $ & $ -0.0497 \pm 0.0518 $  \cite{DiLuzio:2019jyq,HFLAV:2022}   \\
		$\mathcal{B}(D_s^- \to \mu^- \bar{\nu}_\mu)$ & $(5.51 \pm 0.16) \times 10^{-3}$ \cite{HFLAV:2019} & 	$ R_b $ & $ 0.21629 \pm  0.00066 $ \cite{PDG:2022}     \\
		$\mathcal{B}(D_s^- \to \tau^- \bar{\nu}_\tau)$ & $(5.52 \pm 0.24) \times 10^{-2}$ \cite{HFLAV:2019}  &   	$ R_c $ & $ 0.1721 \pm  0.0030 $ \cite{PDG:2022}    \\
		$\mathcal{B}(D^- \to \mu^- \bar{\nu}_\mu)$ & $(3.77 \pm 0.18) \times 10^{-4}$  \cite{HFLAV:2019}  & 	$ R_e $ & $ 20.804 \pm 0.050 $ \cite{PDG:2022}     \\
		$\mathcal{B}(D^- \to \tau^- \bar{\nu}_\tau)$ & $(1.20 \pm 0.27) \times 10^{-3}$ \cite{HFLAV:2019} & 	$ R_{\mu} $ &$ 20.784 \pm 0.034 $ \cite{PDG:2022}  \\
		$\mathcal{B}(K^- \to e^- \bar{\nu}_e)$ & $(1.582 \pm 0.007) \times 10^{-5}$ \cite{PDG:2022}  & $ R_{\tau} $ & $ 20.764 \pm 0.045 $  \cite{PDG:2022} \\
		$\mathcal{B}(K^- \to \mu^- \bar{\nu}_\mu)$ & $0.6356 \pm 0.0011$ \cite{PDG:2022} & $ A_e  $ & $ 0.1515 \pm 0.0019 $ \cite{PDG:2022}    \\
		$\mathcal{B}(\tau^- \to K^- \bar{\nu}_\tau)$ & $(0.6986 \pm 0.0085) \times 10^{-2}$ \cite{HFLAV:2019} & $ A_{\mu} $ & $ 0.142 \pm 0.015 $  \cite{PDG:2022}   \\
		$\frac{\mathcal{B}(K^- \to \mu^- \bar{\nu}_\mu)}{\mathcal{B}(\pi^- \to \mu^- \bar{\nu}_\mu)}$ & $1.3367 \pm 0.0029$ \cite{PDG:2022} & 	$ A_{\tau} $ & $ 0.143 \pm 0.004 $ \cite{PDG:2022}   \\
		$\frac{\mathcal{B}(\tau^- \to K^- \bar{\nu}_\tau)}{\mathcal{B}(\tau^- \to \pi^- \bar{\nu}_\tau)}$ & $(6.467 \pm 0.84) \times 10^{-2}$ \cite{HFLAV:2019}  &  	$ A_{s} $ & $ 0.90 \pm 0.09 $ \cite{PDG:2022}  \\
		$\mathcal{B}(B_s \to \mu^+ \mu^-)$ & $(3.09 ^{+0.46 ~~ + 0.15}_{-0.43 ~~-0.11}) \times 10^{-9}$ \cite{LHCb:2021vsc} &  	$ A_c $ & $ 0.670 \pm 0.027 $ \cite{PDG:2022}  \\
		$\mathcal{B}(B_0 \to \mu^+ \mu^-)$ & $(0.12^{+0.08}_{-0.07}\pm 0.01)\times 10^{-9} $ \cite{LHCb:2021vsc} &  	$ A_b $ & $ 0.923 \pm 0.020 $ \cite{PDG:2022} \\			
		$|V_{cd}|f_+^{D \to \pi}(0)$ & $0.1426 \pm 0.0018$ \cite{HFLAV:2022} & 	$ \delta(\Delta r) \times 10^{3} $ & $ -0.191137 \pm 0.691496  $ \cite{CMS-PAS-SMP-23-002} \\ 
		$|V_{cs}|f_+^{D \to K}(0)$ & $0.7180 \pm 0.0033$  \cite{HFLAV:2022} & Re($V_{\text{R}}$)  & $[-0.11, 0.16]$ \cite{ATLAS:2016fbc,CMS:2016asd,CMS:2020ezf,CMS:2020vac}\\ 		
		\hline
	\end{tabular}
	\caption{List of observables from FCNC, FCCC, and the Z-pole observables used in the global fit. }
	\label{tab:CKM-updated-obs}
\end{table}
In this section, we present the results of our parameter space scan, incorporating all key observables from flavour physics, EWPOs, and the dipole moments discussed earlier. A summary of the main flavour and EWPO observables used in this analysis is provided in Table~\ref{tab:CKM-updated-obs}. In addition, we include the measured values and upper limits on the branching fractions of several rare and invisible decay modes, listed in Table~\ref{tab:rareinputs}. 
All these observables have been discussed in detail in the preceding sections.

In addition to the flavour and EWPT observables, we have analysed the impact of the available inputs on (c-)EDMs and (c-)MDMs. In the sec.~\ref{subsec:lepton_sector}, we have calculated the contributions in these observables from our scenario involving a spin-0 mediator. The SM predictions and the respective experimental bounds on the (c-)EDMs and (c-)MDMs which could be relevant to this analysis are shown in table~\ref{tab:EDM_MDM_values}. We have noted that among these observables, only the electron EDM imposes a stringent constraint; the others are comparatively weaker and do not significantly restrict the allowed parameter spaces relevant to our analysis. We have discussed the details of this numerical check in the appendix~\ref{secappndix:edm} for a few benchmark scenarios allowed by electroweak precision test and flavour physics observables discussed in the earlier subsections. For the electron EDM, the values obtained at the benchmark points are around $\mathcal{O}(10^{-30})$, which is close to the current experimental limit. Therefore, we focus on four observables: the electron EDM, the top EDM, the top quark cEDM, and the top quark cMDM. We have taken into account these constraints and obtained the allowed parameter space from a scan, the results of which are presented in the figures discussed below.
As outlined earlier, we will report results for two distinct mass ranges of $M_{S}$.
\vspace{0.5cm}
\paragraph{\underline{\bf Mass region $ M_{S}> 3 $ GeV :} }
This paragraph will show the allowed parameter spaces corresponding to $ M_{S} > 3 $. We will vary $ M_{S} $ in the region: $ 3 ~\text{GeV}<  M_{S} \leq 10 \rm ~GeV $.
	\begin{figure}[t]
		\centering
		\subfloat[]{\includegraphics[scale=0.17]{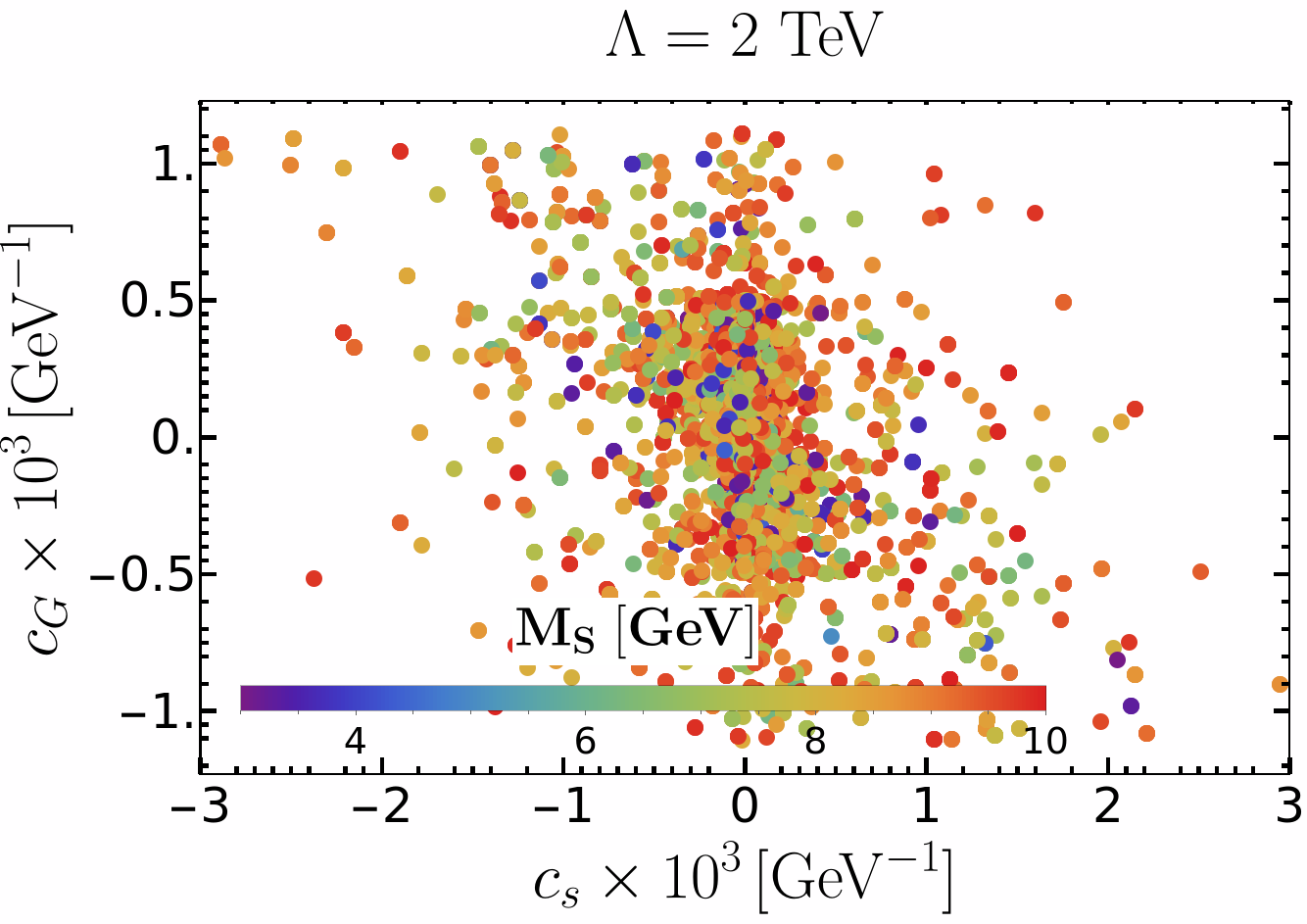}}\hspace{0.0001cm}
		\subfloat[]{\includegraphics[scale=0.17]{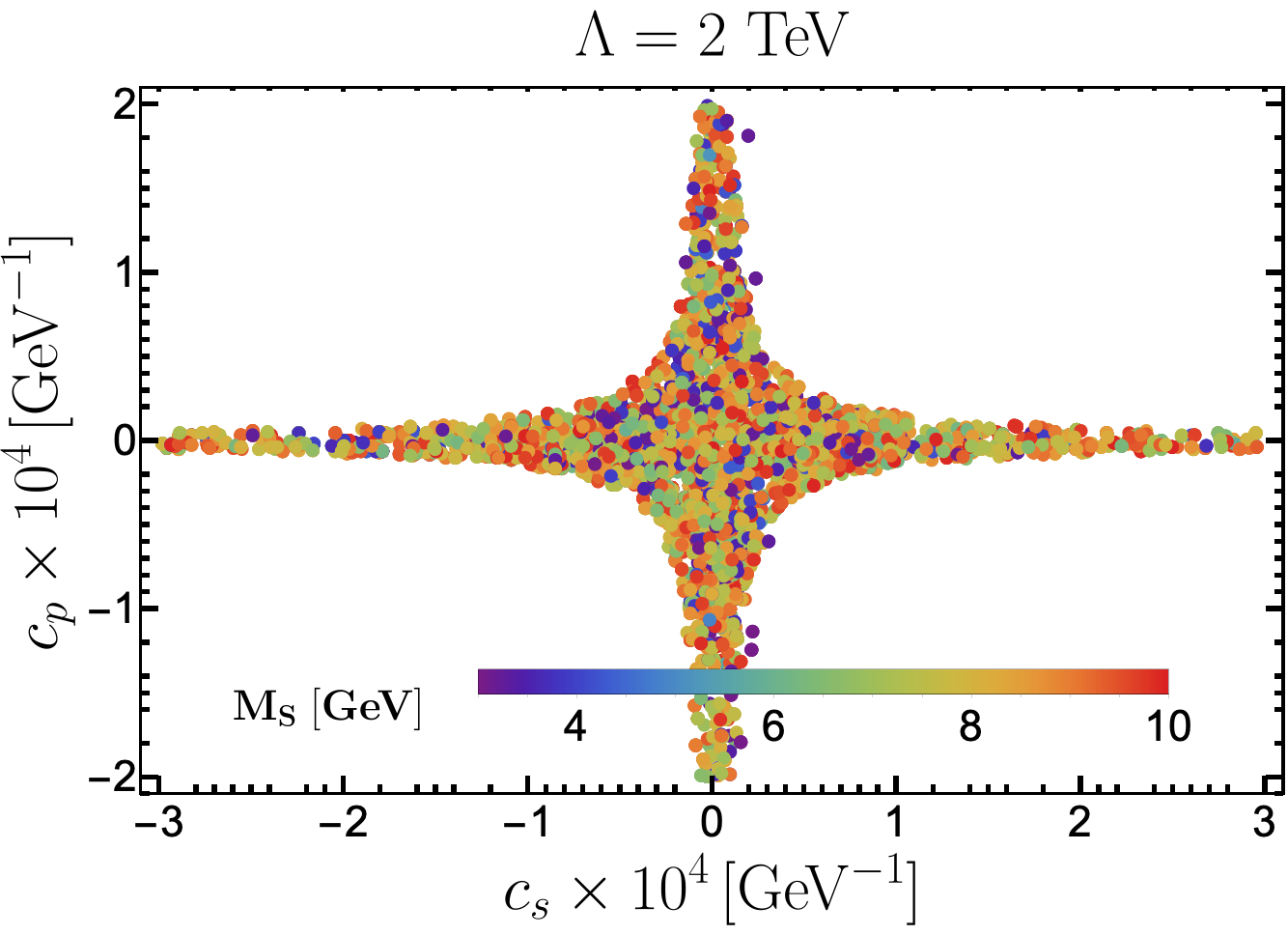}\label{fig:flavor_cscp_higher}}\hspace{0.0001cm}
		\subfloat[]{\includegraphics[scale=0.17]{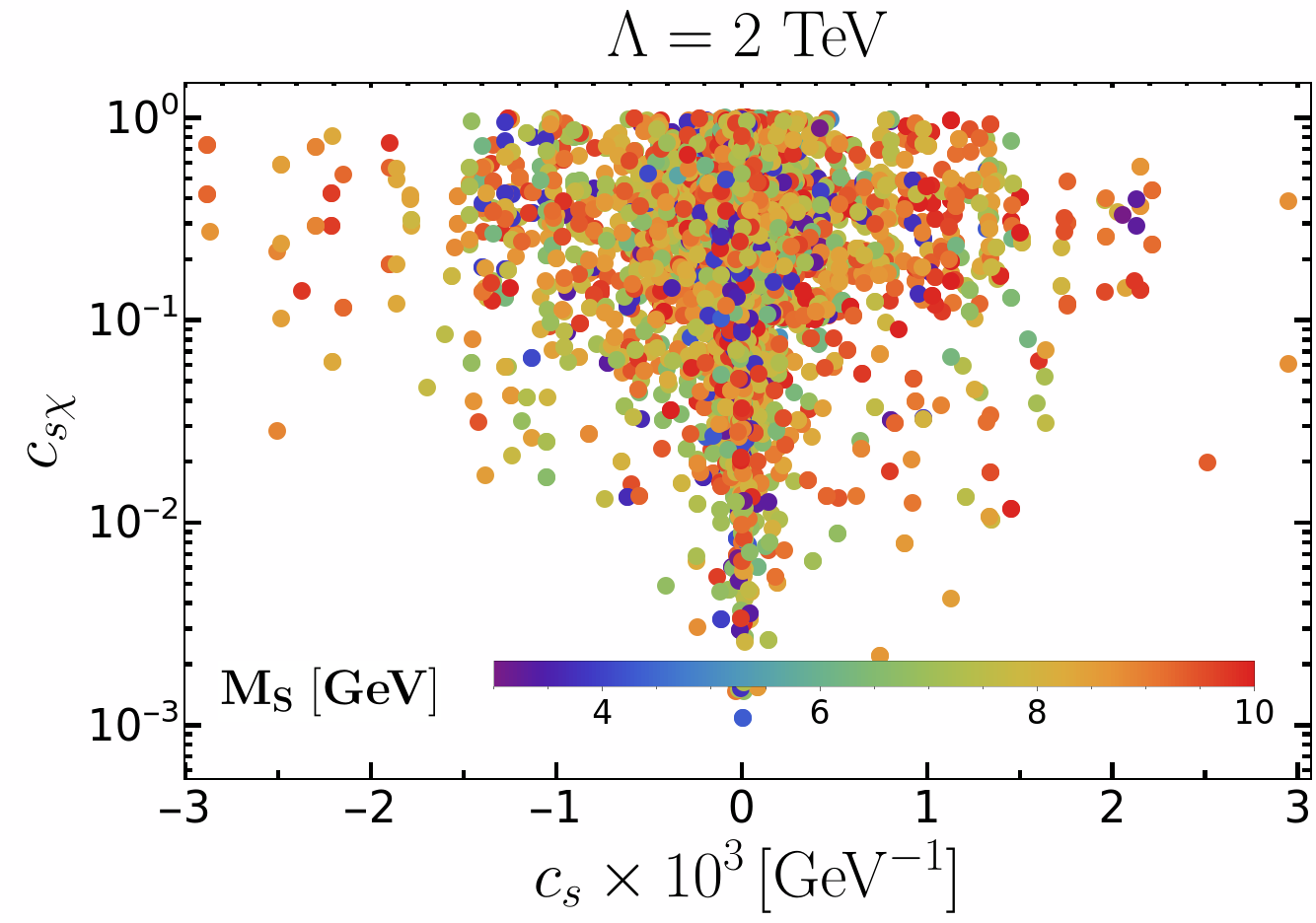}\label{fig:flavor_cscsX_higher}}\hspace{0.0001cm}
		\subfloat[]{\includegraphics[scale=0.17]{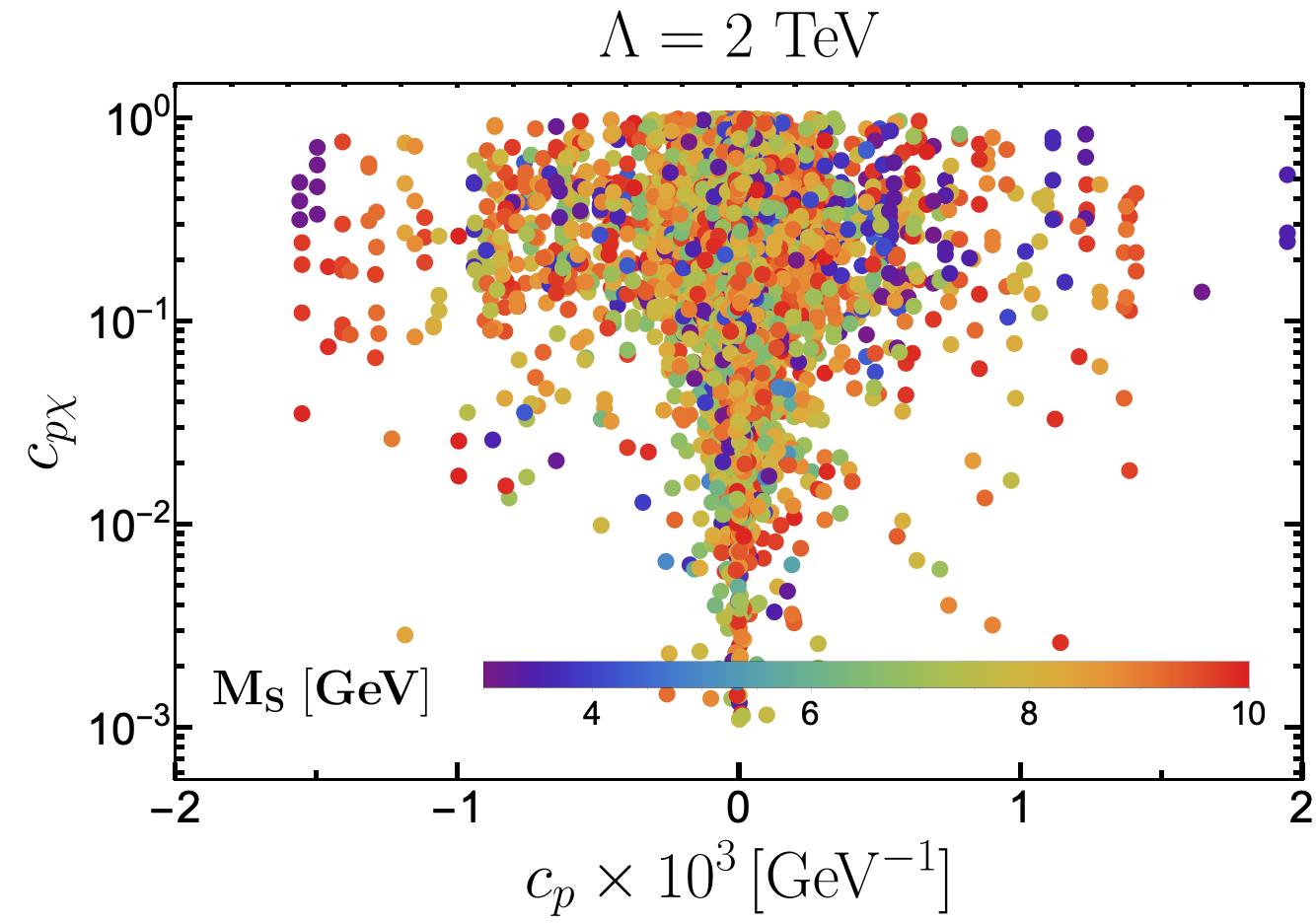}\label{fig:flavor_cpcpX_higher}}\\
		\subfloat[]{\includegraphics[scale=0.17]{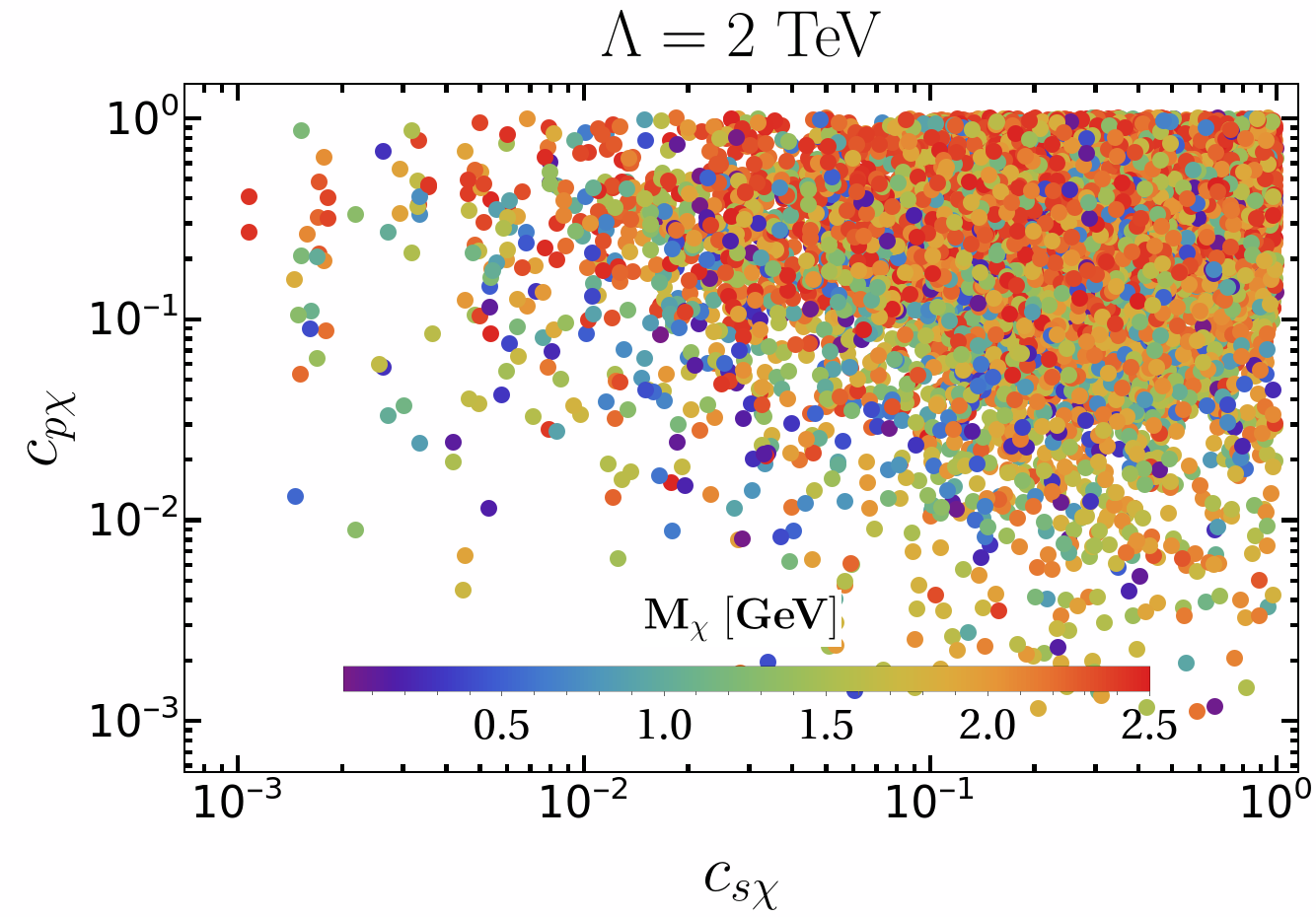}\label{fig:flavor_csXcpX_higher}}
\caption{The correlation and allowed parameter space of different couplings and the mediator mass are determined based on all previously discussed constraints (flavour observables, EWPOs, dipole moments and meson invisible decays). The parameter space is evaluated for the scale $\Lambda = 2~\text{TeV}$. We use the $1\sigma$ experimental uncertainty (or the upper limit, where applicable) of the relevant observables. The analysis focuses on the region where $M_S > 3~\text{GeV}$.} \label{fig:flavor_highermass}
\end{figure}
In this analysis, we have incorporated all the constraints listed in tables~\ref{tab:CKM-updated-obs}, \ref{tab:EDM_MDM_values} and \ref{tab:rareinputs}, respectively. Due to kinematical constraints, we cannot obtain meaningful bounds from fixed-target experiments for $M_S > 3$~GeV, as discussed earlier. The constraints on the parameters $c_{s,p,G}$ from fixed target experiments are discussed separately in appendix~\ref{sec:fixedtarget}. 

Among the various inputs used, the invisible decay channels such as $P \to P' \chi \bar{\chi}$ play an essential role in restricting the allowed mass of the dark matter $M_{\chi}$. As discussed, for $ P \to P' \chi \bar{\chi}$ decay, $ S $ works as a virtual particle. So here, the mass of $ \chi $ is allowed upto $ M_{\chi} \leq \frac{M_{P}-M_{P'}}{2}$. Hence, for the rare decays $B\to K\chi\bar{ \chi}$, the kinematically allowed value of $M_{\chi} \lesssim 2.5$ GeV. Based on this observation, for the parameter scan, we vary the model parameters within the following ranges:
\[
\big( |c_{s}|, |c_{p}|, |c_{G}| \big) \leq 0.1~\text{GeV}^{-1}; \quad 3~\text{GeV} \leq M_{S} \leq 10~\text{GeV}; \quad c_{s(p)\chi} \leq 1; \quad M_{\chi} \leq 2.5~\text{GeV}.
\]
To ensure comprehensive coverage of the parameter space, we generated random sets of benchmark points spanning the entire ranges specified above. All parameters were allowed to vary simultaneously without any imposed bias. This procedure was implemented in \emph{Mathematica\textsuperscript{\textregistered}} using machine precision. For each set of random points, we computed the predicted values of the relevant observables. The parameter points for which these predictions satisfy the experimental constraints are presented in the plots shown in fig.~\ref{fig:flavor_highermass}.
	
The plots in fig.~\ref{fig:flavor_highermass} show the allowed parameter space and the respective correlations in the coupling and mass planes. 
The allowed ranges of the parameters $c_s$, $c_p$, and $c_G$ satisfy
\begin{equation}
	|c_{s(p)}| < 5 \times 10^{-4}\ \text{GeV}^{-1}, 
	\qquad 
	|c_G| \lesssim 10^{-3}\ \text{GeV}^{-1},
\end{equation}
placing them at or below the order of magnitude $10^{-4}$. In particular, the phenomenologically viable values of $c_s$ and $c_p$ are strongly concentrated around magnitudes of order $10^{-5}$.

It is important to note that the couplings $c_s$ and $c_p$ are highly correlated. This correlation arises mainly from the stringent constraints imposed by electric and magnetic dipole moments, since in our model the contributions to (c-)EDMs and (c-)MDMs scale as the product $c_s c_p$. Consequently, relatively large values of $c_s$ are allowed only when $c_p$ is very small, and vice versa. This restricts the allowed parameter space to a narrow region where the product $c_s c_p$ remains sufficiently suppressed to satisfy dipole-moment bounds.

Also, we have obtained bounds on $ c_{s(p) \chi} $ since the invisible decays will be sensitive to these couplings. The respective bounds or the correlations are shown in figs.~\ref{fig:flavor_cscsX_higher}, \ref{fig:flavor_cpcpX_higher} and \ref{fig:flavor_csXcpX_higher}, respectively. As we can see, there are more concentrations of solutions for $ c_{s (p)\chi} > 1\times 10^{-2}$. Also, the concentration of these solutions are for $ c_{s(p)} \lesssim 1\times 10^{-4}$. These lower limits are due to the tight upper limits on $ c_{s(p)}$. The correlation between $ c_{s(p)} $ and $ c_{s(p)\chi} $ comes from the branching ratios of the invisible decays, which vary as $ \propto c_{s(p)}^2  c_{s(p)\chi}^2  $, as shown in eq.~\eqref{eq:Br_inv_virtual}. Moreover, within the viable ranges of the couplings, the dark--matter mass $M_{\chi}$ is driven toward values $\gtrsim 1.5~\text{GeV}$, as illustrated in fig.~\ref{fig:flavor_csXcpX_higher}. The correlations shown in fig.~\ref{fig:flavor_highermass} further indicate that solutions tend to cluster in the region with mediator masses $M_S \geq 5~\text{GeV}$. As noted earlier, the bounds from dipole moment measurements impose restrictions on the couplings $c_s$ and $c_p$, although these observables are largely insensitive to the mediator mass $M_S$. Consequently, given the constraints on $c_s$ and $c_p$, the dominant constraints on the mediator and dark matter masses originate from rare pseudoscalar meson decays and invisible channels such as $P \to P' \chi \bar{\chi}$.

\begin{figure}[t]
\centering
\subfloat[]{\includegraphics[scale=0.17]{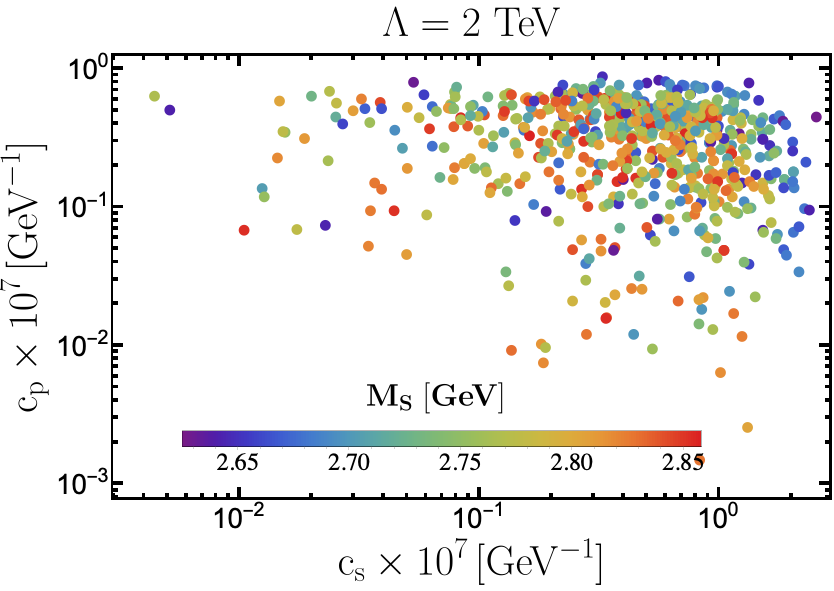}}\hspace{0.0001cm}
\subfloat[]{\includegraphics[scale=0.17]{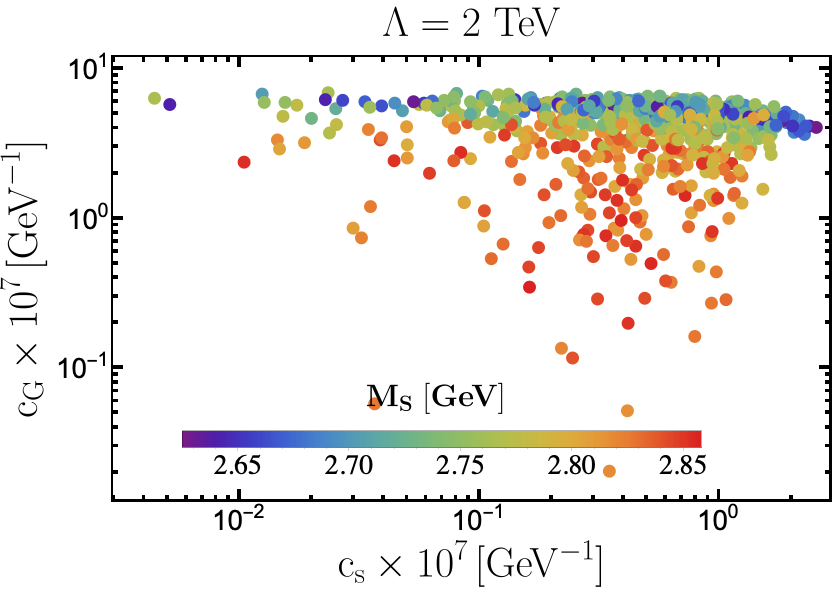}}\hspace{0.0001cm}
\caption{Allowed ranges and correlations among the model parameters, i.e., the mediator mass and couplings, are shown. The parameter space is constrained by all the discussed observables from flavour physics, electroweak precision tests, invisible decays, and fixed-target experiments. The analysis is performed for mediator masses $M_{S} \leq (M_{B} - M_{K})$, in order to allow invisible decays of the type $P \to P' S$. The plot corresponds to the $1\sigma$ allowed values of the decays $B \to K S$ as given in \cite{Altmannshofer:2023hkn}. Here, we set $\Lambda = 2~\text{TeV}$.}\label{fig:flavor_lowermass_1sig}
\end{figure}
\vspace{0.5cm}
\paragraph{\underline{\bf Mass region $ M_{S}\leq 3$ GeV :}}
This paragraph shows the allowed parameter space for masses and couplings with $M_{S} \leq 3$ GeV. For the analysis, as inputs, we have taken all the flavour observables (FCNCs) and EWPOs, which we have presented in tables ~\ref{tab:rareinputs}, \ref{tab:EDM_MDM_values} and \ref{tab:CKM-updated-obs}, respectively. As mentioned, for these mass regions, the invisible decays like $ P \to P' S $ with $ P= B, K $ and $ P' = K^{(*)},\pi $ mesons will play an important role in constraining the parameter space. For the decay $ B \to K^{(*)} + S $, we have taken the inputs from \cite{Altmannshofer:2023hkn} where the variation of $\mathcal{B}(B \to K^{(*)} + S )$ with $ M_{S} (\lesssim 3  $ GeV) presented. For this process, we have analysed the distribution in branching ratios with the mass $M_S$ considering the errors in $\mathcal{B}(B \to K^{(*)} + S )$ at their $ 1\sigma  $ and $ 2 \sigma  $ confidence intervals. This is because the $ \mathcal{B}(B \to K + S)$ at its 1$\sigma$ range is inconsistent with zero. However, at the 2$\sigma$ range, this branching fraction is consistent with zero. On the other hand, the $ \mathcal{B}(B \to K^* + S)$ is consistent with zero even within its 1$\sigma$ error bar. Hence, the bounds on the parameters will depend on whether we are considering $ \mathcal{B}(B \to K + S)$ at its $1\sigma$ or $2\sigma$ range. 

Furthermore, constraints from the CHARM fixed‑target experiment, which we have discussed in sec.~\ref{subsec:fixtarget}, are particularly relevant in this mass range. When $M_S \leq (M_P - M_{P'})$, the decay $P \to P'\chi\bar{\chi}$ can proceed resonantly through an on‑shell spin-0 mediator, $P \to P' + S$, followed by $S \to \chi\bar{\chi}$.
Therefore, to constrain the allowed parameter spaces, the bounds from fixed‑target experiments must be included together with the other experimental constraints. The observable relevant for CHARM is the number of detected events, $N_{\text{det}}$, defined in eq.~\eqref{eq:fix_target_N}. In our framework, $N_{\text{det}}$ receives contributions from the branching fractions $\mathcal{B}(K\to \pi S)$ and $\mathcal{B}(B\to K^{(*)} S)$, and from the decay width $\Gamma(S\to \chi\bar{\chi})$. Since the CHARM experiment did not observe any events, it imposes an upper bound: $N_{\text{det}} < 2.3$.

In appendix~\ref{sec:fixedtarget}, we present the regions of parameter space excluded solely by fixed-target constraints. For scalar masses $M_S < 0.1~\text{GeV}$, these bounds force the relevant couplings to be extremely small, typically $\lesssim 10^{-8}~\text{GeV}^{-1}$; see fig.~\ref{fig:fix_target_scan}. For $M_S > 0.1~\text{GeV}$, however, the fixed-target limits become considerably weaker (fig.~\ref{fig:fix_target_scan}), allowing couplings of order $\gtrsim 10^{-7}~\text{GeV}^{-1}$.

\begin{figure}[t]
\centering
\subfloat[]{\includegraphics[scale=0.17]{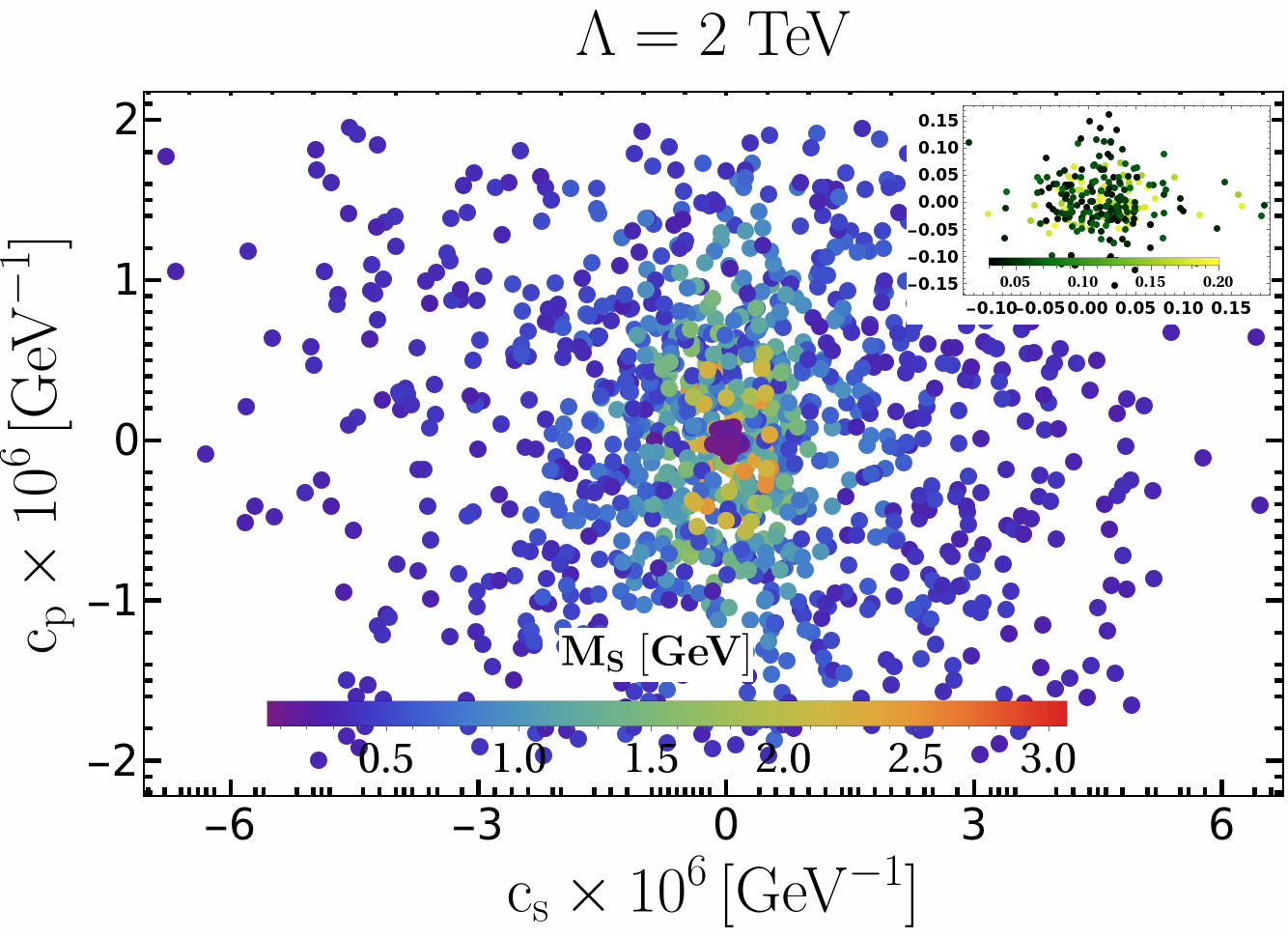}}\hspace{0.0001cm}
\subfloat[]{\includegraphics[scale=0.17]{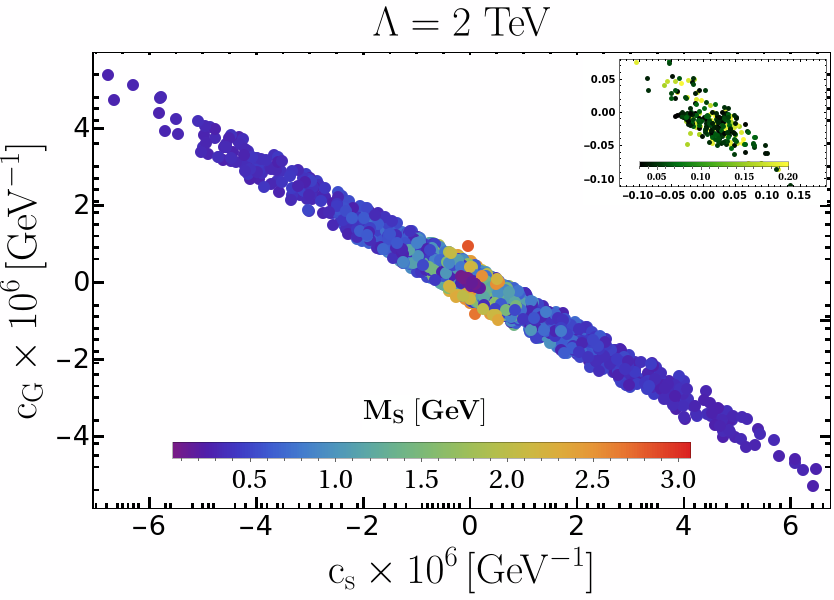}}\hspace{0.0001cm}
\subfloat[]{\includegraphics[scale=0.17]{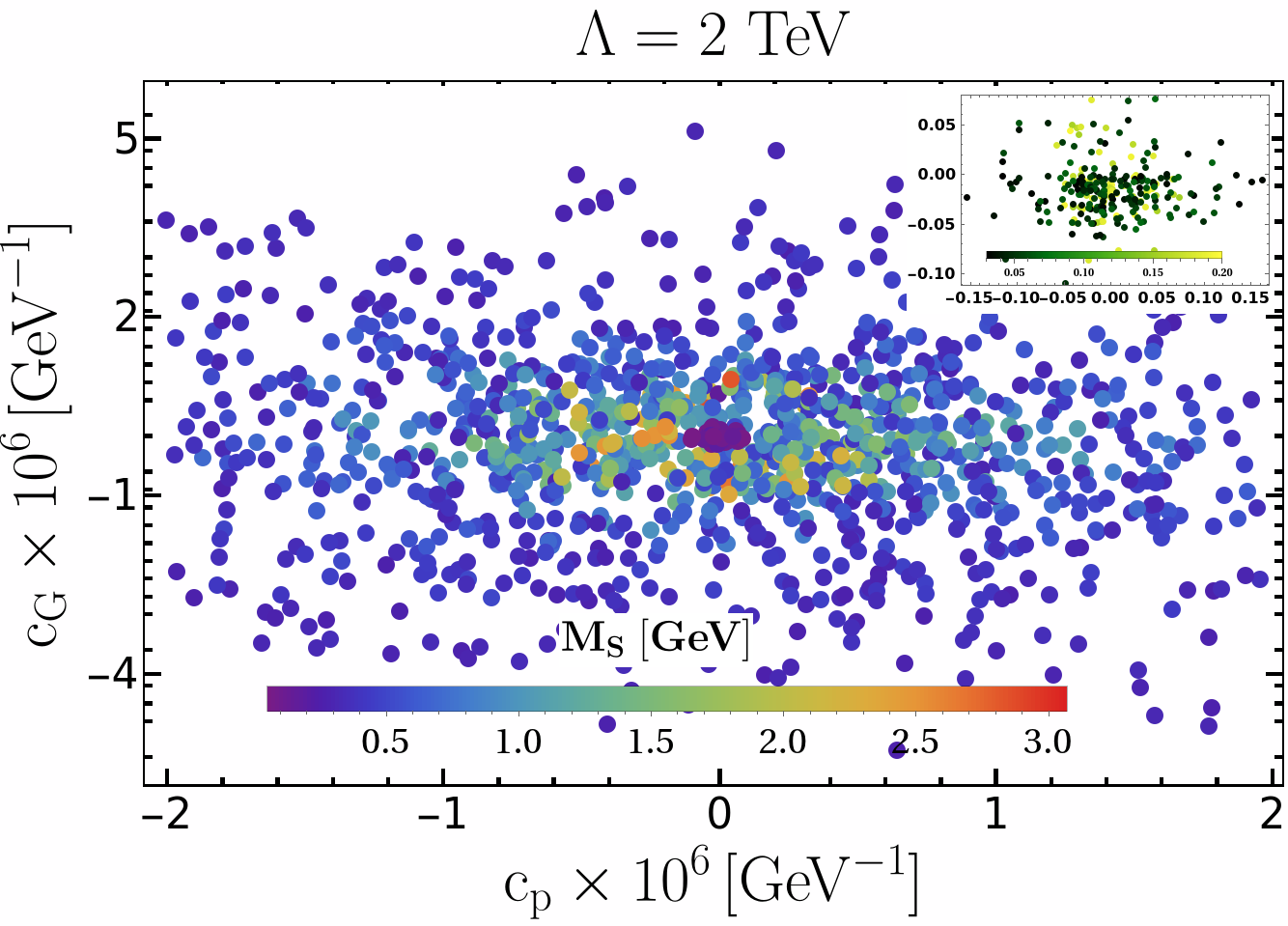}}
\caption{Similar to the previous plots in fig.~\ref{fig:flavor_lowermass_1sig}, here we take the $2\sigma$ allowed values for the invisible decays $B \to K S$ as given in \cite{Altmannshofer:2023hkn}. This analysis also focuses on the region $M_{S} \leq 3 \,\text{GeV}$.}\label{fig:flavor_lowermass_2sig}
	\end{figure}

In fig.~\ref{fig:flavor_lowermass_1sig}, we have shown the allowed parameter space in $ c_{s} $-$ c_{p} $-$ c_{G} $ planes for $M_{S} \leq 3 $ GeV for 1$\sigma$ error band in the distribution in $ \mathcal{B}(B \to K^{(*)} + S)$ which is not consistent with zero. As discussed above, the data from the fixed-target experiment and the $\mathcal{B}(B \to K^{(*)} + S )$ impose strong constraints on the parameter spaces. From the parameter scan, we get solutions only for $M_S \gtrsim 2.5$ GeV and more concentration of allowed points for the values of couplings in the regions: 
$$ 10^{-9}\, \rm GeV^{-1} \lesssim  \left( |c_{s}|, |c_{p}| , |c_{G}| \right)  \lesssim 10^{-7} \, \rm GeV^{-1}.$$ Values of $M_S < 2.5$~GeV are excluded because, in this mass range, it is not possible to simultaneously satisfy the measured values of $\mathcal{B}(B \to K^{(*)} + S)$ and the fixed-target constraint. The $1\sigma$ ranges of $\mathcal{B}(B \to K^{(*)} + S)$ require the couplings to lie in the interval $10^{-9} - 10^{-7}$~GeV$^{-1}$. However, for $M_S < 2.5$~GeV, couplings within these ranges are ruled out by the fixed-target bounds (see fig.~\ref{fig:fix_target_scan}). Consequently, there is no region of overlap between the parameter space allowed by flavor observables and that permitted by the fixed-target experiment when $M_S < 2.5$~GeV. As a result, the combined constraints yield substantially tighter bounds on all three coupling planes than in the case with $M_S > 3$~GeV.
	
If we instead apply the $2\sigma$ bounds on the processes $B \to K^{(*)} + S$, then viable regions of parameter space appear across the entire mass range $0.01 \leq M_S \leq 3~\text{GeV}$\footnote{The choice of the lower limit on $M_S$ is motivated by the fact that, below this value, fixed-target experiments allow only highly suppressed couplings.}. In this case, the allowed solutions remain consistent with zero because the $2\sigma$ limits on $\mathcal{B}(B \to K^{(*)} + S)$ are themselves compatible with zero. As a result, the constraint effectively provides only an upper bound on the decay rate. Consequently, even very small coupling values satisfy the $2\sigma$ requirement, leading to allowed parameter space throughout the entire kinematically accessible region of $M_S$. The resulting correlations among the parameters are illustrated in fig.~\ref{fig:flavor_lowermass_2sig}. A strong correlation among the parameters $c_{s} $ and $ c_{G} $ is obtained, and their magnitudes are positively correlated. Furthermore, the positive values of $c_s$ prefer the negative values of $c_G$ and vice versa. 
The bounds we obtained for this case are:
\[ |c_{s}| \lesssim 5 \times 10^{-6} \, \text{GeV}^{-1} ;  \ \ |c_{p}| \lesssim 2 \times 10^{-6}\, \text{GeV}^{-1} ;  \ \ |c_{G}| \lesssim 4 \times 10^{-6}\, \text{GeV}^{-1} ;    \]  
For $M_S < 0.2~\text{GeV}$, the allowed values of the couplings become negligibly small (below $5 \times 10^{-8}~\text{GeV}^{-1}$). This region is therefore shown in the zoomed-in inset in the upper-right corner of the figures to make the fine structure visible. As discussed above, the bounds in this mass range of $M_S$ arise predominantly from fixed-target experiments.

In the previous case ($ M_{S} > 3 $) GeV, we have seen that the bound gets relaxed with increasing the mass of the mediator. In this case( $ M_{S} \leq 3 $ GeV), the most stringent bound is obtained for a relatively higher mass of the mediator. This nature is coming from the CHARM constraint, which again can be realised from fig.~\ref{fig:fix_target_scan}.
\subsection{Allowed parameter spaces including the data on dark sector} \label{subsec:onlyDM}
In the preceding subsections, we examined the constraints on the parameter space arising from flavour observables, EWPOs, and electric and magnetic dipole moment measurements. These analyses revealed correlated regions among the relevant couplings and masses. For mediator masses $M_{S} > 3~\mathrm{GeV}$, the typical allowed values lie around $c_{s,p} \lesssim 10^{-4}~\mathrm{GeV}^{-1}$, whereas for lighter mediators with $M_{S} \leq 3~\mathrm{GeV}$, the bounds become significantly more stringent, yielding $c_{s,p} \lesssim 10^{-6}~\mathrm{GeV}^{-1}$.

To investigate the implications of current dark matter data within this framework, we therefore divide our analysis into two representative mass regimes: $M_{S} > 3~\mathrm{GeV}$ and $M_{S} \leq 3~\mathrm{GeV}$. Since our primary interest lies in the portal interactions connecting the mediator, the SM, and the dark sector, we fix the coupling $\lambda_{1} = 0.001$ (eq.~\ref{eq:modelpotential}) throughout the analysis.

\subsubsection{\bf Bounds for $ M_{S} > 3 $ GeV} 
\begin{figure}[t]
	\centering
		\subfloat[]{\includegraphics[scale=0.17]{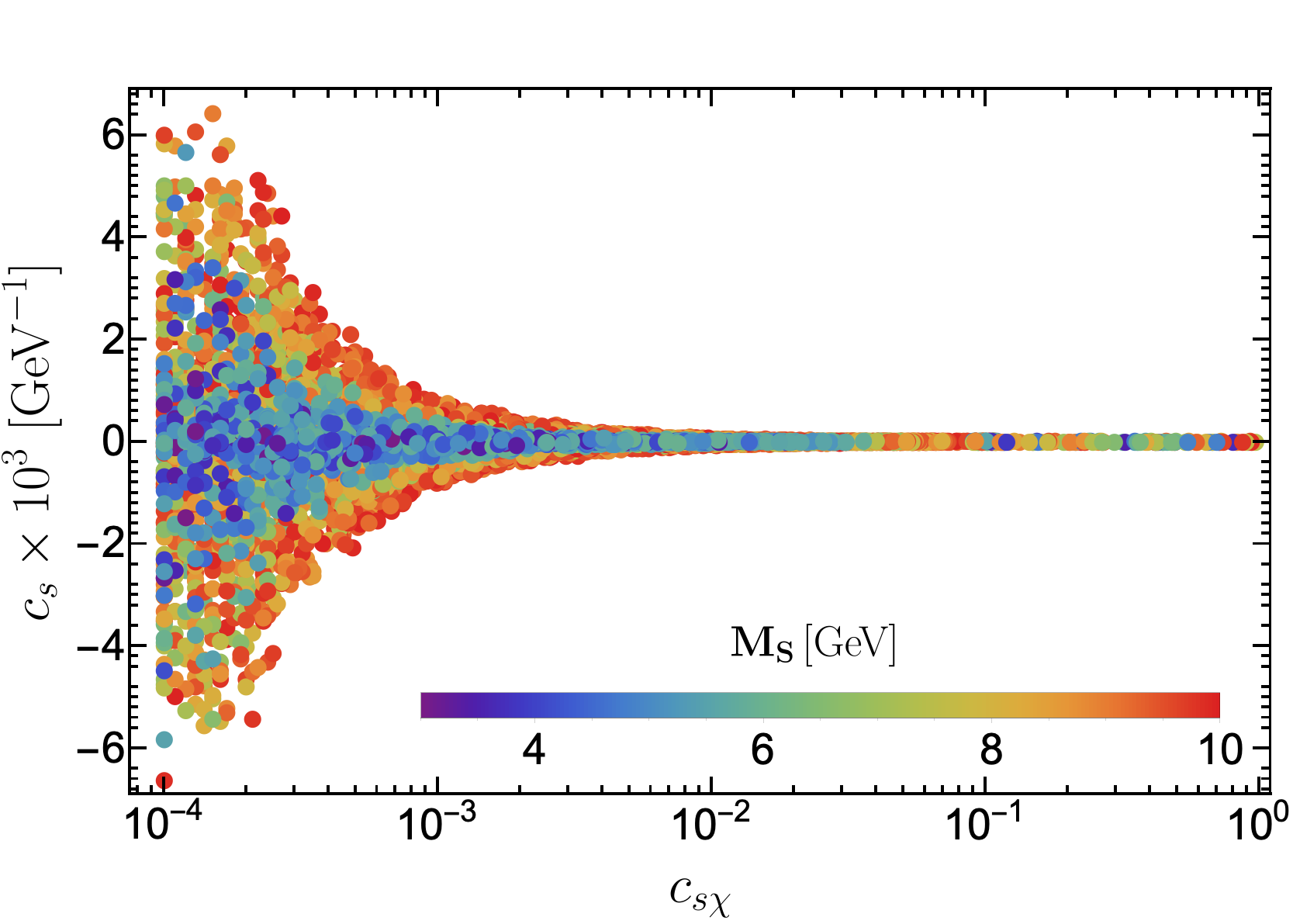}\label{fig:DM_highermass3}}\hspace{0.0001cm}
	\subfloat[]{\includegraphics[scale=0.17]{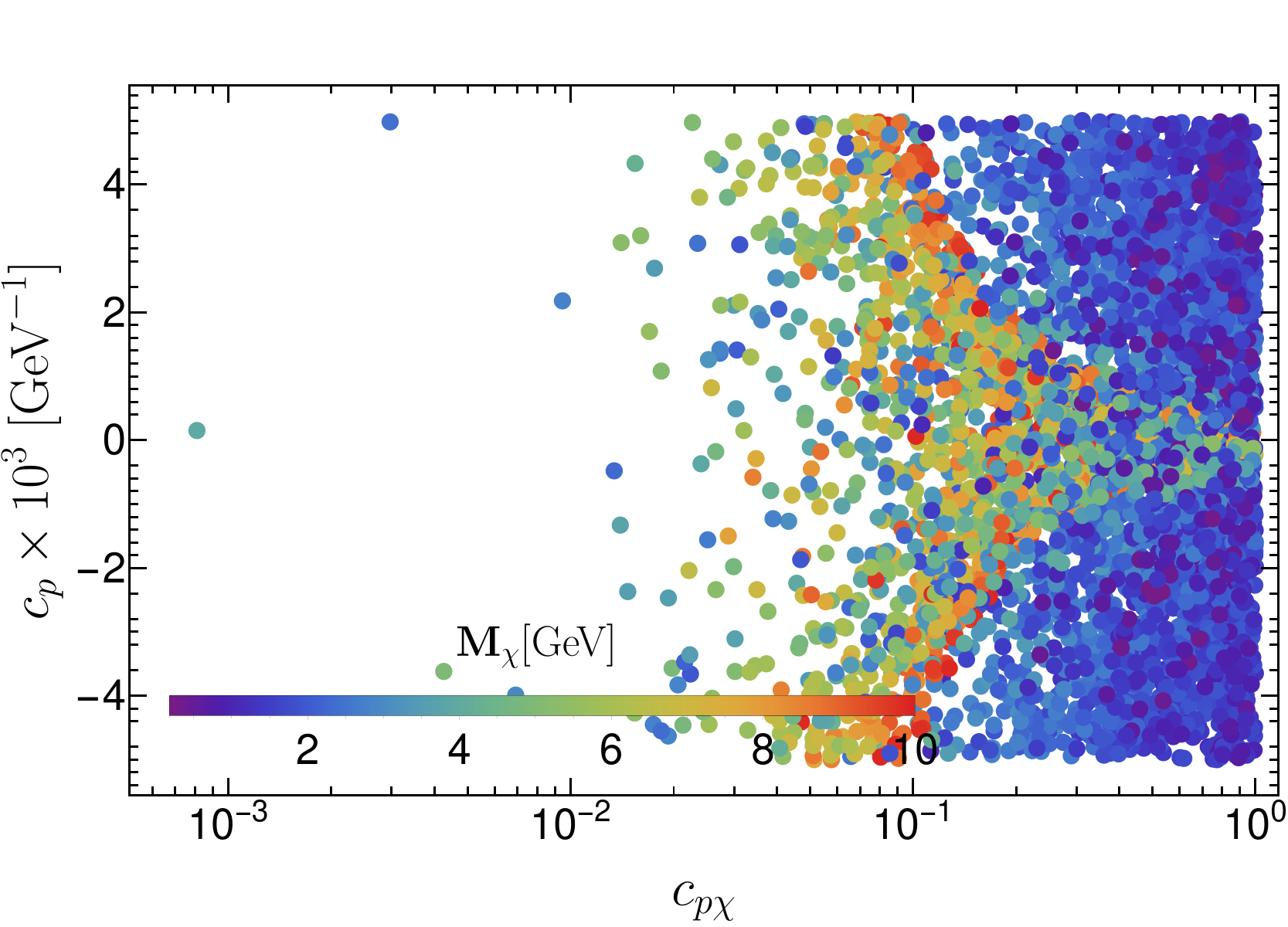}\label{fig:DM_highermass4}}\hspace{0.0001cm}
	\subfloat[]{\includegraphics[scale=0.17]{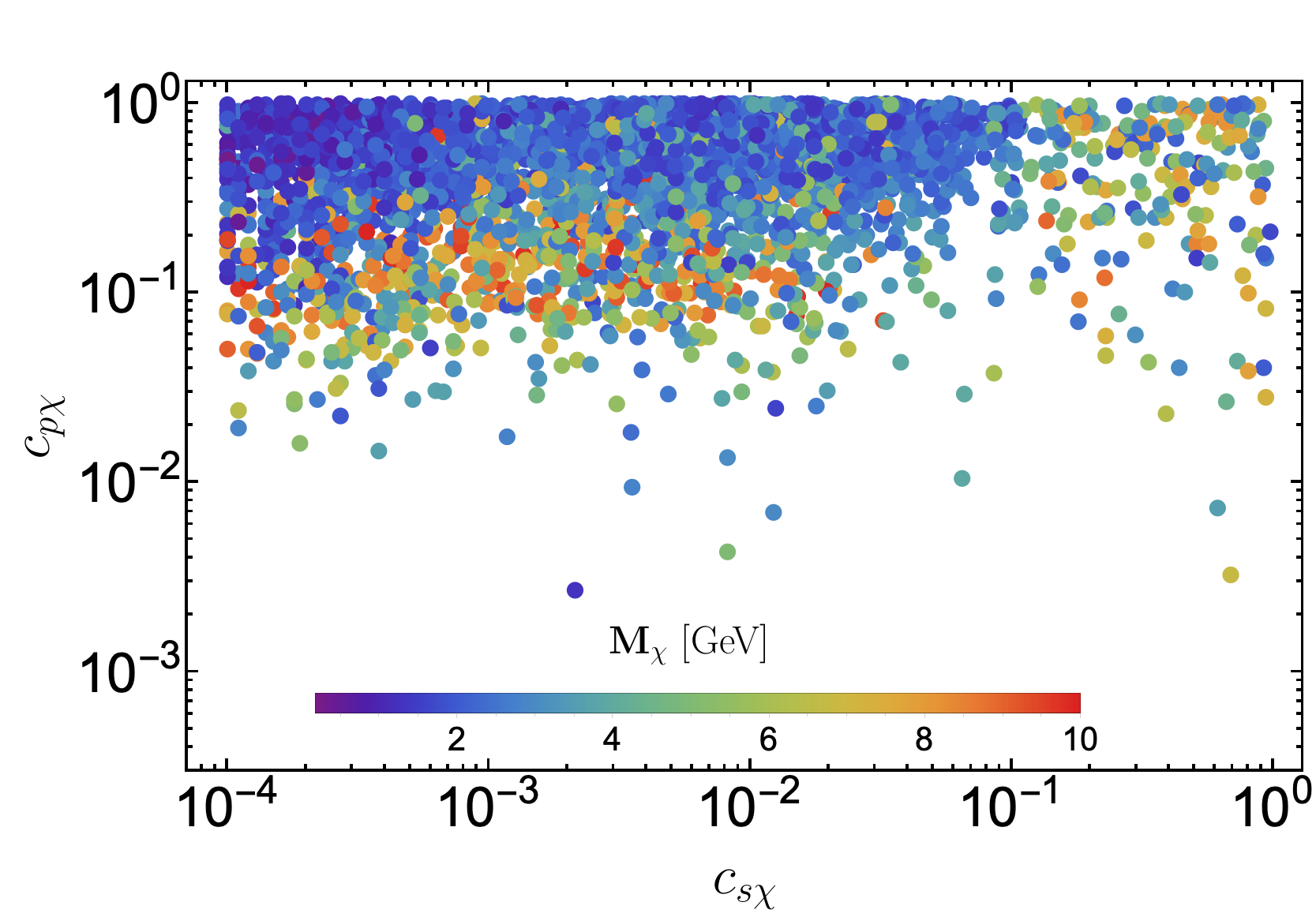}\label{fig:DM_highermass2}}\hspace{0.0001cm}
	\subfloat[]{\includegraphics[scale=0.17]{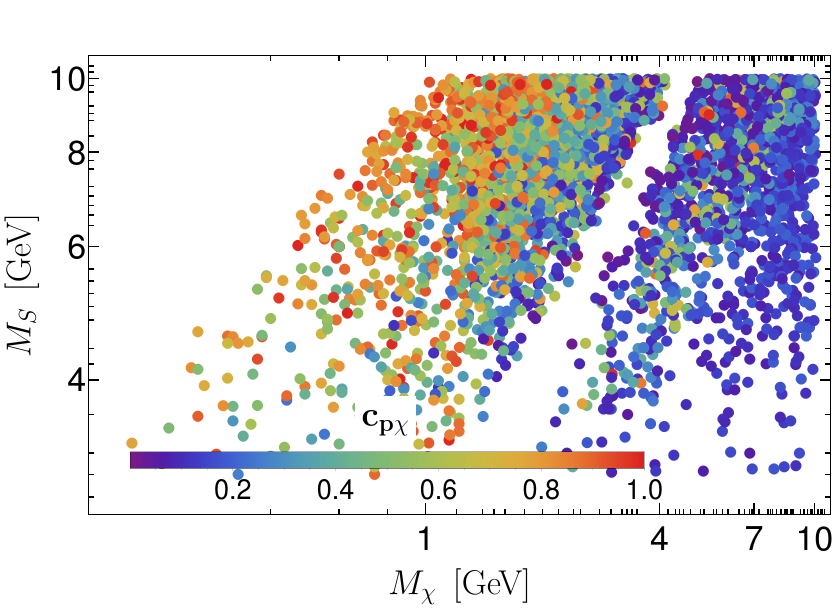}\label{fig:DM_highermass1}}
	\caption{Allowed points from the constraints of dark sectors, i.e., relic density \cite{Planck:2018vyg} and DD \cite{XENON:2025vwd, LZ:2024zvo}, ID cross-sections and correlations among the masses and the couplings for the mediator mass region $ M_{S} > 3 $ GeV. The colorbar is shown with coupling ($c_{p\chi} $ for fig.~\ref{fig:DM_highermass1} and mass of the DM $M_{\chi}$ for the other plots). }\label{fig:DM_highermass}
\end{figure}

We begin by examining the constraints on the parameter space arising from dark-sector observations for mediator masses $M_S > 3~\mathrm{GeV}$. Fig.~\ref{fig:DM_highermass} displays the corresponding allowed regions, incorporating limits from the relic density ($\Omega h^2$) as well as the spin-independent direct-detection cross-sections and indirect detection cross-sections. Guided by the constraints obtained from flavour observables, EWPOs, and EDM measurements, we perform a comprehensive scan of the dark-sector parameter space over the ranges:
$$
(|c_s|, |c_p|) \leq 0.005~\mathrm{GeV}^{-1}; \quad 3~\mathrm{GeV} \leq M_S \leq 10~\mathrm{GeV}; \quad c_{s(p)\chi} \leq 1; \quad M_\chi \leq 10~\mathrm{GeV},
$$
with a minimum step size of $10^{-4}$ for $c_{s(p)\chi}$. The mediator-gauge boson coupling $c_G$ is neglected in this regime due to its irrelevance for low-mass mediators. In the following items, we present the information that we can decode from fig.~\ref{fig:DM_highermass}:
\begin{itemize}
\item Fig.~\ref{fig:DM_highermass3} illustrates the allowed parameter space in the $c_s$-$c_{s\chi}$ plane. For fixed values of $M_S$ and $M_{\chi}$, the combined constraints from relic density and direct detection require that any increase in $c_{s\chi}$ be compensated by a decrease in $c_s$. In particular, for $c_{s\chi} \geq 0.001$, we find that $|c_s| \lesssim 10^{-4}~\mathrm{GeV}^{-1}$. For $c_{s\chi}\sim 1$ region, we get $c_s \lesssim 10^{-5} \, \rm GeV^{-1}$.  Furthermore, the constraints on $c_s$ become increasingly stringent for 
larger $c_{s\chi}$, driven primarily by the bounds from direct-detection experiments.

\item Fig.~\ref{fig:DM_highermass4} presents the correlation between $c_{p\chi}$ and $c_p$, driven by relic density constraints. For $M_\chi \geq 5~\mathrm{GeV}$ and $c_{p\chi} \geq 0.5$, the pseudoscalar coupling is constrained to $|c_p| \lesssim 0.5 \times 10^{-3}~\mathrm{GeV}^{-1}$.
	
\item   Fig.~\ref{fig:DM_highermass2} illustrates the correlation between $c_{s\chi}$ and $c_{p\chi}$, with color indicating the variation of the DM mass $M_\chi$. For lower DM mass, most points accumulate towards higher $c_{p\chi}$. In contrast, for $c_{s\chi}$, the whole region is allowed, and the lower values will be favoured by DD-bounds. The coupling $c_{s\chi}$ is strongly constrained by direct detection. For $M_\chi > 6~\mathrm{GeV}$, $c_{p\chi}$ is limited to $\gtrsim 0.05$.

\item Finally, fig.~\ref{fig:DM_highermass1} shows the allowed region in the $M_S$-$M_\chi$ plane, with color indicating the mediator-DM pseudoscalar coupling $c_{p\chi}$. Near the resonance region $M_S \approx 2 M_\chi$, direct detection constraints exclude many points. For $M_\chi > M_S$, the annihilation channel $\chi\bar{\chi} \to SS$ opens, requiring smaller DM couplings to satisfy relic density bounds. In this region, larger $c_{p\chi}$ values correspond to smaller $c_s$ and $c_p$.

\end{itemize}

Overall, in the $M_S > 3~\mathrm{GeV}$ regime, most of the parameter space remains viable, except for $c_s$ and $c_{s\chi}$, which are tightly correlated for large $c_{s\chi}$ from DD cross-section bounds. Strong correlations among masses and couplings emerge, shaping the final allowed regions.  
%

\vspace{0.5cm}
\paragraph{\underline{ \bf  Combined analysis: Flavour + EWPOs + Dipole moments + DM : } } 

\begin{figure}[t]
	\centering		
	\subfloat[]{\includegraphics[scale=0.17]{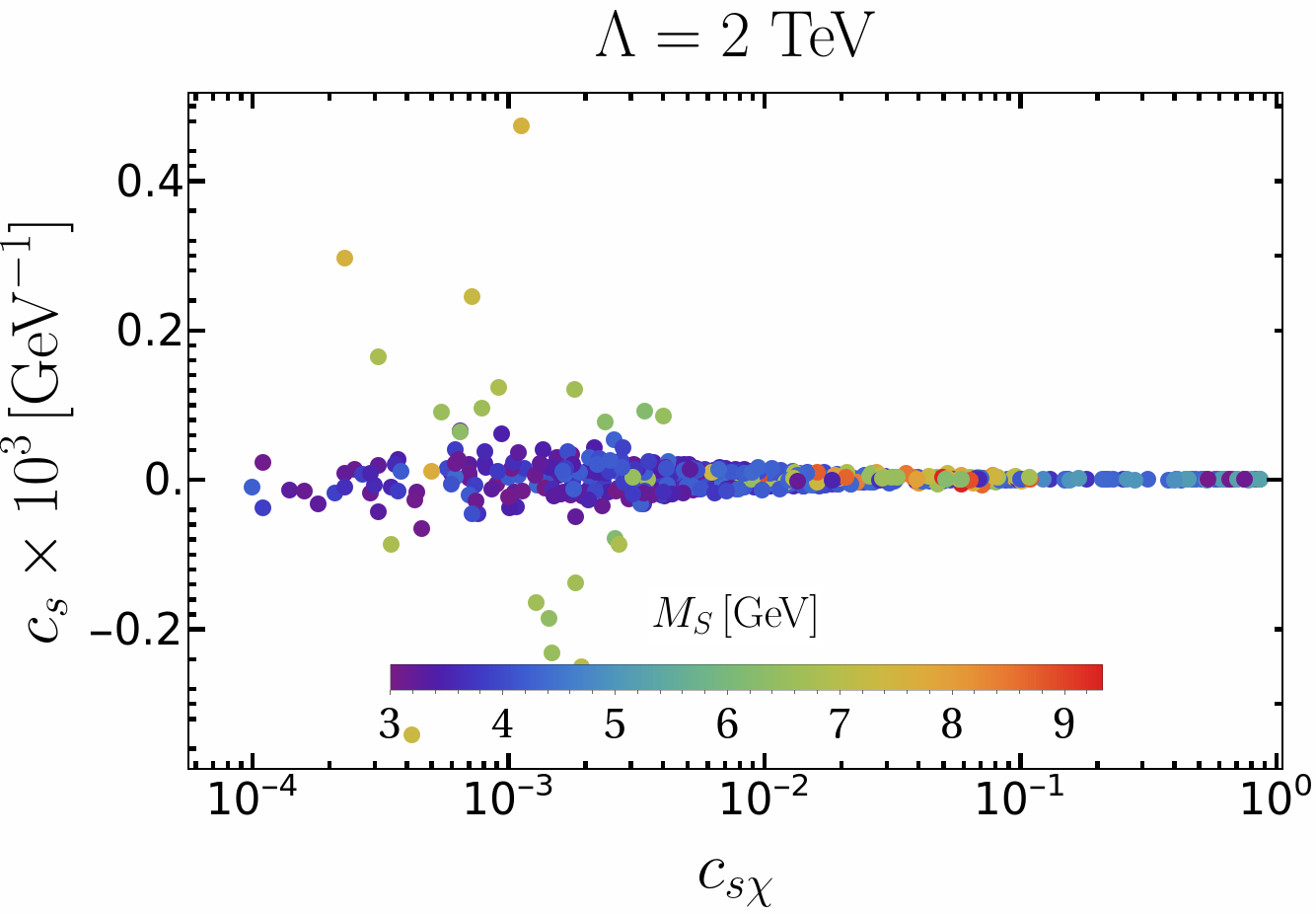}\label{fig:DM_flavor_highermass3}}\hspace{0.0001cm}
	\subfloat[]{\includegraphics[scale=0.17]{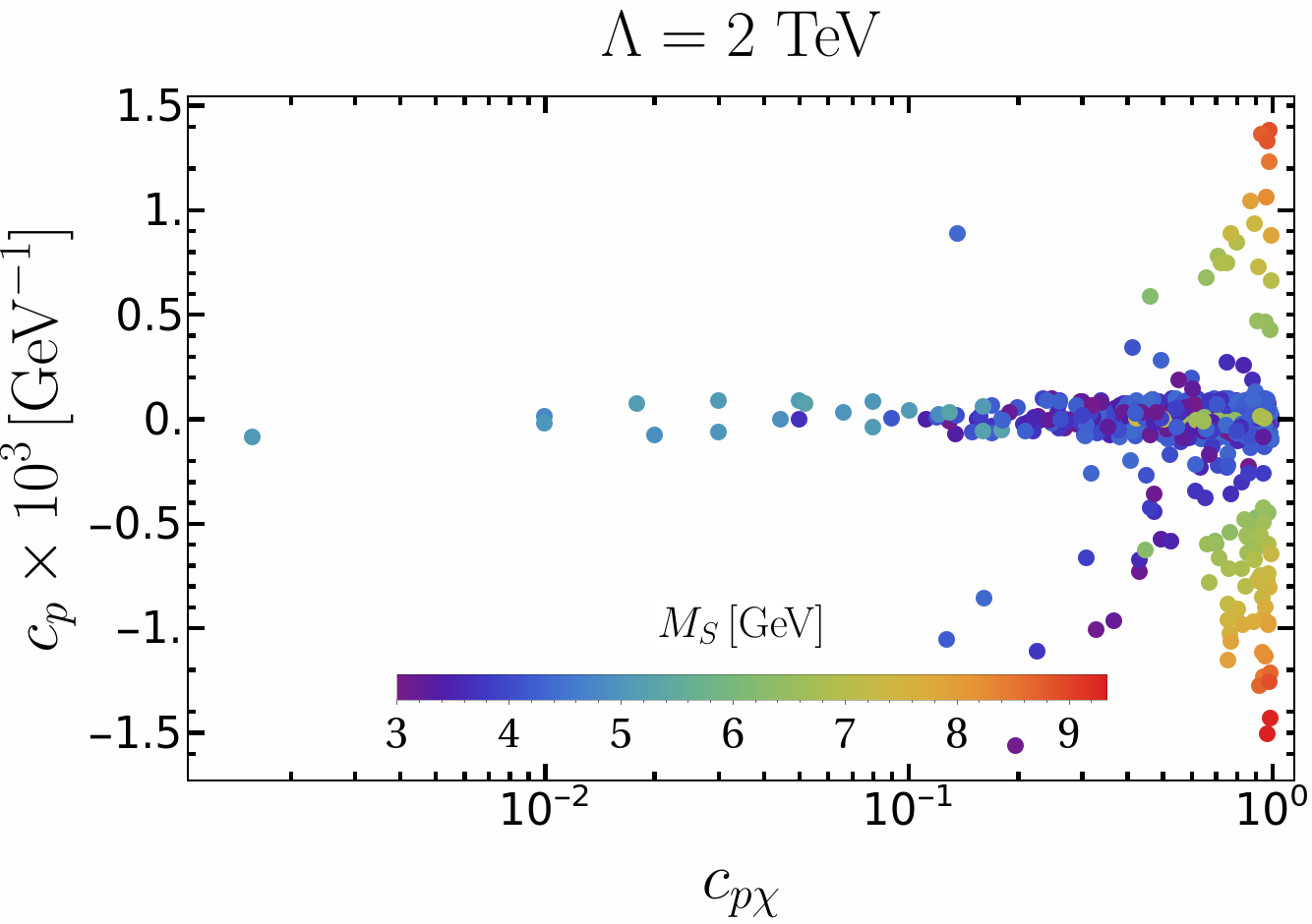}\label{fig:DM_flavor_highermass4}}\hspace{0.0001cm}
	\subfloat[]{\includegraphics[scale=0.17]{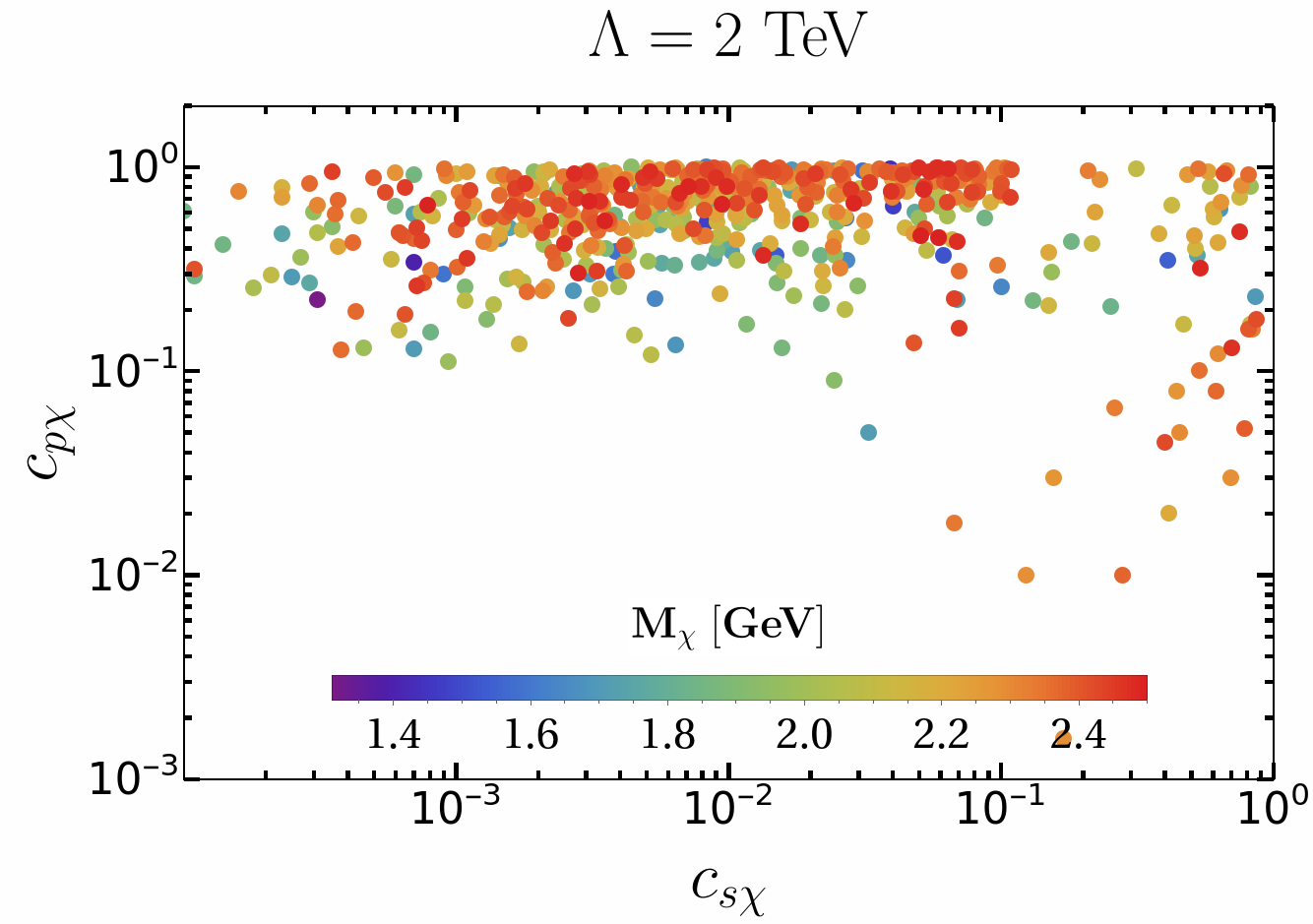}\label{fig:DM_flavor_highermass2}}\hspace{0.0001cm}
	\subfloat[]{\includegraphics[scale=0.17]{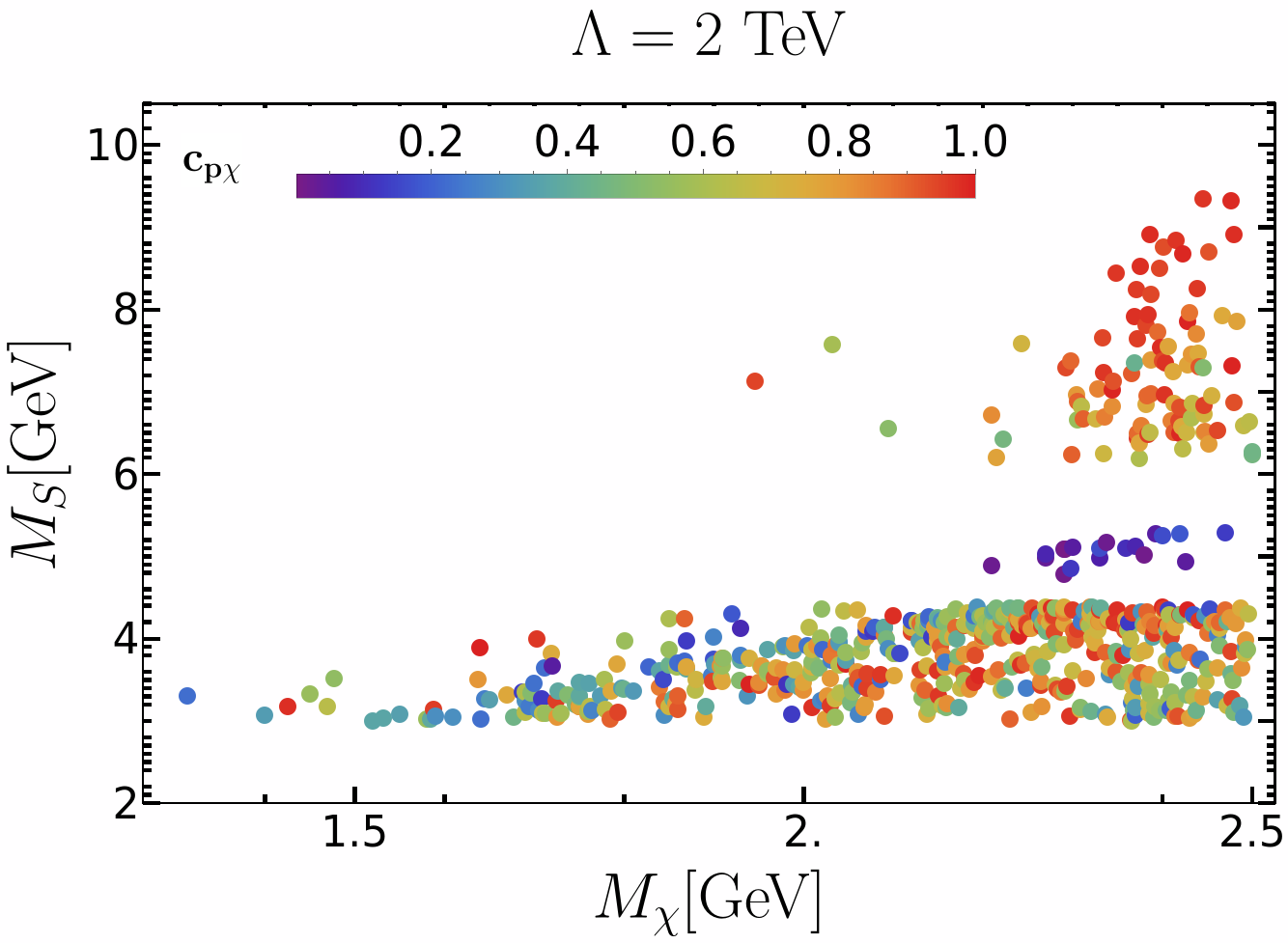}\label{fig:DM_flavor_highermass1}}
	\caption{The allowed parameter space and the correlations among the masses and couplings are obtained by considering all the processes discussed above, including low-energy flavour observables, fermion electric and magnetic dipole moments and their chromo counterparts, as well as EWPOs, together with dark sector constraints such as relic density and DD and ID cross-section bounds. Here, we focus on the region $M_S > 3~\mathrm{GeV}$. }\label{fig:DM_flavor_highermass}
\end{figure}
In the previous sections, we have discussed separately the constraints we are getting from flavour and electroweak processes. We describe above the results of the analysis of the dark sector constraints and note that the constraints are relatively relaxed and perfectly consistent with all the other relevant data. While performing the DM analysis, where relic density and direct detection, indirect detection cross-section and the bounds from CMB are taken into account, we have varied DM mass up to $ 10 $ GeV, as shown in the plots of fig.~\ref{fig:DM_highermass}.

Here, we highlight the regions of parameter space that remain viable after incorporating all relevant constraints, including those from flavour physics, EWPOs, EDM measurements, and dark-matter searches. We present the set of points that satisfy all these bounds in terms of correlated parameter spaces, which are essential for understanding the broader phenomenological implications of the model. In fig.~\ref{fig:DM_flavor_highermass}, we display these allowed correlated regions.

A comparison between the correlations shown in fig.~\ref{fig:DM_flavor_highermass} and those in fig.~\ref{fig:DM_highermass} indicates that a substantial portion of the parameter space permitted by dark-sector data becomes excluded once the EDM and flavour constraints are incorporated. As discussed earlier, for the decay $P \to P'\chi\bar{\chi}$ to occur, the dark-matter mass must satisfy the kinematic requirement $M_{\chi} \leq \frac{M_{P} - M_{P'}}{2}.$
Consequently, the viable dark-matter masses are limited to $M_{\chi} \lesssim 2.5~\text{GeV}$. 

The allowed regions for the couplings $c_s$ and $c_p$ are significantly constrained, primarily due to stringent EDM bounds. Furthermore, from Figs.~\ref{fig:DM_flavor_highermass3}, \ref{fig:DM_flavor_highermass4}, and \ref{fig:DM_flavor_highermass2}, we observe the corresponding allowed ranges for $c_{p\chi}$ and $c_{s\chi}$. In particular, while values of $c_{s\chi} \gtrsim 10^{-3}$ remain allowed, the parameter space with $c_{p\chi} \lesssim 0.1$ appears increasingly disfavored. In our DM analysis, no constraints are derived for the couplings $ c_{p} $ and $ c_{p\chi} $ since DD is independent of these couplings, and relic conditions are satisfied across the parameter space. However, in invisible decays, the branching ratio varies as  $ \propto c_{p}^2 \, c_{p\chi}^2 $, necessitating lower $ c_{p} $ values for higher $ c_{p\chi} $ to match experimental upper bounds.

Fig.~\ref{fig:DM_flavor_highermass1} illustrates the allowed parameter space in the $M_S$--$M_{\chi}$ plane, with the coupling $c_{p\chi}$ represented through the color gradient. We observe that the full range $3 \lesssim M_S \lesssim 10~\text{GeV}$ remains viable only for dark-matter masses $M_{\chi} \gtrsim 2~\text{GeV}$. For lighter dark-matter masses, $M_{\chi} < 2~\text{GeV}$, the mediator mass is restricted to $M_S \lesssim 4.5~\text{GeV}$.

\subsubsection{\bf Bounds for $ M_{S} \leq 3 $ GeV}
Here, we first discuss the dark-matter phenomenology for mediator masses $M_S \leq 3~\text{GeV}$. In this analysis, the couplings $c_{s}$ and $c_{p}$ are varied within the vicinity of the region allowed by flavour constraints. The parameters are scanned over the following ranges:
\begin{equation}
	(|c_{s}|,|c_{p}|) \leq 10^{-5} ~\text{GeV}^{-1} \ ;\ \ \  0.01 ~\text{GeV} \leq M_{S} \leq 3 ~\text{GeV }; \ \ c_{s(p)\chi} \leq 1 \ ; \ \ \ M_{\chi} \leq 10 \text{ GeV.}
\end{equation}
Similar to the previous case, here also we have taken relic and direct detection constraints into account. 

\begin{figure}[t]
	\centering	
	\subfloat[]{\includegraphics[scale=0.17]{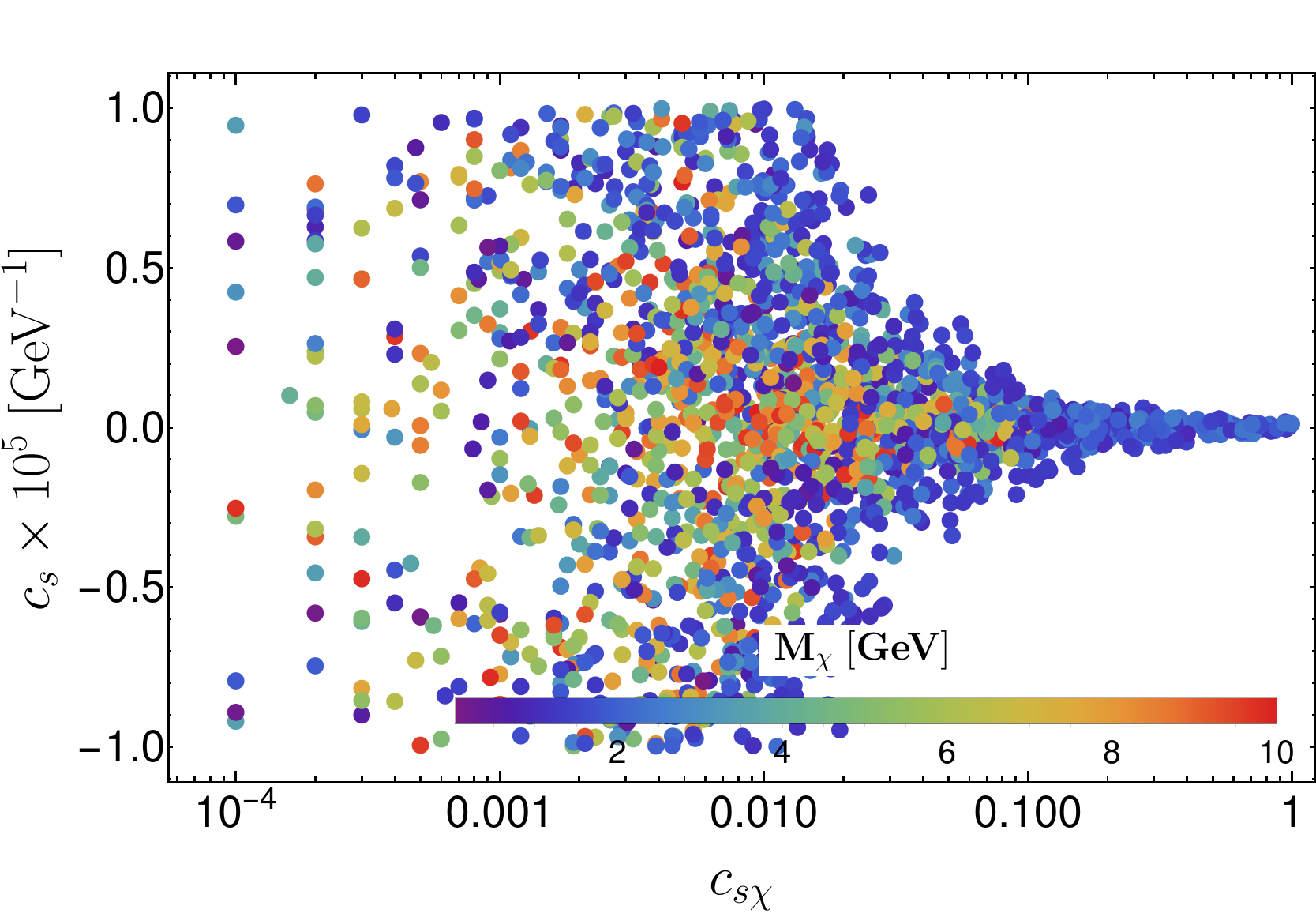}\label{fig:DM_lowermass3}}\hspace{0.0001cm}
	\subfloat[]{\includegraphics[scale=0.17]{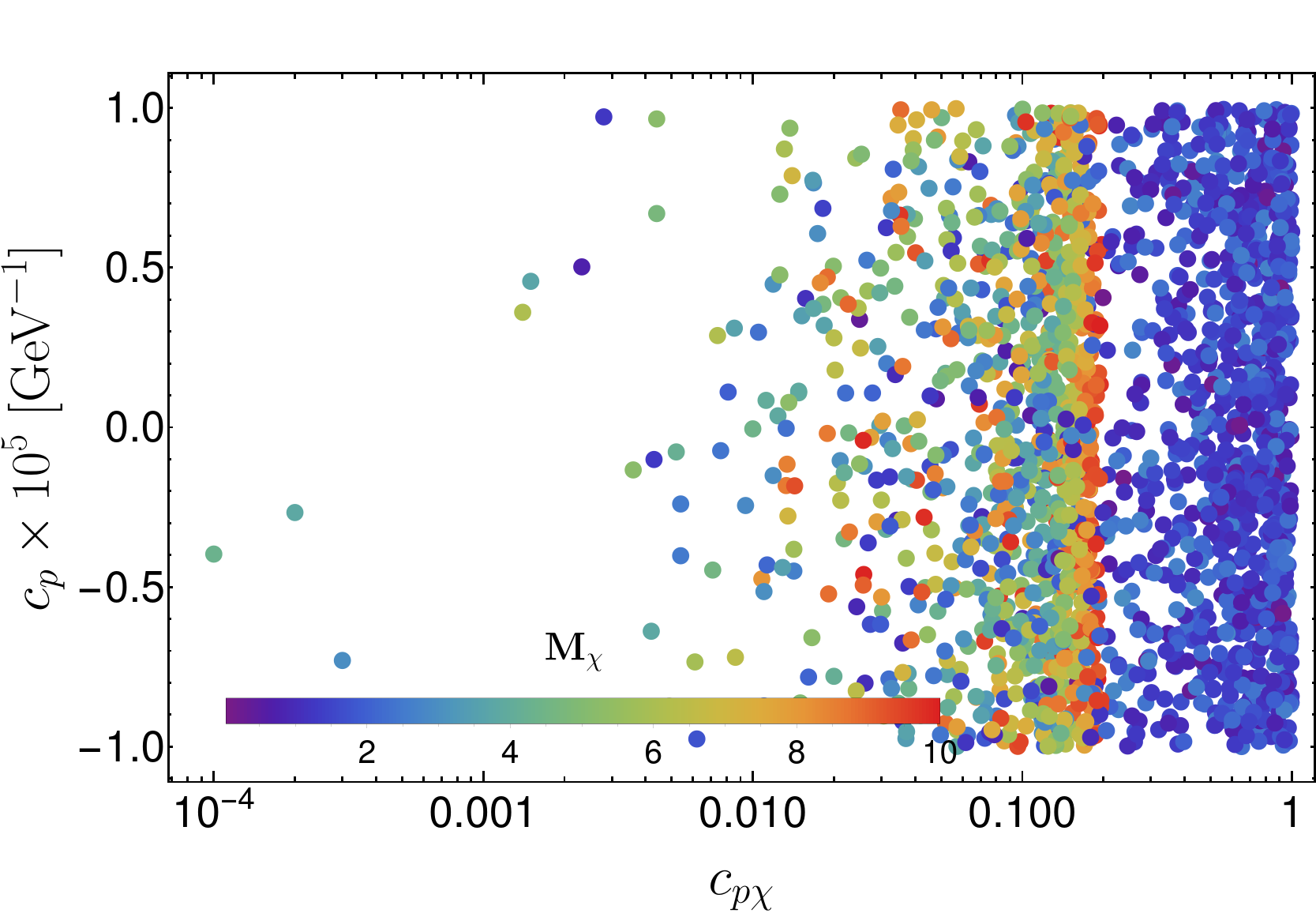}\label{fig:DM_lowermass4}}\hspace{0.0001cm}
	\subfloat[]{\includegraphics[scale=0.17]{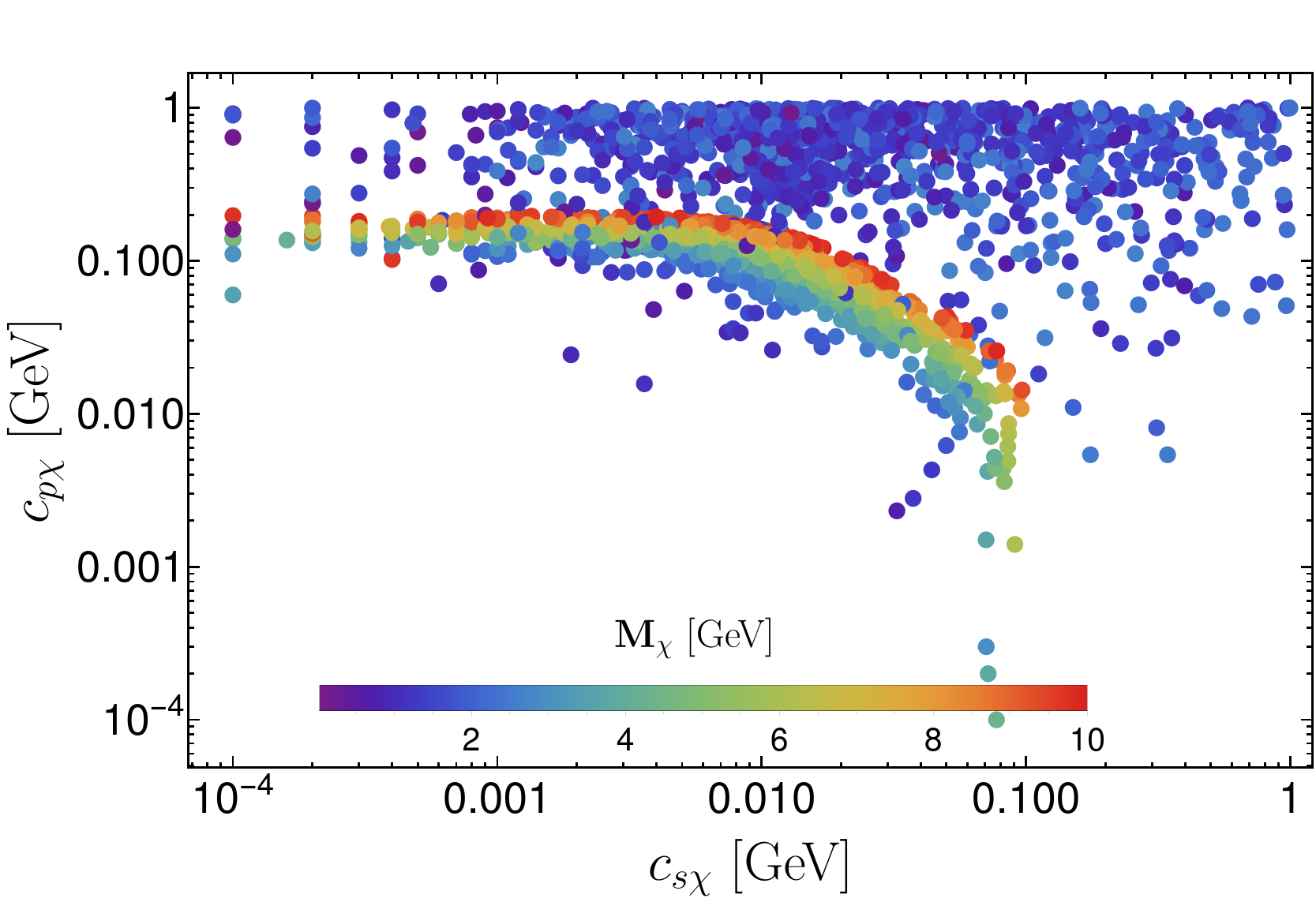}\label{fig:DM_lowermass2}}\hspace{0.0001cm}
	\subfloat[]{\includegraphics[scale=0.17]{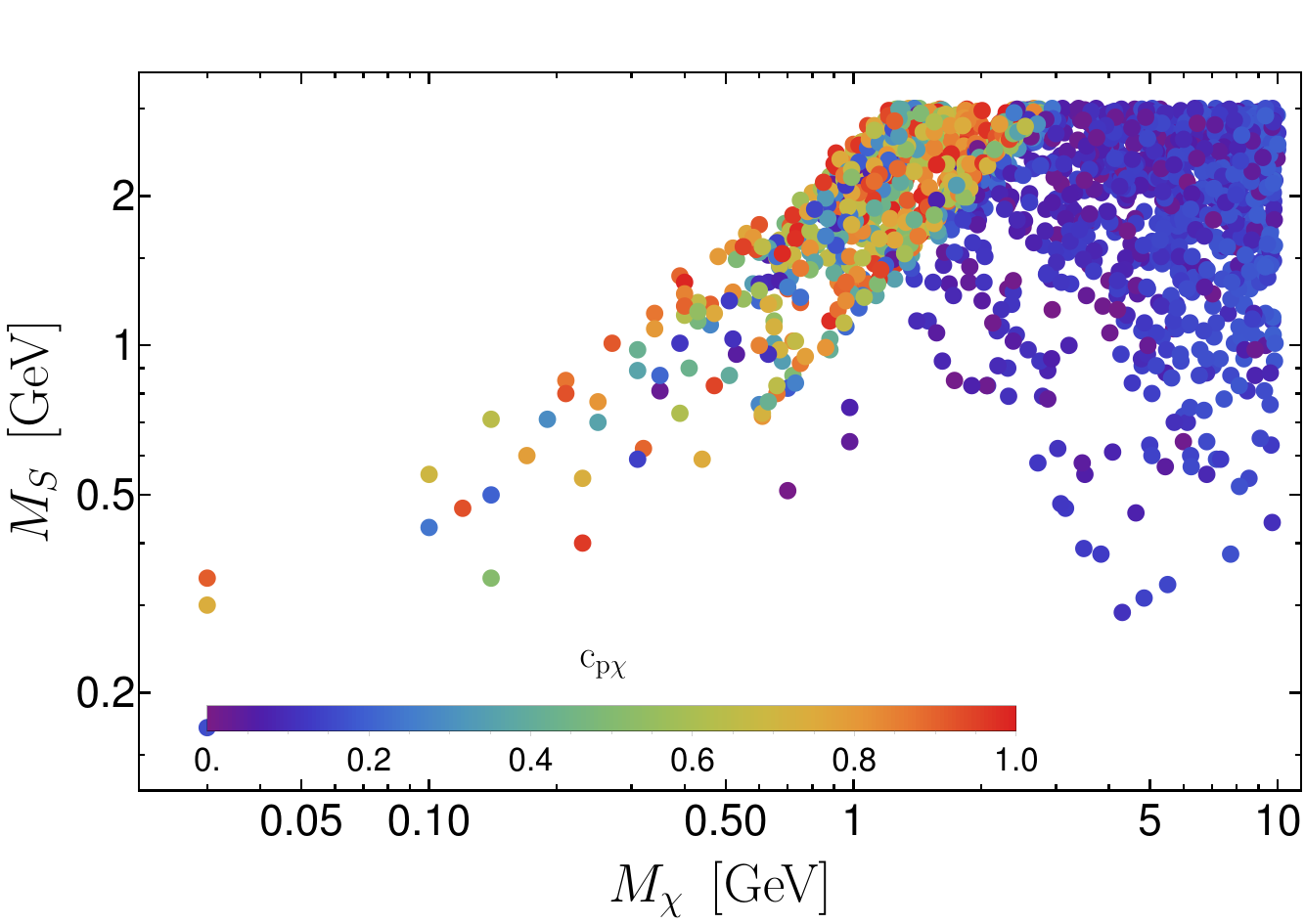}\label{fig:DM_lowermass1}}
	\caption{Allowed points from the constraints of dark sectors, i.e., relic density \cite{Planck:2018vyg} and DD cross-sections \cite{XENON:2025vwd, LZ:2024zvo} and correlations among the masses and the couplings for the mediator mass region $ M_{S} < 3 $ GeV. The colorbar is shown with coupling ($c_{p\chi} $ for fig.~\ref{fig:DM_lowermass1} and mass of the DM $M_{\chi}$ for the other plots).  }\label{fig:DM_lowermass}
\end{figure}

Fig.~\ref{fig:DM_lowermass} shows the correlations and bounds on parameter space from dark sector constraints for $ M_{S} \leq 3  $ GeV. Following bullet items are the key observations from these correlations.

\begin{itemize}
		\item In fig.~\ref{fig:DM_lowermass3}, the allowed parameter space is shown in the $c_{s\chi}$-$c_s$ plane, with the variation in $M_{\chi}$ represented by the color gradient. The viable range of $c_s$ becomes increasingly restricted for larger values of $c_{s\chi}$. In particular, for $c_{s\chi} \geq 0.1$, we obtain the bound $c_s \leq 10^{-6}~\text{GeV}^{-1}$. Moreover, in this region we find that $M_S \gtrsim 1.5~\text{GeV}$. These constraints arise primarily from the direct-detection (DD) cross-section, and this part of the parameter space is accessible only for relatively small dark-matter masses, $M_{\chi} < 3~\text{GeV}$. The dependence of the couplings on $M_S$ exhibits a behavior opposite to their variation with $M_{\chi}$. This is because the correlation with $M_S$ originates from the DD bounds, which permit comparatively larger values of the product $c_s\, c_{s\chi}$ as $M_S$ increases.

	\item Fig.~\ref{fig:DM_lowermass4} shows the correlation among $ c_{p}-c_{p\chi} $ with a color variation of the DM mass $ M_{\chi} $. As mentioned earlier, for higher DM mass, $ c_{p\chi} $ will be constrained up to $ c_{p\chi} \lesssim  0.1$, which can also be seen from here.  We get the whole region of $ c_{p} $ as an allowed solution, as $ c_{p} $ cannot be constrained from the DD cross-section. No correlation with mediator mass $ M_{S} $ is seen from here, as $ c_{p} $ and $ c_{p\chi} $ do not contribute to the spin-independent direct detection cross-section. 
	
	\item In fig.~\ref{fig:DM_lowermass2}, the correlation between DM couplings $ c_{s\chi} $ and $ c_{p\chi} $ is shown. We can see two distinct regions in this plot. For higher values of DM mass, $ M_{\chi} \gtrsim 3 $ GeV, the couplings is restricted to $ c_{s(p)\chi} \lesssim 0.1 $. This region is the effect of the DM annihilation process $ \chi \bar{\chi} \to SS $. Since we have both s and t-channel annihilation diagrams contributing to relic density, to get the observed relic density we need smaller $ c_{s\chi} $ and $ c_{p\chi} $. Also, since the cross-section mostly vary as: $ \sigma_{\chi \bar{\chi} \to SS} \propto (c_{s\chi}^2 + c_{p\chi}^2) $, we get such circle type pattern. The region above the circle is for other annihilation channels except for $ SS $. The higher values of coupling space can only be accessed by the lower mass of the DM.

	\item  Fig.~\ref{fig:DM_lowermass1} shows the correlation between DM mass and mediator mass. The color variation is shown with the coupling $ c_{p\chi}. $ Most of the parameter space is satisfied for the smaller value of the coupling. Higher value of $ c_{p\chi} $ is allowed for the regions: $ M_{\chi} < M_{S} $. The region $ M_{\chi} > 3  $ GeV, is satisfied for lower values of $ c_{s(p)\chi} .$
\end{itemize}
These correlations will further be constrained when doing the combined analysis with all the constraints of flavour, EWPOs, and DM. 

\vspace{0.5cm}
\paragraph{\underline{\bf Co-annihilation :}}
In our scenario, although there is no co-annihilation channel that directly changes the dark matter number density, the process $ S S \to f \bar{f} $ can still indirectly influence the relic abundance. This becomes relevant when the scalar mediator $ S $ is nearly degenerate in mass with the dark matter particle $\chi$, and both species remain in thermal equilibrium during freeze-out. Under such conditions, the annihilation of $S$ can affect the thermal history and, consequently, the effective annihilation cross-section that determines the relic density.

However, within the parameter space allowed by our analysis—specifically under the constraints on the couplings $c_{s,p}$—the contribution of the $S S \to f \bar{f}$ channel is expected to be subdominant. For details, see the discussion in appendix~\ref{secappendix_coannihilation}. We have computed the thermally averaged cross-section for each class of processes (annihilation and co-annihilation) for several benchmark scenarios consistent with the bounds discussed earlier. Our results indicate that the co-annihilation cross-section is significantly suppressed compared to the dominant dark matter annihilation channels. The leading processes remain $\chi \bar{\chi} \to f \bar{f}$ and $\chi \bar{\chi} \to S \bar{S}$, which directly deplete the dark matter population. Therefore, while the $S S \to f \bar{f}$ process is theoretically relevant due to thermal equilibrium considerations, its quantitative impact on the relic density is negligible within the viable regions of parameter space. Consequently, the overall effect of co-annihilation on the relic density is minimal in the explored parameter space.	

\vspace{0.5cm}	
\paragraph{\underline{ \bf Results of the Combined analysis :} } 


\begin{figure}[t]
	\centering		
	\subfloat[]{\includegraphics[width=0.45\textwidth]{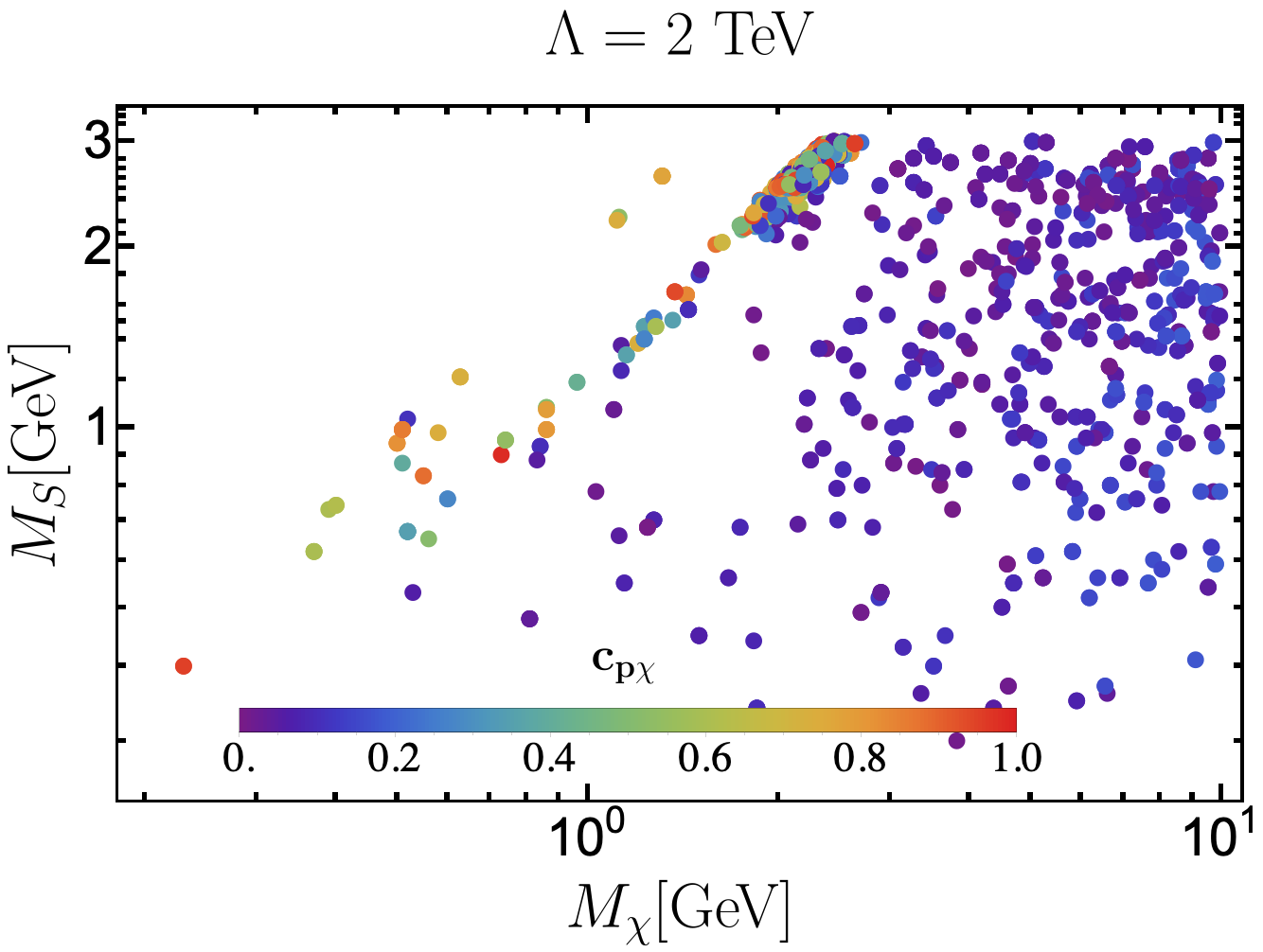}\label{fig:DM_flavor_lowermass1}}\hspace{0.0001cm}
	\subfloat[]{\includegraphics[width=0.48\textwidth]{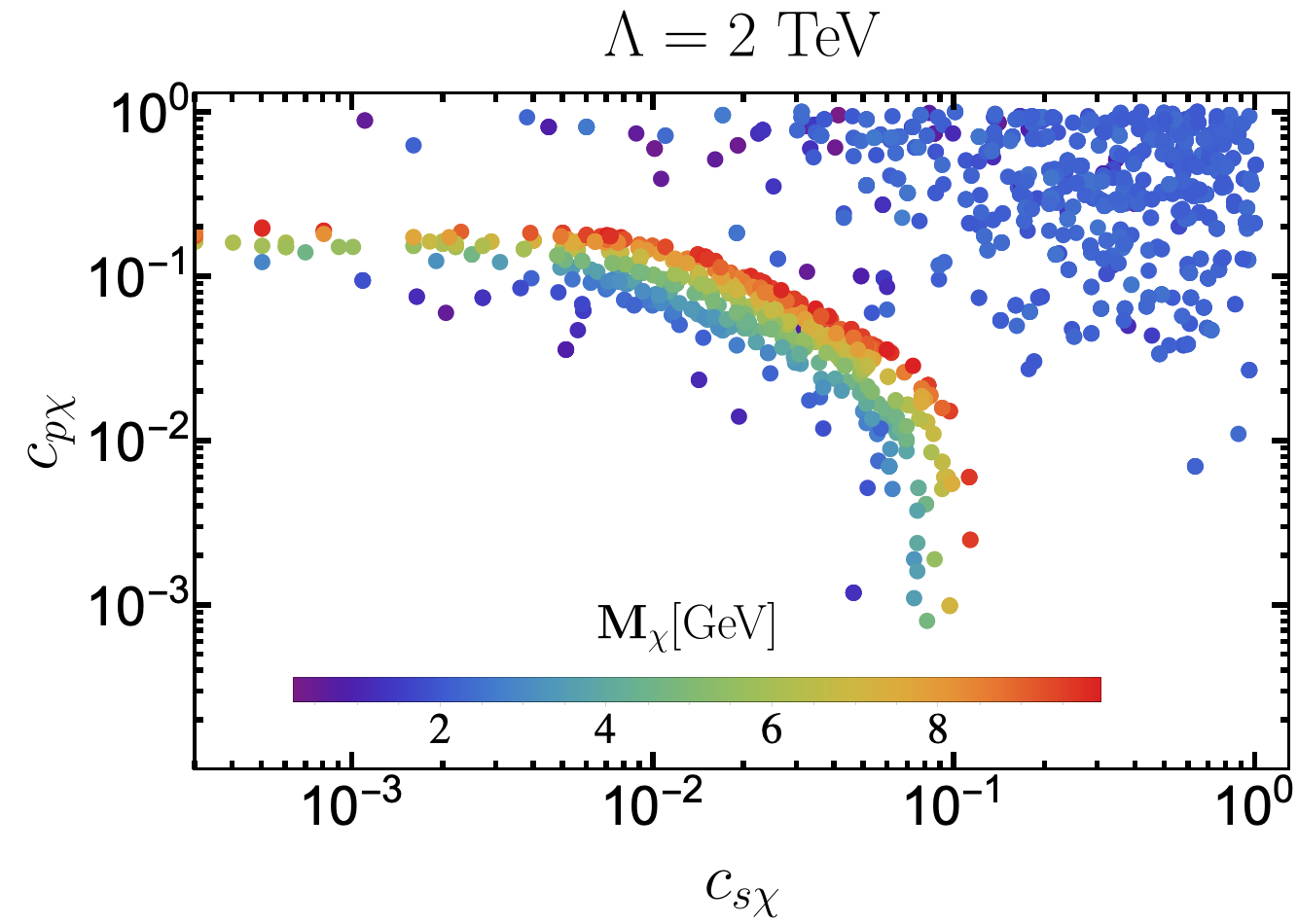}\label{fig:DM_flavor_lowermass2}}\hspace{0.0001cm}
	\subfloat[]{\includegraphics[width=0.47\textwidth]{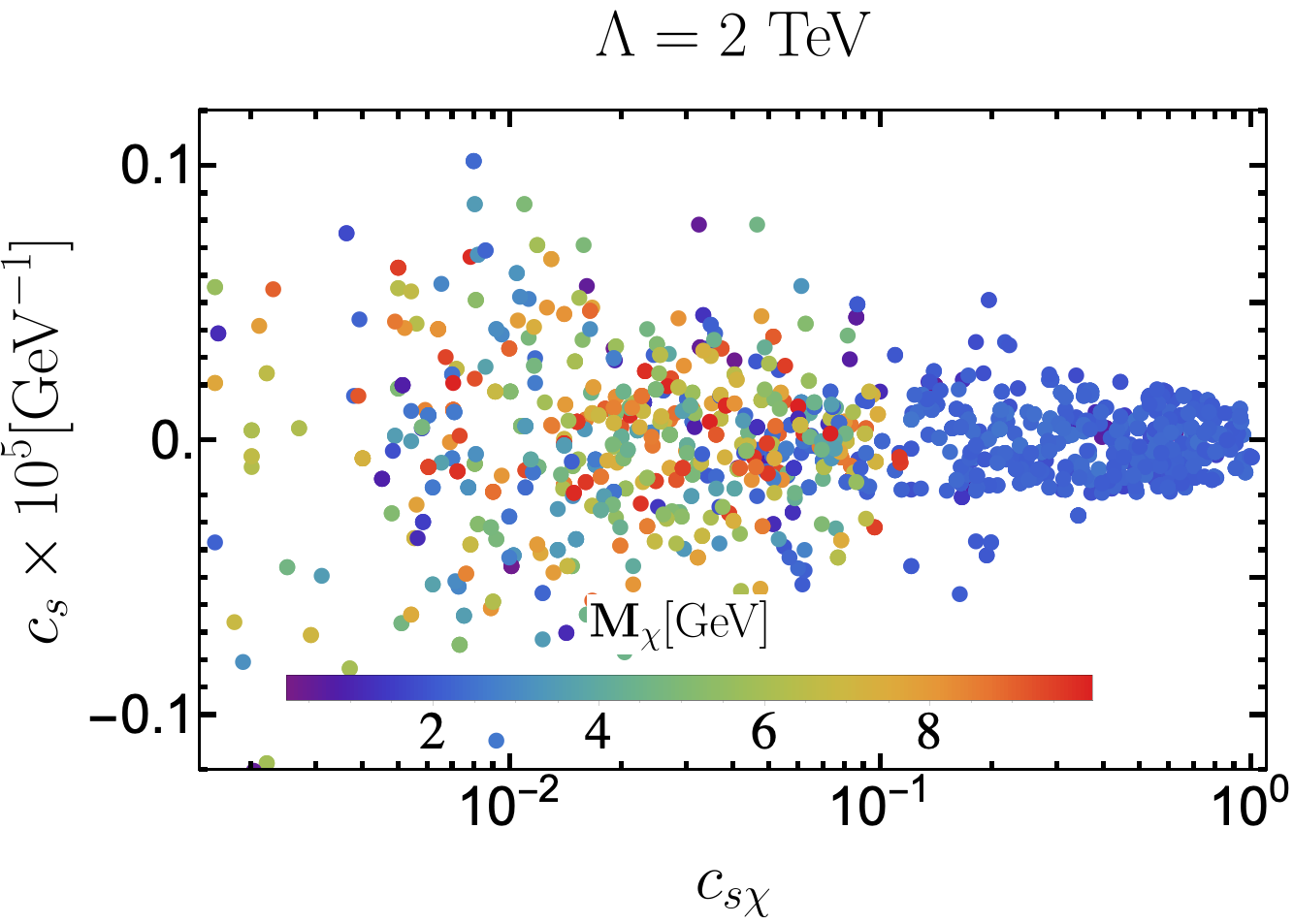}\label{fig:DM_flavor_lowermass3}}\hspace{0.0001cm}~
	\subfloat[]{\includegraphics[width=0.47\textwidth]{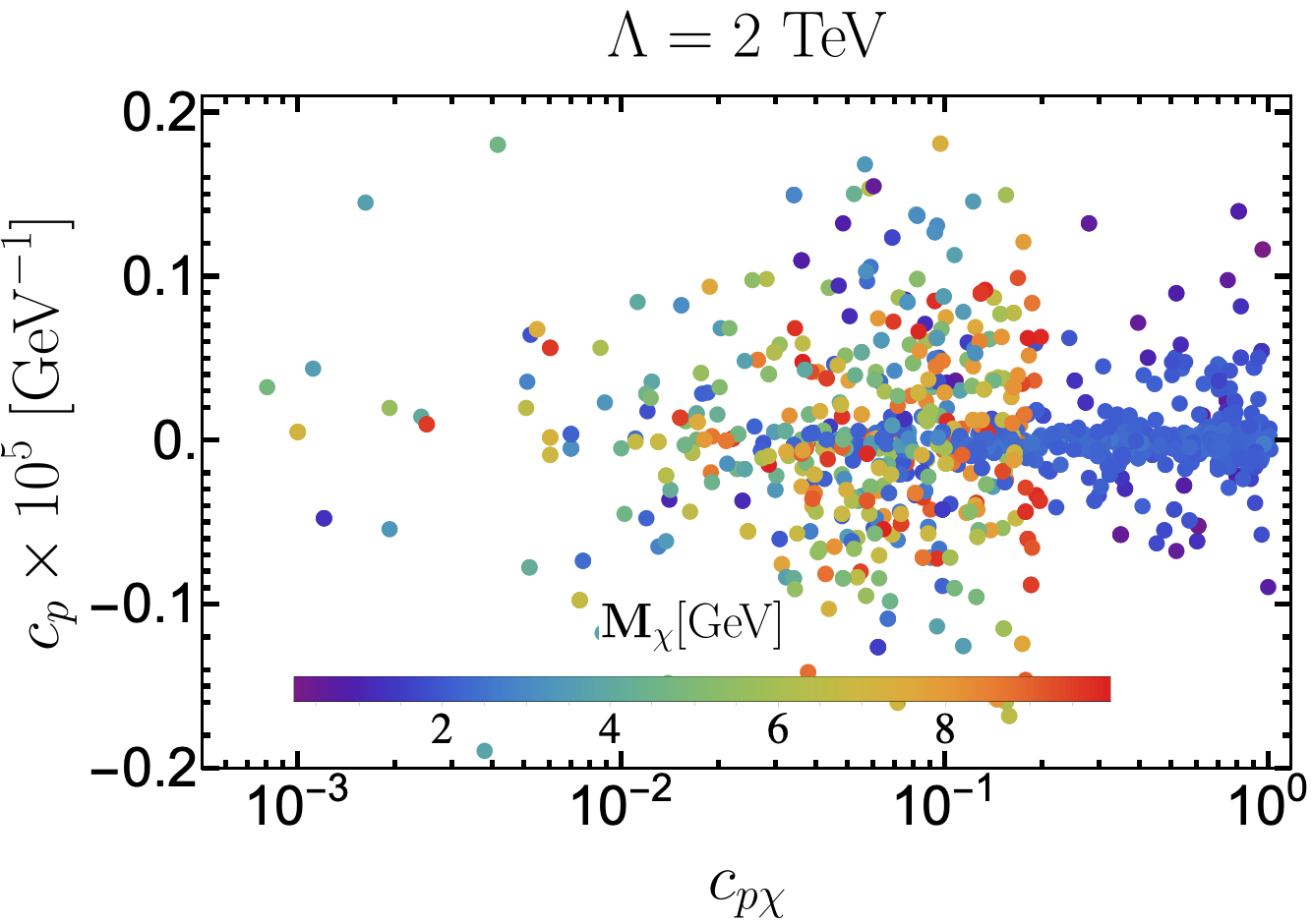}\label{fig:DM_flavor_lowermass4}}
	\caption{Allowed parameter space and correlations among the masses and the couplings by considering all the discussed flavour (considering $ 2\sigma $ error of invisible decays)  and EWPO constraints as well as the dark sector constraints, for the region $ M_{S} \leq 3 $ GeV. }\label{fig:DM_flavor_lowermass}
\end{figure}

In the scenario where the mediator mass satisfies $M_S \leq 3~\text{GeV}$, the parameter space is predominantly constrained by limits from invisible decay channels and fixed-target experiments. The regions allowed by flavour observables, electroweak processes, and dipole moment measurements are shown in fig.~\ref{fig:flavor_lowermass_2sig}. In this mass regime, the contributions from dipole-moment constraints play only a negligible role. In this part of the discussion, we will explore how the DM constraints further modify the parameter space when combined with these data. While discussing the outcome from the analysis of only flavour and electroweak processes, we presented our results for two cases depending upon whether we take the $  1\sigma $ error or $ 2\sigma  $ error of the process $ B \to K^{(*)} S $. For the case of $ 2\sigma  $ error, the resulting interplay between the allowed parameter spaces is illustrated in the plots in fig.~\ref{fig:DM_flavor_lowermass}.

As mentioned, the strongest constraint on this parameter space is coming from invisible decays ($ P \to P' + S $) and fixed target experiment (CHARM). So, the constraint of the flavour sector does not really depend on the mass of the DM (a small dependency is coming from the Breit-Wigner propagator of $ S $). Similar to the previous case, here also we do not get any bound or impact of the coupling $ c_{G} $ from DM analysis. So to get the common parameter space from flavour and DM, we have varied $ c_{G} $ in the range allowed from only flavour observables.

Fig.~\ref{fig:DM_flavor_lowermass1} shows the correlation between $ M_{S}-M_{\chi} $, with color variation of the coupling $ c_{p\chi}.$ Since the flavour sector constraint does not have a noticeable dependency on the DM mass and the couplings, the whole $ M_{\chi} $ region is allowed. However, we could see more concentration of allowed points for $M_{\chi} \gtrsim 1$ GeV and $M_S \gtrsim 1$ GeV. In figs. \ref{fig:DM_flavor_lowermass2}, \ref{fig:DM_flavor_lowermass3}, and \ref{fig:DM_flavor_lowermass4}, we have shown the allowed solutions and the respective correlations for $c_{s(p)}$ and $c_{s(p)\chi}$ with the varying masses. Note that for values of $M_\chi \gtrsim 5$ GeV, the allowed solutions are concentrated around the values of $c_{p\chi}$ and $c_{s\chi}$ in the range 0.01 to 0.15. Values above these ranges are not allowed for $M_{\chi} \gtrsim 5$ GeV. Higher values of these couplings are allowed, but for $M_\chi < 5$ GeV. For $c_{s\chi} < 0.1$, the allowed values of $c_s$ could be as large as $1.5\times 10^{-6}$ GeV$^{-1}$. However, for $c_{s\chi} > 0.1$, we obtain $|c_s| \lesssim 0.5\times 10^{-6}$ GeV$^{-1}$. The coupling $c_p$ does not have noticeable correlations with the other parameters, and its values could be as large as $2 \times 10^{-6}$ GeV$^{-1}$. 

The plots in fig.~\ref{fig:DM_flavor_lowermass} illustrate the correlations among masses and couplings when considering a \(2\sigma\) error for the invisible decays. If we instead take a \(1\sigma\) error for the invisible decays of the \(B\) meson, the parameter space becomes more constrained fig.~\ref{fig:flavor_lowermass_1sig} shows the allowed parameter space from only the flavour observables and EWPOs. A similar parameter space is obtained when combining dark sector constraints with these. The mediator mass is restricted to a very narrow region, as seen in fig.~\ref{fig:flavor_lowermass_1sig}. The mediator-SM couplings are allowed up to \( |c_{s(p)}| \lesssim 10^{-7} \, \text{GeV}^{-1} \). The correlations and allowed regions for $c_{s\chi}-c_{p\chi}$ remains unchanged. Since $M_{S}$ is only allowed in the region $ \approx \left( 2.5   - 3.0 \right) $ GeV, the allowed range for \( M_{\chi} \) is also limited to \( \gtrsim 2 \) GeV, as can be seen from fig.~\ref{fig:DM_flavor_lowermass1}.  


\section{Other Important Bounds}\label{sec:bounds_toy}
	
	\subsection{Bounds on Dimensionless Couplings}
	In our analysis, we have obtained bounds on the couplings \( c_s \), \( c_p \), and \( c_G \), which can be expressed in terms of dimensionless couplings \( g_s \), \( g_p \), and \( g_V \) according to eqs.~\eqref{eq:fermionS} and ~\eqref{eq:gauge_coupling}. In this section, we will discuss how the bounds on \( c_s \), \( c_p \), and \( c_G \) translate to dimensionless couplings $g_{s}, g_{p}, g_{V}$ and the couplings of a toy model.
	
	\paragraph{\underline{Bounds for $ M_{S} > 3 $ GeV} : }In our analysis, we get  (in $ \rm GeV^{-1} $)
	\begin{equation}\label{eq:bounds_cscpcG1}
	|c_{s}| \leq 10^{-4}\,, \ \ \ \ |c_{p}| \leq  2 \times 10^{-3} \,, \ \ \ \ |c_{G}| \leq  1.0 \times 10^{-3}.
	\end{equation}
	which will give us: 
	\begin{equation}\label{eq:bounds_gsgpgv1}
	|g_{s}| \leq 0.017 \,, \ \ \ \ |g_{p}| \leq  0.034 \,, \ \ \ \ |g_{V}| \leq  0.246. 
	\end{equation}
	In connection with the DM phenomenology, the above bounds $g_s$ will be applicable if we take the coupling $c_{s\chi} \lesssim 0.003$. For $c_{s\chi}> 0.1$ the bounds on $g_s$ will be much stronger, like $ |c_{s} | \leq 10^{-5}  \rm \, GeV^{-1}$, which again gives us 
	\begin{equation}\label{eq:bounds_gsgpgv2}
	|g_{s}| \leq 0.0017 \,.
	\end{equation}
	
	\paragraph{\underline{Bounds for $ M_{S} \leq 3 \rm \, GeV$} : } For this region of $ M_{S} $, we get from our analysis (in $ \rm GeV^{-1} $): 
	\begin{equation}\label{eq:bounds_cscpcG2}
	|c_{s}| \leq 0.15 \times 10^{-5}\,, \ \ \ \ |c_{p}| \leq 0.2 \times 10^{-5}\,, \ \ \ \ |c_{G}| \leq  0.2 \times 10^{-5} \,,
	\end{equation}
	which gives us the bounds on dimensionless couplings: 
	\begin{equation}\label{eq:bounds_gsgpgv3}
	|g_{s}| \leq 2.6 \times 10^{-4} \,, \ \ \ \ |g_{p}| \leq 3.5 \times 10^{-4} \,, \ \ \ \ |g_{V}| \leq 5 \times 10^{-4} \,.
	\end{equation}
	The above bound on $g_s$ is applicable when $c_{s\chi} < 0.1$. For higher values, like $c_{s\chi} > 0.1$, the resulting bound on $c_s$ will be smaller. Hence, the bound on $|g_s|$ will also be lesser, for example 
	\begin{equation}
	|g_{s}| \leq 1 \times 10^{-4}. 
	\end{equation}
	
	\subsection{Bounds on a Toy Model Parameters} 
	
	The results of the above analysis are useful to constraint the couplings $\frac{C}{\Lambda}$ and $\frac{C'}{\Lambda}$ of the toy model with dimension five operators defined in eqs. \eqref{eq:toylag_fer} and \eqref{gauge_higherDim}, respectively.  
	The couplings $c_s$, $c_p$ and $c_G$ are related to $\mathbb{C}_s^{S}$,  $\mathbb{C}^{S}_p$ and $\mathbb{C}_{W,Z}^{S}$ which are defined in eqs. \eqref{eq:higherdimeq} and \eqref{eq:dim5gaugedcouplings}, respectively. The parameter $\alpha$ is defined in eq.~\eqref{eq:alpha}.
	The bounds on $c_s$, $c_p$ and $c_G$ will directly put constraints on $\frac{C}{\Lambda}$ and $\frac{C'}{\Lambda}$ alongside mixing angle $\theta$ and the VEV $u$. In this regard, also the data on the decay rate $\Gamma(h_{1}\to f\bar{f})$ ($ f $ = SM fermions) are useful. 
	The particle $h_{1}$ (in eq.~\eqref{eq:higherdimeq}) represents the SM Higgs boson with a mass of $m_{h_{1}} = 125~\mathrm{GeV}$, while $S$ is a BSM spin-0 particle. The Lagrangian shows that $h_{1}$ exhibits both scalar and pseudoscalar interactions, which can be constrained using data on Higgs decay widths to light quarks and leptons. The data on the branching ratios are available for the channels $h_{1} \to (b\bar{b}, \mu^{+}\mu^{-}, \tau^{+}\tau^{-})$, with upper limits on other possible decay channels \cite{ParticleDataGroup:2024cfk,ATLAS:2022vkf}, which are given by
	\begin{subequations}
	\begin{eqnarray}
	\mathcal{B}(h_{1} \to b \bar{b} ) & = & 0.53 \pm 0.08 \,, \\
	\mathcal{B}(h_{1} \to \tau^{+} \tau^{-} ) & =& 0.060 ^{+0.008}_{-0.007} \,,  \\
	\mathcal{B}(h_{1} \to \mu^{+} \mu^{-} ) & =& (2.6 \pm 1.3 ) \times 10^{-4} \,.  
	\end{eqnarray}
	\end{subequations}
	
	\begin{figure}[t]
		\begin{center}
			\subfloat{\includegraphics[scale=0.2]{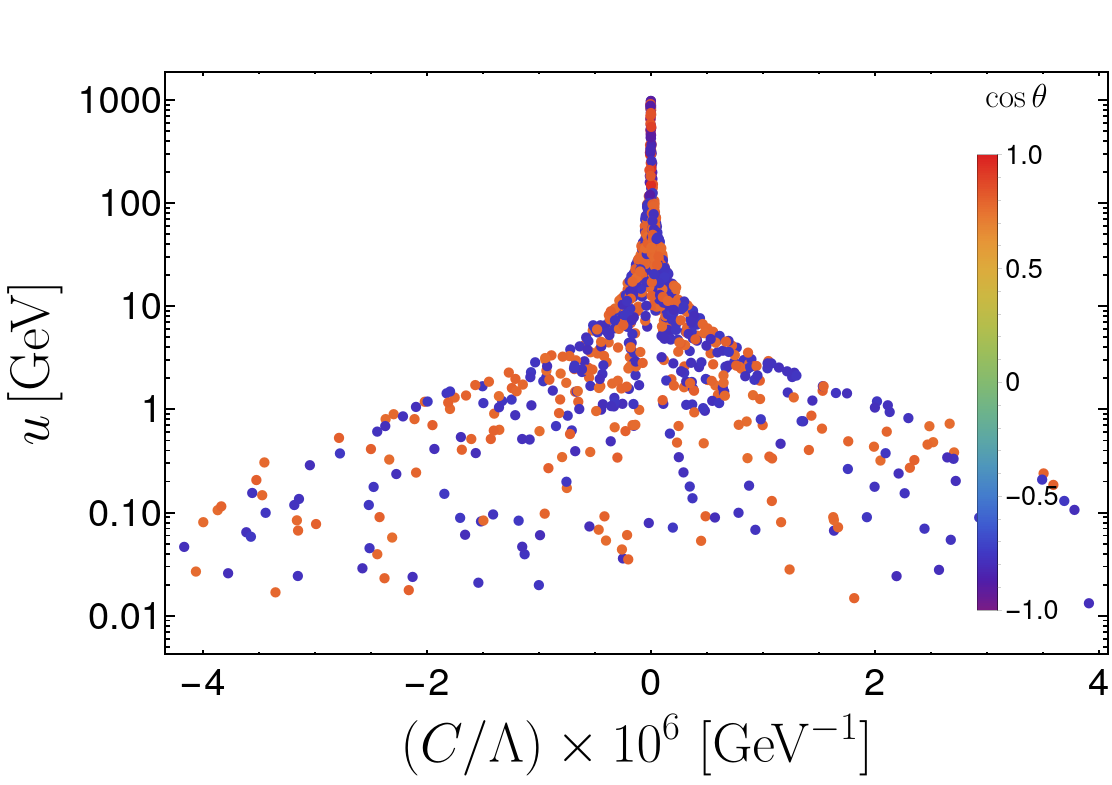}} \hspace{0.0001cm}
			\subfloat{\includegraphics[scale=0.2]{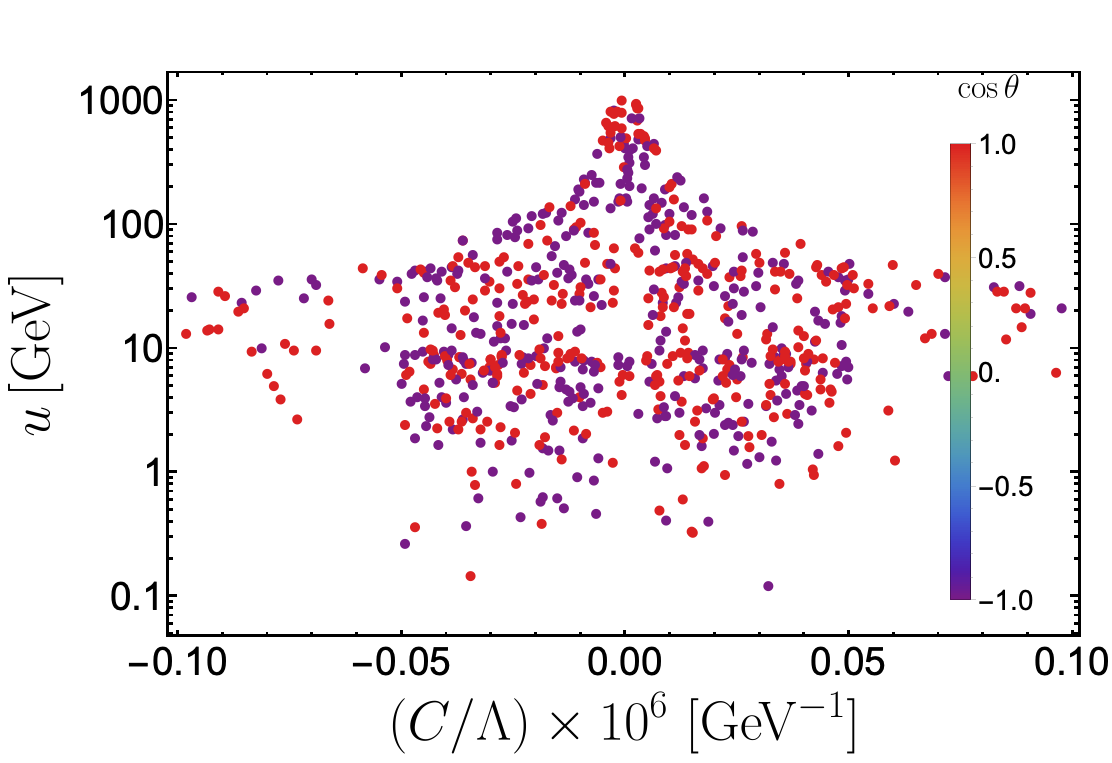}} 
		\end{center}
		\caption{Allowed parameter space in the plane of $u  - C/\Lambda $ (VEV of  $ P $-interaction strength), where the color variation is shown with the cosine of the mixing angle $ \theta $. The left plot is the allowed region for the case when we have taken only the Higgs data. The right plot shows the bound for the combined input of the Higgs decay data along with the bounds on $c_{s}$ and $c_{p}$, which we get from our analysis.} \label{fig:toy_model_bound}   
	\end{figure}
	
	We have analysed the data available on the Higgs decays separately and the respective correlations of $\frac{C}{\Lambda}$ with the VEV $u$ and $\cos\theta$ are shown in the left plot in fig.~\ref{fig:toy_model_bound}. Note that the solutions are allowed only in the regions when $\theta \to 0$ (or $\pi$), and for $u < 10$ GeV, the allowed range of the coupling is given by 
	\begin{equation}\label{eq:boundcLHiggs}
	\big|\frac{C}{\Lambda}\big| < 5 \times 10^{-6}  \, \rm GeV^{-1}.
	\end{equation}
	For values of $u$ around 100 GeV and above, the allowed solutions of $|\frac{C}{\Lambda}|$ will be of order $10^{-8}$. 
	On top of this, we can use the bounds on the $c_s$ and $c_p$ to constrain this coupling further. 
	Note that the bounds $c_s$ and $c_p$ for $ \bf M_{S}  > 3 $ GeV is much more relaxed and will not restrict $|\frac{C}{\Lambda}|$ further than what we have obtained in eq.~\eqref{eq:boundcLHiggs}. However, the allowed solutions on $c_s$ and $c_p$ for $ \bf M_{S}  \leq 3 $ GeV will be relevant. The result of a combined analysis of the Higgs data and the bounds $c_s$ and $c_p$ is shown in the right plot in fig.~\ref{fig:toy_model_bound}. The allowed solutions will be one order of magnitude smaller than what we have shown in eq.~\eqref{eq:boundcLHiggs}.   
	
	The major constraints on $\frac{C'}{\Lambda}$ will mainly come from the bounds we obtain on $c_G$. Following which we have obtained for the case $ \bf M_{S}  > 3 $ GeV: 
	\begin{subequations}
		\begin{eqnarray}\label{eq:toy_couplings1}
		\left| \frac{C'}{\Lambda}\right|& \leq &  0.0020 \, \rm GeV^{-1}\,,
		\end{eqnarray}
	\end{subequations}
	and for $ \bf M_{S}  \leq  3 $ GeV, we get 
	\begin{subequations}
		\begin{eqnarray}\label{eq:toy_couplings2}
		\left| \frac{C'}{\Lambda}\right| &\leq& 0.4 \times 10^{-5} \, \rm GeV^{-1} \,. 
		\end{eqnarray}
	\end{subequations}

	\section{Summary}\label{sec:summary}
	We have examined the allowed parameter spaces of a simplified dark matter model featuring fermionic dark matter and a spin-0 mediator that interacts with SM fermions and gauge bosons. This study focuses on the lower mass range of the spin-0 mediator, i.e., $M_{S} \leq 10$ GeV. We divide the parameter space into two regions depending on the mass of the spin-0 mediator $ S $; one is $ M_{S} >3 $ GeV, and the other one is $ M_{S} \leq 3 $ GeV, depending on whether the bounds of the invisible meson decays $ P \to P' S $ can be used or not. We obtain bounds on the coupling strengths $c_s$, $c_p$, and $c_G$ associated with the interaction of the spin-0 mediator with the SM fermions and gauge bosons, respectively. 
	We analyzed these two regions separately using all flavour observables and EWPOs. Then, we studied the constraints from the DM analysis. Finally, we identified a common parameter space that satisfies all available constraints, including the data on flavour, dipole moment, and EW observables, and DM constraints. 
	The bounds on the couplings we are getting for the lower mass region ($M_{S} \leq 3$ GeV) are very tightly constrained. Also, we have obtained the lower bound on the couplings from the BBN and the condition of DM to be in thermal equilibrium as $ \chi $ is a WIMP-type DM. The BBN constraint is effective up to a very small mass. We get the lower limits on couplings $c_{s(p)}$ of the order of $\sim 10^{-8}\, \rm GeV^{-1}$ (strongly depends on mediator-DM couplings).
	
For mediator masses $M_{S} > 3~\text{GeV}$, the most stringent constraints arise from rare and invisible decays of pseudoscalar mesons, such as $P \to P'\chi\bar{\chi}$, as well as from dipole-moment measurements. Using only the bounds from flavour observables, EWPOs, and dipole‑moment data, we obtain the limits $|c_{G}| \lesssim 1 \times 10^{-3}$ and $|c_{s(p)}| \lesssim 1 \times 10^{-4}~\text{GeV}^{-1}$. In addition, the dipole‑moment constraints induce a strong correlation between the allowed values of $c_s$ and $c_p$, leading to a concentration of viable solutions in the region $|c_{s(p)}| \lesssim 1 \times 10^{-5}$.

Additionally, we have performed a comprehensive scan of the parameter space and identified the correlations among the parameters consistent with constraints from dark‑matter phenomenology, flavour physics, dipole‑moment measurements, and electroweak precision observables. The solution points allowed by the dipole‑moment constraints remain compatible with the bounds arising from the dark sector. 

Our analysis shows that values of $c_{s\chi} \gtrsim 10^{-4}~\text{GeV}^{-1}$ are permitted, whereas regions with $c_{p\chi} \lesssim 0.1~\text{GeV}^{-1}$ are disfavored by the combined constraints. Furthermore, the allowed regions in the $(M_S, M_{\chi})$ plane indicate that for $M_{\chi} \lesssim 2~\text{GeV}$, viable solutions exist only for $M_S < 4.5~\text{GeV}$. In contrast, for $M_{\chi} > 2~\text{GeV}$, solutions are found across the full range $3 < M_S < 10~\text{GeV}$. 

For $ M_{S} \leq 3  $ GeV, the parameter space will mostly be constrained from invisible decay and fix-target experiment constraints. The most stringent bound for invisible decay comes from the decay $ B \to K S $. The bounds on the couplings $ |c_{p}|, |c_{G}|$ are $\lesssim 5 \times 10^{-6} \, GeV^{-1} $. We observe similar limit on $|c_s|$ if we take $ c_{s\chi} < 0.1$. For $ c_{s\chi} \geq 0.1$, we get solutions only for $ M_{\chi} \lesssim 3  $ GeV and $ |c_{s}| \lesssim 10^{-6} \, \rm GeV^{-1}$. 
For $ M_{\chi} \geq 5 $ GeV, we get $ (c_{p\chi}, c_{s\chi} ) \lesssim 0.1 $. 
	
Using the bounds on $|c_{s}|$, $| c_{p}|$, $|c_{G}|$, we have obtained bounds on the dimensionless couplings $|g_{s}|$, $| g_{p}|$ and $|g_V|$, respectively. For $M_S > 3$ GeV the approximate bounds on $g_{s(p)}$ and $g_{V}$ will be $\lesssim 0.017$ and $\lesssim 0.25$, respectively. For $M_S \lesssim 3$ GeV the respective approximate bounds are $g_{s/p/V} \lesssim 0.0004 $ when we take the $|c_{s\chi}| < 0.001$. The allowed values of $g_s$ will be less than $ 0.0001 $ when we consider $|c_{s\chi}| > 0.001$. These results suggest that the benchmark scenarios used in the collider analysis should be carefully chosen following the correlations among the parameters and their limits. 
	
Finally, we have obtained bounds on the couplings of the dimension-5 operators, which could lead to the type of interactions we are considering between the SM fermions or gauge bosons with the spin-0 mediator $S$. Since the mediator $S$ mixes with the SM Higgs ($h_1$), the measured values on the branching fractions $h_1 \to f\bar{f}$ ($f$ = fermions) will play an important role in the constraint of the mixing parameter ($\sin\theta$) and the other relevant couplings. We have obtained points that are allowed only for small mixing angles from a combined study of all the relevant inputs. In addition, we have obtained strong bounds on the couplings normalized by the scale of NP ($\Lambda$). 
	
\section*{Appendix}	
\appendix

\section{Bounds from Fixed Target Experiment}\label{sec:fixedtarget}

\begin{figure}[htbp]
	\begin{center}
		\subfloat[]{\includegraphics[height=4cm,width=4.9cm]{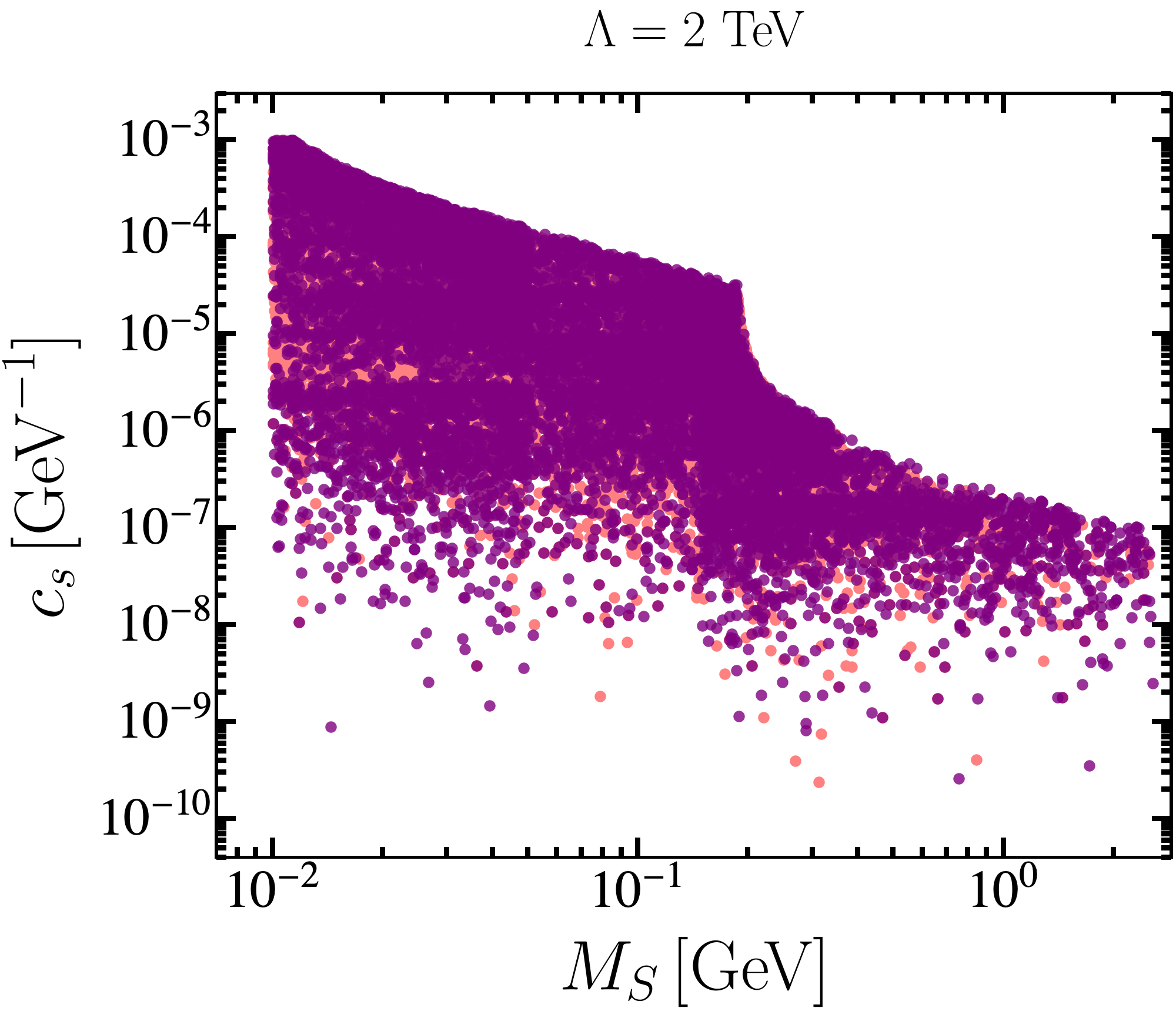}}\hspace{0.00001cm}
		\subfloat[]{\includegraphics[height=4cm,width=4.9cm]{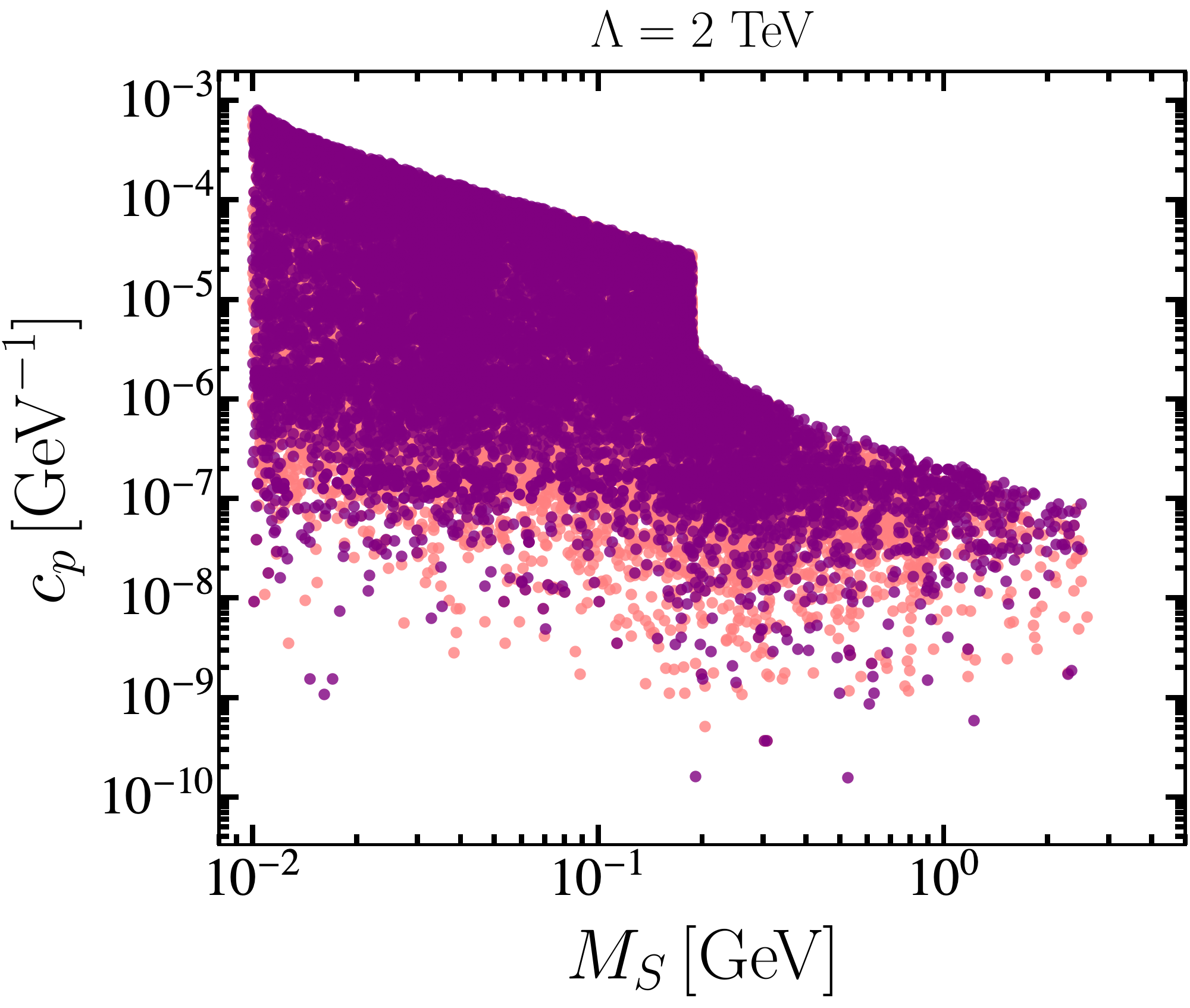}}\hspace{0.00001cm}
		\subfloat[]{\includegraphics[height=4cm,width=5cm]{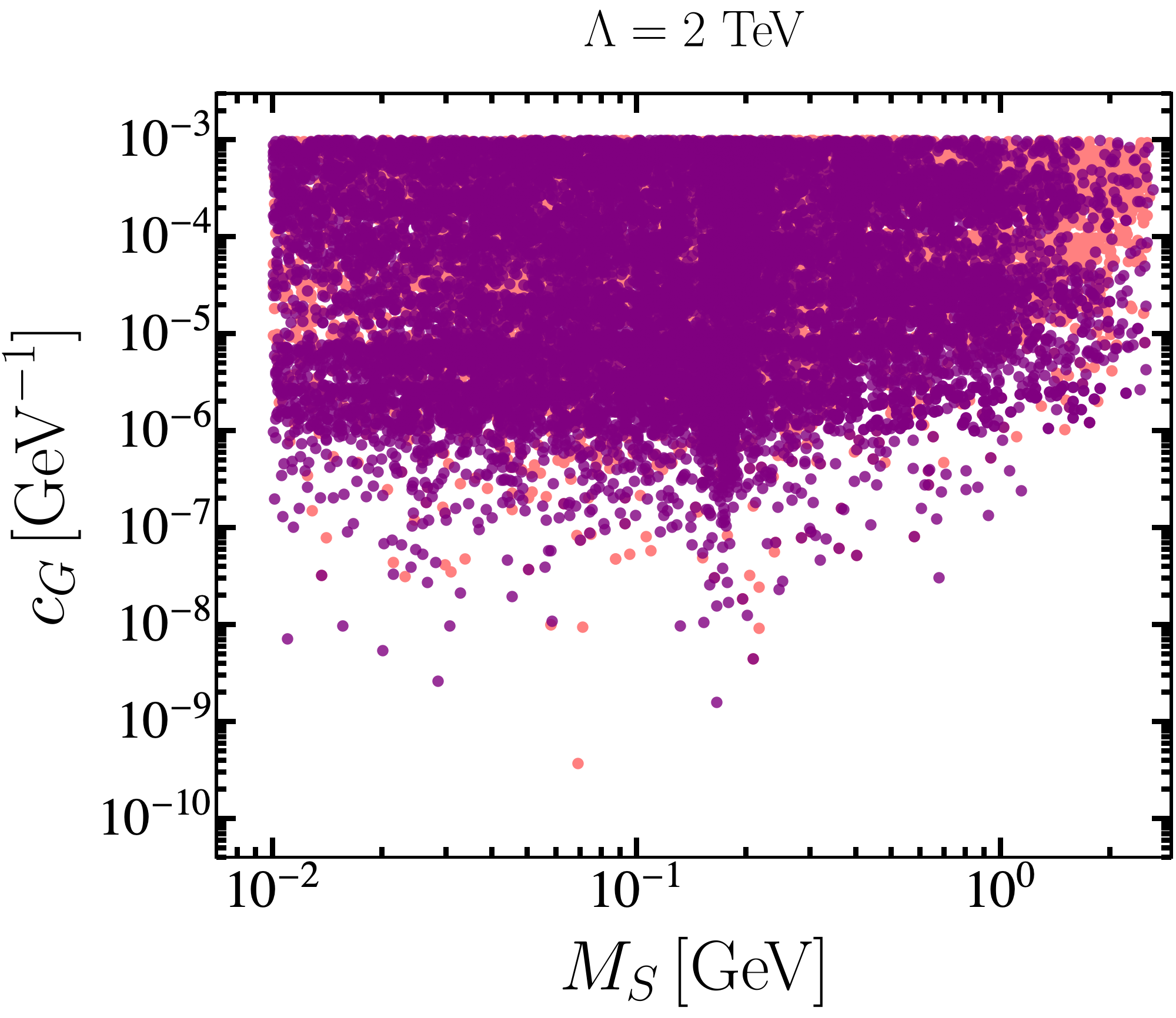}} 
	\end{center}
	\caption{Excluded parameter space in the mass-coupling ($c_i$-$M_S$) plane from CHARM fixed-target constraints (colored points indicate excluded regions). The three panels show the variation of the excluded region when scanning over the couplings $c_s$, $c_p$, and $c_G$ simultaneously, with the mediator mass $M_S$ displayed on the x-axis. Pink points correspond to scenarios with $M_{\chi} > M_S$. Purple points correspond to $M_{\chi} < M_S$, where the dark-sector couplings $c_{s\chi}$ and $c_{p\chi}$ are also varied simultaneously.
 } \label{fig:fix_target_scan}   
\end{figure}

In sec.~\ref{subsec:fixtarget}, we have discussed the fixed target experiment where the spin-0 mediator $S$ can be produced from meson decay, which further decays into pairs of electrons, muons, and photons. So far, no event(s) are detected in the CHARM detector \cite{Clarke:2013aya}, and the available upper bound is $ N_{det}<2.3 $. We have used this bound as an input in our analysis and obtained the plots in fig.~\ref{fig:fix_target_scan}, which shows the excluded parameter space (colored regions) in the coupling-mass plane. The detailed theoretical expressions can be seen from the sec.~\ref{subsec:fixtarget}. While generating the plot, we kept all the parameters free except the cutoff scale, which was set at $\Lambda = 2 $ TeV. The pink points in the plots correspond to the case where $M_{\chi}> M_{S}$, i.e., the mediator cannot decay to the DM particle. The purple points are for when $M_{\chi} = 1$. To generate the plots, the couplings are varied in the full region $(|c_{s}|, |c_{p}|, |c_{G}|) \leq 10^{-5} \rm \, GeV^{-1}$.  The mass of the mediator is varied up to $M_{S} \leq 5 $ GeV. We perform a random scan over the parameter space, selecting only those points that satisfy the imposed constraints. The excluded points are shown in color, while the white regions correspond to parameter points consistent with all constraints and are therefore allowed.

As we can see from eq.~\eqref{eq:fix_target_N},  $ N_{det} $ depends on the total decay width of the mediator, so the mass and the coupling of DM will play a role here. For example, for a fixed value of the coupling $N_{det}$ will decrease with the increasing values of $M_S$. The extreme left plot represents the variation of $M_S$ with the scalar coupling $c_{s}$, and the middle and right plot is for the variation with $c_{p}$ and $c_{G}$, respectively. We can see that the constraints on the couplings are more severe for $M_{S} \lesssim 0.2 $ GeV. Conservatively, the allowed values of all three couplings will $\lesssim$ $10^{-8}$ (in GeV$^{-1}$). We have checked that this data will not provide any strong correlations between the couplings $c_{s}$, $c_{p}$, and $c_{G}$. Furthermore, the data from CHARM will not constrain the DM-mediator couplings $c_{s\chi}$ and $c_{p\chi}$, nor will it restrict the allowed region of $M_{\chi}$. From the fig.~\ref{fig:fix_target_scan}, we can see a dip region starting from $ M_S \sim 2 m_{\mu} $.  

\section{Few results: Electric and magnetic dipole moments}\label{secappndix:edm}

\begin{table}[t]
	\centering
	\rowcolors{1}{cyan!10}{lime!5}
	\renewcommand{\arraystretch}{1.7}
	\resizebox{0.55\textwidth}{!}{%
		\begin{tabular}{|c|c|c|}
			\hline
			Observable  & Value@BP1  & Value@BP2 \\
			\hline\hline
			$\Delta a_{e}$ &$1.99 \times 10^{-21}$ & $1.37 \times 10^{-21}$ \\
			$\Delta a_{\mu}$ & $1.57 \times 10^{-12}$ & $ 9.48 \times 10^{-13}$ \\
			$\Delta a_{\tau}$ & $3.84 \times 10^{-8} $ & $1.58 \times 10^{-8}$ \\
			$\Delta a_{t}$ & $3.77 \times 10^{-3}$ & $6.59 \times 10^{-4}$ \\
			\hline
			$*|\mu_{e}^{E}|$ & $6.95 \times 10^{-32}$ e-cm& $3.41 \times 10^{-32}$ e-cm \\
			$\mu_{\mu}^{E}$ & $2.49 \times 10^{-25}$ e-cm & $1.07 \times 10^{-25}$ e-cm \\
			$\mu_{\tau}^{E}$ & $3.12 \times 10^{-22}$ e-cm & $8.93 \times 10^{-23}$ e-cm \\
			$*\mu_{t}^{E}$ & $2.01 \times 10^{-19} $ e-cm & $2.77 \times 10^{-20}$ e-cm \\
			$*|d_{n}| $ & $ 1.47 \times 10^{-27}$ e-cm & $ 6.42 \times 10^{-28}$ e-cm \\
			\hline
			$*\mu_{t} $ & $ 4.83 \times 10^{-19} \, \rm g_{s}\text{-}cm $ & $ 8.46 \times 10^{-20} \, \rm g_{s}\text{-}cm $ \\
			$*d_{t}  $ & $4.53 \times 10^{-19}\, \rm g_{s}\text{-}cm $ & $ 6.24 \times 10^{-20} \, \rm g_{s}\text{-}cm $ \\
			\hline
	\end{tabular}}
	\caption{Values obtained for different dipole moments of leptons, quarks, and neutrons for two benchmark points, defined in eq.~\eqref{eq:dipole_BPs}.Here, we consider the contributions to EDMs at the one-loop level. In this table, we define $\Delta a_{f}$ as the new physics contributions in the respective anomalous magnetic moments, which we can compare with the difference $a_{f}^{\rm exp} - a_{f}^{\rm SM}$ wherever applicable. The observables with the asterisks are taken to perform a parameter space plot (shown in fig.~\ref{fig:flavor_highermass}).}
	\label{tab:EDM_MDM_BPs}
\end{table}

Here, we present the values of several observables for two benchmark points (BPs), defined as follows:
\begin{gather}\label{eq:dipole_BPs}
	\text{BP1}~:~~ c_{s} = 5 \times 10^{-3} \, \text{GeV}^{-1},~ c_{p} = 3 \times 10^{-3} \, \text{GeV}^{-1},~ M_{S} = 8 \, \text{GeV}\,, \nonumber \\
	\text{~~~~~BP2}~:~~ c_{s} = 2 \times 10^{-3} \, \text{GeV}^{-1},~ c_{p} = 1 \times 10^{-3} \, \text{GeV}^{-1},~ M_{S} = 4 \, \text{GeV}\,.
\end{gather}
These benchmark points were selected based on an analysis of constraints from low-energy flavour physics and electroweak precision observables, as discussed in sec.~\ref{sec:flavour_all_paramspcs}. The corresponding values of the (c-)EDMs and (c-)MDMs for leptons and quarks at these two benchmark points are presented in table~\ref{tab:EDM_MDM_BPs}, where both (c-)EDMs and (c-)MDMs are considered at one-loop level. For all these observables, we have presented only the respective NP contributions. For the anomalous magnetic moments, we report the new physics contribution, denoted by $\Delta a_{f}$, which we can compare with the difference $a_{f}^{\rm exp} - a_{f}^{\rm SM}$ wherever applicable. The SM predictions and experimental limits, provided in table~\ref{tab:EDM_MDM_values}, serve as a reference to assess the magnitude of potential new physics contributions for each of these observables. 

Note that in the predictions of the anomalous magnetic moment, we don't see much improvement compared to the respective SM predictions, and the current experimental limits are not useful for constraining the NP parameter spaces. Hence, for the combined analysis, we have not considered these data. Furthermore, the model makes significant contributions to all the EDMs, c-EDMs, and c-MDMs, and largely deviates from the respective SM predictions. However, among all these observables only  $|\mu_{e}^{E}|$, $|\mu_t^E|$, $|d_n|$, $\mu_t$ and $d_t$ are relevant to cosntraint the new physics parameter spaces. The rest are several orders of magnitude smaller than current experimental bounds. Hence, to constrain the NP parameter space, we have considered only the above-mentioned four dipole moments, which we have discussed in the main text.    

\section{Contribution in the Co-annihilation Channel}\label{secappendix_coannihilation}

In our scenario, the co-annihilation channel $ S S \to f \bar{f} $ can still indirectly influence the relic abundance. Details on co-annihilations can be found in \cite{Baker:2015qna, Edsjo:1997bg}. Following this, we have given the numerical contribution from each process to understand the degree of suppression. As pointed out in \cite{Baker:2015qna}, when the DM $\chi $ and the mediator $S$ are in chemical equilibrium, the effective cross-section of DM annihilation to be used in the Boltzmann equation can be found as: 
\begin{equation}
	\sigma_{\rm eff} = \frac{g_{\chi}}{g_{\rm eff}} \left[ \sigma_{\chi \bar{\chi} \to \text{all}} + \sigma_{SS \to \text{ all} } \frac{g_{S}^2}{g_{\chi}^2} \left(1+\Delta \right)^3\, \text{e}^{-2 x \Delta}\right]\,.
\end{equation}
The parameter $\Delta$ defines the mass splitting between the DM and the mediator and is defined as: $\Delta = \frac{ M_{S} - M_{\chi}}{ M_{\chi}}$. $g_{\rm eff} $ is given by: $g_{eff} = g_{\chi} + g_{S} (1 + \Delta)^{3/2} \text{e}^{-x \Delta}$.
Note that the contribution from co-annihilation depends on the mass difference between DM and the mediator $\Delta$. So, the effect of co-annihilation will only be effective when $\Delta$ is very small. 
We have calculated the thermal averaged cross-section for each type of process (annihilation and co-annihilation) and in the following comparison for a few benchmark points.
\begin{table}[htbp]
	\centering
	\begin{tabular}{|c|c|c|}
		\hline \hline
		& \multicolumn{2}{c|}{Cross-section ($\rm GeV^{-2}$)} \\ 
		\multirow{-2}{*}{Benchmark Points} & $\langle \sigma \, v\rangle_{\chi \bar\chi \to (SS + f \bar{f} )} $ & $\langle \sigma \, v\rangle_{S S \to f \bar{f} }$  \\ \hline
		$(M_{\chi},M_{S}) = (2,\, 2.5 ) $ GeV,   & & \\
		$(c_{s},c_p) = (0.005, 0.003) \rm \, GeV^{-1}$, $ (c_{s \chi} , c_{p\chi}) = (0.015,0.01) $ & \multirow{-2}{*}{$9.34 \times 10^{-11}$} & \multirow{-2}{*}{$3.92 \times 10^{-15}$} \\
		\hline
		$(M_{\chi},M_{S}) = (1.5,\, 5 ) $ GeV,   & & \\
		$(c_{s},c_p) = (0.006, 0.005) \rm \, GeV^{-1}$, $ (c_{s \chi} , c_{p\chi}) = (0.001,0.001) $ & \multirow{-2}{*}{$1.57 \times 10^{-13}$} & \multirow{-2}{*}{$1.76 \times 10^{-57}$} \\ \hline
		$(M_{\chi},M_{S}) = (2.5,\, 3 ) $ GeV,   & & \\
		$(c_{s},c_p) = (0.006, 0.005) \rm \, GeV^{-1}$, $ (c_{s \chi} , c_{p\chi}) = (0.001,0.01) $ & \multirow{-2}{*}{$1.27 \times 10^{-10}$} & \multirow{-2}{*}{$6.94 \times 10^{-14}$} \\ \hline
	\end{tabular}
	\caption{Comparison of the DM annihilation cross-section to the co-annihilation cross-section numerically. }
	\label{tab:DM_annihilation_coannihilation}
\end{table}

Table~\ref{tab:DM_annihilation_coannihilation} presents a comparative analysis of the annihilation and co-annihilation cross-sections for several benchmark points. It is evident from the table that, for a given benchmark scenario, the co-annihilation cross-section is significantly suppressed relative to the dark matter annihilation channels.

\section{DM-electron cross-section}\label{appendix:DM_electron}
In the low-mass region of the DM, another significant direct detection process would be the DM-electron scattering cross-section. As mentioned in sec.~\ref{sec:DM_theory}, the bound and projected bound on the scattering cross-section is provided by various current and future experiments. The amplitude square to calculate the scattering cross-section is given as follows:
\begin{align}\label{eq:DM_elec_ampsq}
|\mathcal{M}|^2=&\frac{m_e^2 \left(c_p^2 \left(-2 m_e^2-2 M_{\chi }^2+s+t\right)+c_s^2 \left(2 m_e^2-2 M_{\chi }^2+s+t\right)\right) }{\left(-2 m_e^2+M_S^2-2 M_{\chi }^2+s+t\right){}^2} \times  \nonumber \\& 
\left(\left(c_{p \chi }^2+c_{s \chi }^2\right) \left(-2 m_e^2+s+t\right)+2 M_{\chi }^2 \left(c_{s \chi }^2-c_{p \chi }^2\right)\right).
\end{align}
The cross-section will be given by \cite{Essig:2011nj,Kopp:2009et}:
\begin{equation}\label{eq:DM_elec_crsstn}
\sigma_{\chi e \to \chi e } = \frac{\mu_{\chi e}^2 }{16 \pi M_{\chi}^2 m_{e}^2   } \overline{ \left| \mathcal{M}_{\chi e} (q)\right|^2 }_{q^2 = \alpha^2 m_{e}^2  }\,,
\end{equation}
where, 
\begin{equation}\label{eq:DM_e_Mbar}
\overline{ \left| \mathcal{M}_{\chi e} (q)\right|^2 } =\overline{ \left| \mathcal{M}_{\chi e} (q)\right|^2 }_{q^2 = \alpha^2 m_{e}^2  } \, \times \, \left| F_{DM}(q) \right|^2 \,,
\end{equation}	
$ \mu_{\chi e} $ is the reduced mass of DM and electron. $ \overline{ \left| \mathcal{M}_{\chi e} (q)\right|^2 } $ is the squared matrix element averaged over the initial and summed over the final particle spin. The upper bound is provided on either the observable $ \sigma_{\chi e \to \chi e} \left| F_{DM}(q) \right|^2  $ or by taking $ F_{DM}(q)  =1 $. 
\begin{table}[htb!]
 \centering
    \rowcolors{1}{cyan!10}{lime!5}
    \renewcommand{\arraystretch}{1.9}
    \resizebox{0.8\textwidth}{!}{%
    \begin{tabular}{|c|c|c|}
    \hline
     & BP  & $\sigma_{e} \, \rm (cm^2)$   \\
    \hline
    \hline
    1. & $M_{\chi} = 1 {\, \rm GeV}, M_{S} = 8 { \, \rm GeV} , c_{s(p)} = 3 (4) \times 10^{-3} \, {\rm GeV^{-1}}, \, c_{s(p)\chi} = 1 $    & $2.55 \times 10^{-44} \rm \, cm^{2}$ \\
    \hline
    2. & $M_{\chi} = 2.5 {\, \rm GeV}, M_{S} = 6 { \, \rm GeV} , c_{s(p)} = 1 (2) \times 10^{-3} \, {\rm GeV^{-1}}, \, c_{s(p)\chi} = 1 $    & $1.43 \times 10^{-43} \rm \, cm^{2}$ \\
    \hline
    3. & $M_{\chi} = 2.5 {\, \rm GeV}, M_{S} = 1 { \, \rm GeV} , c_{s(p)} = 0.1 (0.1) \times 10^{-3} \, {\rm GeV^{-1}}, \, c_{s(p)\chi} = 1 $    & $ 3.23 \times 10^{-44} \rm \, cm^{2}$ \\
    \hline
    4. & $M_{\chi} = 0.8 {\, \rm GeV}, M_{S} = 1 { \, \rm GeV} , c_{s(p)} = 0.5 (1) \times 10^{-6} \, {\rm GeV^{-1}}, \, c_{s(p)\chi} = 1 $    & $ 2.48 \times 10^{-47} \rm \, cm^{2}$ \\
    \hline
    \end{tabular}}
    \caption{DM-e scattering cross-section for a few benchmark points. }
    \label{tab:DM_e_scattering}
\end{table}

\bibliographystyle{ieeetr} 
	\bibliography{spin0_lowmass.bib}		
	

\end{document}